\documentclass[aps,pre,twocolumn,eqsecnum]{revtex4} 
\usepackage{graphicx}
\graphicspath{{Fig/dc12/}{Fig/}{.}}
\usepackage{amsmath}
\usepackage{amssymb}
\usepackage{CJK}
\usepackage{rev2002a}
\usepackage{macro2012o}
\usepackage{chi4s1301}
\newcommand{\nd}{{n_{\text{d}}}}
\newcommand{\Vmax}{{V_{\text{max}}}}
\newcommand{\tauB}{{\tau_{\text{B}}}}
\newcommand{\dtF}{\Delta{t}_{\text{f}}}
\newcommand{\Tw}{\mathcal{T}_{\text{w}}}
\newcommand{\tmax}{t_{\text{max}}}
\newcommand{\diScalar}{\tilde{d}}
\newcommand{\xirr}{\chi^{\text{irr}}}
\renewcommand{\Ms}{M_{\text{S}}}
\begin{document}
\begin{CJK*}{JIS}{}  
\title{Analytical Calculation of Four-Point Correlations\\
for a Simple Model of Cages Involving Numerous Particles}
\author{\surname{Ooshida} Takeshi}
\email[E-mail:~]{ooshida@damp.tottori-u.ac.jp}
\affiliation{%
   Department of Mechanical and Aerospace Engineering,
   Tottori University, Tottori 680-8552, Japan}
\author{Susumu \surname{Goto}}
\affiliation{%
   Graduate School of Engineering Science,  
   Osaka University, 
   \relax{Toyonaka, Osaka 560-8531, Japan}}
\author{Takeshi \surname{Matsumoto}}
\affiliation{%
   Division of Physics and Astronomy, Graduate School of Science, 
   Kyoto University, 
   \relax{Kyoto 606-8502, Japan}}
\author{Akio \surname{Nakahara}}
\affiliation{%
   Laboratory of Physics, College of Science and Technology,
   Nihon University, 
   \relax{Funabashi, Chiba 274-8501, Japan}}
\author{Michio \surname{Otsuki}}
\thanks{%
   Present address:
   Department of Materials Science, 
   Shimane University, 
   Matsue 690-8504, Japan}
\affiliation{%
   Department of Physics and Mathematics, Aoyama Gakuin University, 
   \relax{Sagamihara, Kanagawa 229-8558, Japan}}

\date{\today}

\begin{abstract}
  Dynamics of a one-dimensional system of Brownian particles
    with short-range repulsive interaction (diameter $\sigma$)
    is studied with a liquid-theoretical approach.
  The mean square displacement, the two-particle displacement correlation,
    and the overlap-density-based generalized susceptibility
    are calculated analytically
    by way of the Lagrangian correlation of the interparticulate space,
    instead of the Eulerian correlation of density
    that is commonly used in the standard mode-coupling theory.
  In regard to the mean square displacement,
    the linear analysis reproduces the established result
    on the asymptotic sub\-diffusive behavior of the system.
  A finite-time correction is given
    by incorporating the effect of entropic nonlinearity  
    with a Lagrangian version of mode-coupling theory.
  The notorious difficulty in derivation of the mode-coupling theory
    concerning violation of the fluctuation-dissipation theorem
    is found to disappear by virtue of the Lagrangian description.
  The Lagrangian description also facilitates
    analytical calculation of four-point correlations in the space-time,
    such as the two-particle displacement correlation.
  The two-particle displacement correlation,
    which is asymptotically self-similar in the space-time,
    illustrates how the cage effect
    confines each particle within a short radius on one hand
    and creates collective motion of numerous particles
    on the other hand.
  As the time elapses,
    the correlation length grows unlimitedly,
    and the generalized susceptibility based on the overlap density
    converges to a finite value
    which is an increasing function of the density.
  The distribution function
    behind these dynamical four-point correlations
    and its extension to three-dimensional cases,
    respecting the tensorial character
    of the two-particle displacement correlation,
    are also discussed.
\end{abstract}

\maketitle
\end{CJK*}


\section{Introduction}
\label{sec:intro}

Confined dynamics of Brownian particles
  has been studied for many reasons,
  such as its relevance to micro\-fluidic devices 
  \cite{Squires.RMP77,Hale.NanoL1}, 
  molecular biology \cite{Kusumi.BPJ65,Cell5.Book2007}, 
  and energetics of micro\-machines 
  \cite{Sekimoto.Book2010,Hanggi.RMP81}.
Most notably, 
  the problems are intriguing 
  because the confinement makes
  even the simplest cases non-trivial,
  not to speak of more challenging cases 
  in which the particle--particle interaction 
  comes into play.
The simplest and apparently easier cases 
  are exemplified 
  by diffusion of non-interacting Brownian particles 
  in a cylindrical pore with a varying cross section
  \cite{Burada.CPhC10}.
The diffusive dynamics is then described 
  by a spatially one-dimensional Fokker--Planck equation 
  for particles in a rugged free-energy landscape
  \cite{Zwanzig.JPhysChem96}.
Due to the entropic nature 
  of this free-energy landscape,
  the diffusion under an external driving force
  exhibits a peculiar temperature dependence 
  \cite{Reguera.PRL96};
  it is also sensitive to the particle size, 
  which can be applied 
  to a design of a device for sorting particles 
  \cite{Reguera.PRL108}.

Interaction among the particles 
  makes the problem of confined dynamics a real challenge.
It means that the confinement 
  is caused by the particles themselves
  and the motion is thus slowed down,
  as if each particle is constrained 
  in a cage that consists of its neighbors.
This kind of mutual hindrance of motion,
  which has been studied 
  in connection with the glass transition 
  \cite{Berthier.RMP83}
  and now in a broader context \cite{Liu.Book2001},
  is known by the name of the \emph{cage effect}.  
    
To see how the slowdown of the dynamics due to the cage effect 
  is studied quantitatively,
  let us consider a dense colloidal suspension
  modeled as a system of interacting Brownian particles,
  denoting the position vector of the $j$-th particle 
  with $\mb{r}_j(t)$.
The slow dynamics is studied 
  by defining the (particle-scale) density field as 
  \begin{equation}
    \rho(\mb{r},t) 
    = \sum_j \delta(\mb{r}-\mb{r}_j(t))
    = \rho_0 
    + \sum_{\mb{k}}\hat\rho(\mb{k},t) e^{-\II\mb{k}\cdot\mb{r}}
    \notag 
  \end{equation}
  and focusing on its correlations,
  such as the intermediate scattering function
  (the dynamical structure factor),
  $F(k,t) 
  \propto \Av{{\hat\rho(\mb{k},t)}{\hat\rho(-\mb{k},0)}}$,
  with $\hat\rho$ 
  denoting the Fourier component of the density field.
As the mean density $\rho_0$ increases 
  or the temperature $T$ decreases, 
  the cages have stronger effect,
  which results in the extremely slow relaxation 
  of $F(k,t)$.
This behavior of $F(k,t)$ 
  has been reproduced theoretically,
  at least to some extent, 
  by the \emph{mode-coupling theory} (MCT) 
  \cite{Gotze.RPP55,Goetze.Book2009,Reichman.JStat2005},
  which consists in the derivation 
  of an equation for $F(k,t)$
  in the form of an integro-differential equation,
\begin{equation}
  \left(\dt + \Dc k^2\right)  F(k,t) 
  = -\int_0^t \D{t'} M(k,t-t') \partial_{t'} F(k,t') 
  \label{MCT.F},
\end{equation}
  with $\Dc$ denoting the collective diffusion constant;
  the cage effect is incorporated 
  via the memory kernel $M$
  which is a quadratic functional of $F$.
In spite of this success, however,
  MCT suffers from several difficulties
  and has its own limitations
  \cite{Reichman.JStat2005,Berthier.RMP83}.
Since theoretical understanding of glassy dynamics 
  still remains far from being resolved,
  a methodological insight 
  into kinetic approaches to glassy systems,
  which will permit an improvement 
  over the existing theories such as MCT,
  is highly desired.

It is one aspect of the cage effect 
  that each particle is confined within a short radius, 
  while it has another aspect 
  that concerns long length scales.
The slowdown of the glassy dynamics
  is now regarded as attributable 
  to dynamical heterogeneity 
  \cite{Yamamoto.PRE58,Berthier.Book2011},
  which refers to the presence of collective motion 
  with some lifetime and correlation length.
Unfortunately, kinetic-theoretical treatment 
  of this collective motion is formidably difficult, 
  as its correct description requires 
  a four-point space-time correlation,
  such as $\chi_4$
  that will be explained later 
  [see Eq.~(\ref{chi4=}) in Sec.~\ref{sec:chi}].
As far as we know,
  analytical calculation of four-point correlations
  has been infeasible 
  except for some special cases
  such as linear elastic bodies
  and kinetically constrained models 
  on a lattice~\cite{Toninelli.PRE71}.
In regard to MCT, 
  we must emphasize here 
  that MCT targets on the dynamical structure factor $F(k,t)$
  and not on four-point correlations.
As long as the standard variables 
  such as $\hat\rho(\mb{k},t)$ are used, 
  a \emph{four-point} correlation function 
  implies a \emph{four-body} correlation.
Since MCT is a closure theory 
  in which quadruple (four-body) correlations 
  are approximated by products of $F$, 
  it is unlikely to describe four-point correlations
  accurately.
Although 
  a calculation of three-point correlation
  within the MCT approximation
  was recently reported \cite{Biroli.PRL97},
  still $\chi_4$ remains insurmountable.
  
A breakthrough may be found, 
  through a profound study of a simpler system,
  by developing a method 
  that can describe 
  the two aspects of the cage effect simultaneously,
  namely the short-ranged particle interaction 
  and the long-ranged dynamical correlation.
To make progress in this direction,
  here we develop a nonlinear theory 
  for a one-dimensional system 
  of interacting Brownian particles:
\begin{equation}
 m \ddot{X}_i
 = -\mu \dot{X}_i 
 - \frac{\partial }{\partial X_i} \sum_{j<k} V(X_k - X_j) 
 + \mu f_i(t)
 \label{Langevin.X},
\end{equation}
  whose behavior is known by the name of 
  \emph{single-file diffusion} (SFD)~%
  \cite{Harris.JAP2,Jepsen.JMP6,%
  Fedders.PRB17,Alexander.PRB18,van-Beijeren.PRB28,%
  Hahn.JPhA28,Kollmann.PRL90,%
  Taloni.PRL96,Barkai.PRL102,Barkai.PRE81,Ooshida.JPSJ80,%
  Delfau.PRE85},
  and which has been studied 
  also as a model of glassy dynamics with ideal cages
  \cite{Rallison.JFM186,Lefevre.PRE72,Abel.PNAS106}.
In the Langevin equation (\ref{Langevin.X}),
  $X_i$ represents the position of the $i$-th particle,
  and the meaning of the other symbols should be self-evident.
Without the interaction ($V=0$),
  the statistically averaged or coarse-grained density field 
  would be governed by the simple diffusion equation,
  with the diffusion constant $D=\kT/\mu$.
Here we adopt for $V$
  a \emph{short-ranged} repulsive potential,
  such as Eq.~(\ref{V=}) 
  in Appendix~\ref{app:num},
  so that the system exhibits a liquid-like structure factor
  \cite{Lutz.PRL93}.
The statistics of the random forcing, $\mu f_i(t)$,
  are given by Eq.~(\ref{f1}).

As will be clarified below, 
  we propose to contribute two things 
  to the theory of SFD.
Firstly, we establish a framework 
  for systematic treatment of the nonlinear effect 
  of density fluctuations
  in the form of an MCT-like equation,
  which gives a correction to the asymptotic theory.
In other words, 
  we demonstrate how to take into account
  the free-energy landscape of the system
  beyond the linear (harmonic) approximation~\cite{Lizana.PRE81}.
Secondly, 
  we calculated 
  some four-point space-time correlations in SFD, 
  hoping that they will be useful 
  as tools to quantify collective dynamics.
These two concepts, 
  namely MCT and four-point space-time correlations,
  are imported from the theory of glassy liquids,
  but in the original context 
  of three-dimensional systems
  it has been difficult to combine them.
In a simpler problem of SFD,
  we can demonstrate how they should be combined,
  so that the result will be exported 
  back to the theory of glassy liquids in the near future.

\begin{figure}  
 \includegraphics[clip,width=0.4\textwidth]{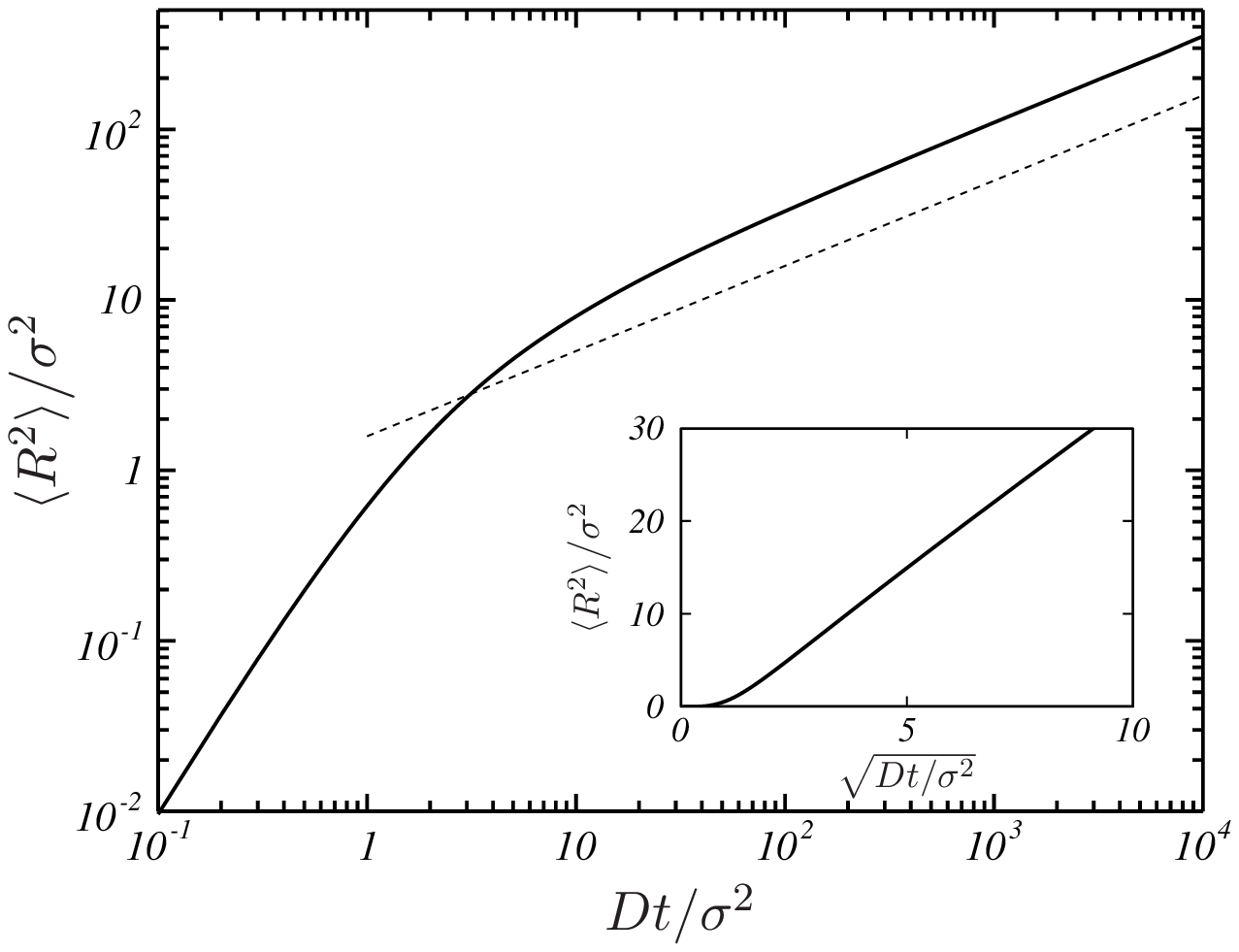}
 \caption{\label{Fig:MSD}%
   The mean square displacement $\Av{R^2}$ in SFD,
   plotted as a function of $t$.
   The dotted line indicates the slope for $t^{1/2}$.
   The system size is specified as $N = 2^{15} = 32768$
   and $\rho_0 = N/L = 0.25\,\sigma^{-1}$.
   In the inset, the same $\Av{R^2}$
   is plotted against $\sqrt{t}$
   for $0 < \sqrt{Dt} < 10\,\sigma$.
 }
\end{figure}

Before clarifying the key idea for these tasks,
  we need to notice 
  that the four-point correlation 
  indicates collective motion
  associated with the slowness of SFD. 
By ``slow'' 
  we mean that SFD is sub\-diffusive~%
  \cite{Alexander.PRB18,Hahn.JPhA28,Kollmann.PRL90}:
  in regard to long-time behavior 
  of the mean square displacement (MSD),
  denoted by $\Av{R_j^2}$ with 
\begin{equation}
  R_j = R_j(t) = X_j(t) - X_j(0)  \label{R=},
\end{equation}
  it is known exactly 
  that $\Av{R_j^2}$ for SFD behaves like $\sqrt{t}$
  [see Eq.~(\ref{Kollmann}) shown later],
  which is slower 
  than $\Av{R_j^2} \propto t$ 
  expected for the normal diffusion.
The subscript in $\Av{R_j^2}$ can be omitted 
  as we assume that the system is statistically uniform.
The sub\-diffusive law, $\Av{R^2} \propto \sqrt{t}$,
  is readily confirmed 
  by simulation of a system with $N$ particles
  in a periodic box of the size $L$.
In the calculations shown in Fig.~\ref{Fig:MSD},
  the mean density, $\rho_0 = N/L$,
  equals $0.25\,\sigma^{-1}$,
  and the system is statistically steady.
For computational details,
  see Appendix~\ref{app:num}.
The finite size effect is eliminated 
  by taking sufficiently large $L$ 
  and interpreting the word ``long-time''  
  as $\rho_0^{-2} \ll Dt \ll L^2\to \infty$.
For the long-time regime in this sense, 
  the asymptotic MSD is given by
\begin{equation}
  \Av{R^2} 
  = \frac{2S}{\rho_0} \sqrt{\frac{{\Dc}t}{\pi}} 
  \propto {t^{1/2}}  
  \label{Kollmann},
\end{equation}
  where 
  $S = S(0)$ denotes the long wave limiting value 
  of the static structure factor $S(k)$.
For particles with a well-defined diameter $\sigma$,
  an equation equivalent to Eq.~(\ref{Kollmann})
  was derived by Hahn and K{\"a}rger \cite{Hahn.JPhA28}.
Later, Kollmann~\cite{Kollmann.PRL90} demonstrated 
  that Eq.~(\ref{Kollmann}) holds 
  for systems with arbitrary interaction potential,
  as long as the range of the interaction is finite.
We note 
  that Kollmann also needed to calculate 
  a kind of four-point correlation
  ($\kappa^{(2)}$ in his notation)
  in derivation of Eq.~(\ref{Kollmann}).

The slowness of SFD 
  is ultimately due to the presence 
  of the repulsive potential term in Eq.~(\ref{Langevin.X}).
Then there is a question:
  Does a straightforward application of MCT 
  to Eq.~(\ref{Langevin.X})
  reproduce the sub\-diffusive law~(\ref{Kollmann})?
Unfortunately, 
  the answer is negative.
The MCT equation for the tagged particle,
  in any spatial dimension, 
  reads as
\begin{equation}
  \left(\dt + D k^2\right)  \Fs(k,t) 
  = -\int_0^t \D{t'} \Ms(k,t-t') \partial_{t'} \Fs(k,t') 
  \label{MCT.Fs},
\end{equation}
  where 
\[
  \Fs(k,t) 
  = \Av{ \hat\rho_j(\mb{k},t) \hat\rho_j(-\mb{k},0) } 
  = \Av{ e^{-\II\mb{k}\cdot\mb{R}_j} },
\]
  with $\mb{R}_j = \mb{r}_j(t) - \mb{r}_j(0)$,
  is the self part 
  of the intermediate scattering function $F$,
  and $\Ms$ is the memory kernel
  which is a bilinear functional of $F$ and $\Fs$.
The MSD is given by expanding $\Fs$ in power series of $k$, 
  as $\Fs = 1 - {\frac12}k^2 \Av{R^2} + \cdots$.
The asymptotic behavior 
  of MCT equations (\ref{MCT.F}) and (\ref{MCT.Fs})
  with standard memory kernels,
  in any spatial dimension,
  is mostly the same as the schematic MCT equation,
  whose asymptotic solutions
  are either arrested or subject to an exponential decay
  \cite{Goetze.Book2009}.
This implies
  that the dynamical constraint 
  by the long-lived cages in SFD
  is not described accurately enough by the standard MCT.
Essentially the same difficulty occurs 
  in MCT for rod polymers \cite{Miyazaki.JCP117},
  as it predicts 
  that the self-diffusion coefficient becomes isotropic 
  for large aspect ratio
  and therefore fails to describe the entanglement effects.
However, 
  in this particular case \cite{Miyazaki.JCP117}, 
  an alternative version of MCT-like kinetic theory 
  can be developed.
Using SFD as an illustrative example,
  Miyazaki \cite{Miyazaki.Bussei88,Miyazaki.JPS08s} suggested 
  that a scheme for improved treatment of four-point correlation 
  should be sought for.

The key idea of the present study 
  for improved treatment of slow dynamics  
  is adoption of the \emph{label variable}
  \cite{Ooshida.JPSJ80}.
This is essentially 
  an application of the Lagrangian description
  in fluid mechanics 
  \cite{Landau.fluid,Bennett.Book2006}, 
  as opposed to the Eulerian, 
  to the Langevin equation for the density field.
The label-variable method allows us,
  on one hand,
  to calculate four-point correlations explicitly.
On the other hand, 
  we demonstrate 
  that an MCT-like nonlinear theory 
  for the fluctuation of $1/\rho$
  can be developed in a field-theoretical style,  
  without violating 
  the fluctuation--dissipation theorem (FDT).
This is possible 
  because the problem of the multiplicative noise 
  is naturally resolved by virtue of the label variable.
Using this version of MCT,
  we find the contribution of the memory term 
  to give a finite-time correction 
  to the asymptotic 
  Hahn--K{\"a}rger--Kollman law~(\ref{Kollmann}), 
  visible as a finite intercept
  of the asymprotic straight line on the $\sqrt{t}$-axis
  in the inset in Fig.~\ref{Fig:MSD}.

We obtain four-point correlations  
  by generalizing the calculation of MSD
  \cite{Ooshida.JPSJ80} 
  to the \emph{two-particle displacement correlation} (2pDC),
\begin{equation}  
  \Av{{R_i}{R_j}} 
  = \Av{
  \left[ X_i(t) - X_i(0) \right]
  \left[ X_j(t) - X_j(0) \right]\vphantom{0^0}}
  \label{RR=}.
\end{equation} 
Quantities analogous to 2pDC 
  have been studied by a number of researchers 
  with numerical data from molecular dynamics 
  of glassy liquids
  \cite{Muranaka.PRE51,Hiwatari.JNCS235,Donati.PRL82,Doliwa.PRE61}
  and with linear theories of generalized elastic model
  for systems such as fluctuating membranes
  \cite{Taloni.PRE82,Taloni.EPL97},
  and 2pDC is also the main ingredient 
  of the theory of $\chi_4$ for elastic waves 
  by Toninelli \textit{et al.} \cite{Toninelli.PRE71}.
Nevertheless, 
  analytical calculations of 2pDC 
  for ``bond\-less'' particle systems 
  have never been reported.
Here we calculate 
  2pDC \emph{analytically}
  in terms of generalizable and liquid-oriented concepts,
  so that the theory could be extended to truly bond\-less systems 
  in the near future.
At first, we calculate 2pDC 
  as a function of the elapsed time $t$ 
  and some properly defined label distance 
  which coincides with $|i-j|$ in the one-dimensional cases;
  and subsequently, 
  we show it to be re-expressible 
  as a function of $t$ 
  and the initial distance $\tilde{d} = X_j(0) - X_i(0)$,
  which we denote with $\chiR(\tilde{d},t)$.
For $i=j$,
  Eq.~(\ref{RR=}) reduces to MSD.
The two-particle correlation, 
  $\Av{{R_i}{R_j}}$ with $i\ne j$, 
  provides 
  an intuitive form 
  of four-point correlation 
  in comparison to $\chi_4$;
  the 2pDC with $i\ne j$ 
  vanishes for free Brownian particles 
  and, for SFD, evidences the cluster size 
  that behaves like $\rho_0 \sqrt{{\Dc}t}$,
  accounting for the slow diffusion. 
In addition, 
  from the knowledge of $\Av{{R_i}{R_j}}$,
  we can perform a fully analytical calculation 
  for the self part of $\chi_4$, 
  denoted with $\chiS$.
Reflecting 
  the eternity of the one-dimensional cages, 
  the long-time limiting value, $\chiS(+\infty)$, 
  is finite,
  as will be shown in Eq.~(\ref{chi4s.coll}).

The paper 
  is organized as follows:
After summarizing in Sec.~\ref{sec:cont}
  the idea of the continuous label-variable method 
  and some of its results, 
  we apply it to the calculation 
  of the two-particle displacement correlation 
  in Sec.~\ref{sec:4p}.
Strictly speaking,
  what we present in Sec.~\ref{sec:4p}
  is not $\Av{{R_i}{R_j}}$ itself 
  but its continuum equivalent, 
  calculated theoretically for the long-time regime.
Subsequently,
  in Sec.~\ref{sec:NL}
  we demonstrate 
  a systematic and FDT-preserving derivation 
  of an MCT-like equation 
  in the Lagrangian description.
The ``Lagrangian'' MCT equation 
  provides us with the finite-time correction to MSD
  and the two-particle displacement correlation.
With this finite-time correction taken into account,
  two different forms of four-point correlation functions
  are calculated in Sec.~\ref{sec:chi}:
  one is $\chiR(\tilde{d},t)$ 
  and the other is $\chiS(t)$.
We will discuss in Sec.~\ref{sec:discuss}
  how the collective dynamics is represented by $\chiS(t)$,
  what is the relation between the theories of SFD in the past 
  and the present one,
  and whither the method of the Lagrangian MCT
  may guide us in the future.
Section~\ref{sec:conc} 
  is allotted for concluding remarks.

\section{Continuum Theory of Single-File Diffusion}
\label{sec:cont}

Let us begin with summarizing 
  our previous results on MSD for SFD \cite{Ooshida.JPSJ80}.
By ``continuum theory'' we mean 
  that the theory is formulated 
  in terms of some hydrodynamic quantity 
  such as the density $\rho$, 
  rather than 
  direct treatment 
  of the particles. 
Our idea consists in adoption 
  of the continuous label variable $\xi$,
  which we take instead of the position $x$
  as the independent variable,
  and we also change the dependent variable
  from the density $\rho(x,t)$ 
  to the fluctuation of the particle interval,
  denoting it with $\psi(\xi,t)$.
On the basis of the correlation of $\psi$
  calculated for the long-time regime,
  we can re-derive Eq.~(\ref{Kollmann}).

Since our label-variable method 
  is intended as a reformulation of MCT,
  we start from essentially the same Langevin equation 
  as in the field-theoretical formulation of MCT 
  for dense colloidal suspension \cite{Dean.JPhAMG29}.
The Langevin equation,
  derived from Eq.~(\ref{Langevin.X})
  for the density $\rho(x,t) = \sum_j \rho_j$ 
  with $\rho_j = \delta(x-X_j(t))$
  and its flux $Q = \sum_j \rho_j \dot{X}_j$,
  is given as follows:
\begin{subequations}%
  \begin{align}
    & \dt\rho + \dx{Q} = 0   
    \label{cont}, %
    \\
    & Q 
    = -D\,\left( \dx\rho + \frac{\rho}{\kT}\dx{U} \right)
    + \sum_j \rho_j(x,t) f_j(t) 
    \label{*Q}, 
    \\ 
    & U = U[\rho](x) = \int\!\D{x'}\,V(x-x')\rho(x')    
    \label{U2}. %
  \end{align}%
  \label{eqs:rho}%
\end{subequations}%
By eliminating $Q$ and introducing
\[
  \fRho(x,t) = -\dx \sum_j \rho_j(x,t) f_j(t),  
\]
  we write down the equation for $\rho(x,t)$ as
\begin{equation}
 \dt\rho(x,t) 
 = D\dx\left( \dx\rho + \frac{\rho}{\kT} \,\dx{U} \right) 
 + \fRho(x,t)  
 \label{*rho}
\end{equation}
  with $\dx{U} = \dx{U[\rho](x)}$ and
\begin{equation}
 \Av{\fRho(x,t) \fRho({x'},{t'})}
 = 2D \dx \partial_{x'} \rho(x,t) \delta(x-x') \delta(t-t')   
 \label{f2}.
\end{equation}
Note the presence of $\rho(x,t)$  
  on the right-hand side of Eq.~(\ref{f2}):
  the noise is multiplicative
  \cite{Dean.JPhAMG29,Miyazaki.JPA38,Andreanov.JStat2006,%
  Kim.JPhA40,Kim.JStat2008}.
A linear version of Eq.~(\ref{*rho})
  is sometimes referred to as the diffusion-noise equation 
  \cite{van-Kampen.Book2007},
  which was used to describe collective dynamics 
  of a single-file system \cite{Taloni.PRE78,note.fRho}.

As the particles have the excluded volume effect
  and therefore cannot overlap,
  the barrier expressed by $V$ must be infinitely high.
In the one-dimensional system,
  this barrier acts as a topological constraint,
  or the ``no overtaking'' rule in plain words,
  which keeps the order of the particles.
In MCT formulated for coarse-grained $\rho$, however, 
  $V$ must be replaced 
  with a finite effective potential,
  in a manner analogous to the dynamical density-functional theory 
  \cite{Kawasaki.PhysicaA208,Kawasaki.JStatPhys93}.
As a result, 
  MCT fails to incorporate 
  the ``no overtaking'' effect of $V$ properly.
An asymptotic analysis of MCT for the long-time limit 
  shows that a certain mathematical feature 
  of the MCT memory kernel (the presence or the absence 
  of the zero-frequency singularity 
  in Laplace transform of $\Ms$) 
  determines the asymptotic behavior of $\Av{R^2}$.
The memory kernels, $M$ and $\Ms$, 
  ultimately decay exponentially 
  on the liquid side of the theory,
  though anomalous diffusion 
  may occur transiently
  near the MCT transition point
  \cite{Krakoviack.PRE79,Schnyder.JPhysCM23}.
As a corollary,
  the one-dimensional version of Eq.~(\ref{MCT.Fs})
  gives the normal diffusion,
  contradicting to Eq.~(\ref{Kollmann})
  and suggesting 
  that the mathematical feature of the MCT kernel
  is not compatible with the ``no overtaking'' rule
  \cite{Miyazaki.Bussei88,Miyazaki.JPS08s}.

Thus the difficulty is located 
  in the memory kernel $\Ms$ in Eq.~(\ref{MCT.Fs}).
Consequently, 
  for an MCT-based nonlinear theory of SFD,
  there is a choice between two strategies.
The first one consists in amending the memory kernel $\Ms$,
  so that it has a proper singularity;
  this is the strategy 
  developed by Fedders~\cite{Fedders.PRB17} 
  and Abel \textit{et al.}~\cite{Abel.PNAS106}.
Alternatively,
  one may totally dispense with $\Ms$, 
  using only the collective MCT equation
  corresponding to Eq.~(\ref{MCT.F}),
  and replacing the tagged-particle MCT equation (\ref{MCT.Fs}) 
  by something that gives $\Av{R^2}$ without time integral.
Here we choose this alternative,
  which becomes possible 
  by importing the concept of Lagrangian correlation 
  from the theory of fluid turbulence 
  \cite{Kraichnan.PhF8,Kaneda.JFM107,Kida.JFM345}.
The idea consists in adoption 
  of the continuous label variable,
  which we denote with $\xi$.

\begin{figure}
  \includegraphics[clip,width=0.60\linewidth]{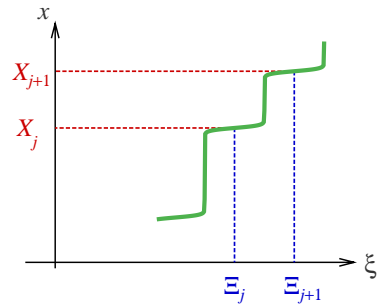} 
  \caption{\label{Fig:xi}%
    (Color online)
    A schematic view of the mapping from $\xi$ to $x$,
    given by inverting the function $\xi = \xi(x,t)$ 
    in Eq.~(\ref{xi.int}), with $t$ fixed arbitrarily.
    The label and the position of the $j$-th particle
    is denoted by $\Xi_j$ and $X_j$, respectively.
  }
\end{figure}

Though it is popular in continuum mechanics 
  to take the initial position of each material element 
  to label it,
  here we define $\xi$ in a different way, 
  avoiding to trace the whole history of the system 
  back to the initial configuration.
We construct the label function $\xi = \xi(x,t)$ 
  so as to satisfy the following three requirements:
\begin{enumerate}
 \item The label should satisfy the convective equation
       \begin{equation}
         (\rho \dt + Q\dx) \xi(x,t) = 0
         \label{rho*DtXi=0}.
       \end{equation}
 \item The label should be related 
       to the snapshot of $\rho$ and $Q$ 
       in such a way 
       that the continuity equation (\ref{cont}) is satisfied.
 \item The function needs to be invertible,
       in the sense 
       that a mapping from $(\xi,t)$ to $x=x(\xi,t)$ 
       should exist.
\end{enumerate}
To satisfy the second requirement, 
  we utilize the fact 
  that Poincar{\'e}'s lemma \cite{Spivak.Book1965}
  is applicable to the continuity equation (\ref{cont}),
  which guarantees the existence of $\xi$ such that  
\begin{equation}
  \dx\xi(x,t) = \rho, \quad
  \dt\xi(x,t) = -Q
  \label{L3+4};
\end{equation}
  in fact, 
  a solution to Eq.~(\ref{L3+4}) is explicitly given by 
\begin{equation}
 \xi = \xi(x,t) 
  = \int_{X_0(t)}^x \rho(x',t) \D{x'} + \text{const.}
  \label{xi.int}
\end{equation}
Then it is straightforward 
  to verify that all the three requirements are satisfied.
To be precise, 
  since $\rho$ consists of a sum of delta functions,
  the integral in Eq.~(\ref{xi.int}) 
  gives a multiple step function (see Fig.~\ref{Fig:xi}),
  which needs to be slightly smoothed 
  to justify the single-valued\-ness and the invertibility.
Although this smoothing could be interpreted 
  physically as a consequence of coarse-graining,
  here we refrain from involving ourselves 
  with such a delicate matter 
  and regard the smoothing 
  simply as a mathematical regularization.
We define $\Xi_j = \xi(X_j(t),t)$ for $j\in\Z$,
  taking it for granted 
  that the particles are numbered consecutively; 
  then, 
  with the smoothing and the integral constant 
  tuned appropriately, we have $\Xi_j = j$.

Using the label variable $\xi$ 
  as the spatial coordinate,
  now we rewrite the Langevin equation (\ref{eqs:rho}). 
The chain rule for the differential operators
  gives 
\begin{gather}
  \dx = \dd{\xi}{x} \dXi = \rho \dXi    , \quad    
  \left(\dt{~\cdot~}\right)_x 
  = \left(\dt{~\cdot~}\right)_\xi - Q\dXi \notag.  
\end{gather}
As a kinematic relation
  that replaces Eq.~(\ref{cont}),
  from the identity $\dt\dXi{x} = \dXi\dt{x}$
  we find 
\begin{equation}
  \dt\left[ \frac{1}{\rho(\xi,t)} \right]
  = \dXi\left( \frac{Q}{\rho} \right)
  \label{L7}.
\end{equation}
Then, rewriting Eq.~(\ref{*Q}) with $\dXi$
  and substituting it into Eq.~(\ref{L7}),
  we obtain 
\begin{align}
  \dt\left[ \frac{1}{\rho(\xi,t)} \right]
   &= -D\dXi\left( \dXi\rho + \frac{\rho}{\kT} \dXi{U} \right) 
   \notag\\&\qquad{}
   + \dXi \sum_j \delta\left( \xi - \Xi_j \right) f_j(t)
   \relax 
   \notag \\ 
  &= -D \dXi
  \left[ 
    \dXi\rho + 2 \sinh\left( \rho\sigma\dXi \right) \rho
  \right] 
  + \fL(\xi,t)  
  \label{*2+}, %
\end{align}
  where $V$ is replaced with the effective potential 
  as before,
  and $\fL$ satisfies
\begin{equation} 
  \Av{\fL(\xi,t)\fL(\xi',t')}   
  = 2D \dXi\dXiPrime 
   {\sum_i \delta\left( \xi - \Xi_i \right)}  
   \delta(\xi-\xi') \delta(t-t')
  \label{f3}.
\end{equation}

Having changed the independent variables 
  from $(x,t)$ to $(\xi,t)$,
  we change the dependent variable as well.
Introducing  
  the fluctuation of the particle interval,
\begin{equation}
  \psi = \psi(\xi,t) = \frac{\rho_0}{\rho(\xi,t)} - 1  
  \label{psi=},
\end{equation}
  and its Fourier modes 
\begin{equation}
 \check\psi(k,t) = {\frac1N} \int \D\xi\, e^{\II k\xi} \psi(\xi,t) 
 \quad
 \left(  \frac{k}{2\pi/N} \in \Z  \right)
 \label{m1},
\end{equation}
  we rewrite Eq.~(\ref{*2+}) in the form  
\begin{multline}
 \partial_t  \check\psi(k,t) 
  = 
  -D_* k^2 \left( 1 + \frac{2\sin \rho_0\sigma k}{k} \right) 
  \check\psi(k,t)  
  \\ {} 
  + \smash{\sum_{p+q+k=0}} 
  {\Vertex_k^{pq}} \check\psi(-p,t) \check\psi(-q,t)
  + O(\check\psi^3)
  \\ {}
  + \rho_0 \check{f}_{\text{L}}(k,t)
  \label{*3+}
\end{multline}
  with $D_* = \rho_0^2 D$ and
\begin{multline}
 \Vertex_k^{pq} \\{} 
 = D_* k^2 \left(
 1
 + \frac{k}{pq} \sin{\rho_0\sigma k}
 + \frac{p}{kq} \sin{\rho_0\sigma p} 
 + \frac{q}{kp} \sin{\rho_0\sigma q}  
 \right)
 \label{vertex}.
\end{multline}
The summation 
  is taken over all $(p,q)$ 
  satisfying the triad condition $p+q+k=0$ for given $k$
  \cite{note.triad}.
As for the statistics of the random force term, 
  Eq.~(\ref{f3}) is re-expressed as
\begin{equation}
  \rho_0^2 \Av{\CheckFL(k,t) \CheckFL(-k',t')}
  = \frac{2 D_*}{N} k^2 \delta_{kk'} \delta(t-t')  
  \label{f4};
\end{equation}
  see Endnote~\cite{note.f4}.
We also note 
  that the linearization of Eq.~(\ref{*3+}),
  corresponding to harmonization 
  of the effective interaction between the particles
  \cite{Doliwa.PRE61,Toninelli.PRE71,Lizana.PRE81,Delfau.PRE84}, 
  coincides with the one-dimensional version 
  of Edwards--Wilkinson equation 
  \cite{Edwards.PRSLA381,Lefevre.PRE72}
  and also with the Rouse model \cite{Doi.Book1986}.

To calculate MSD without employing $\Ms$, 
  we developed a formula for it
  in terms of the correlation 
  $\Av{\psi(\xi,t)\psi(\xi',0)}$,
  or its Fourier transform 
  $\Av{\check\psi(k,t)\check\psi(-k',0)}$.
In Fourier representation, 
  the formula reads as follows:
\begin{equation}
  \Av{R^2}
  = \frac{L^4}{{\pi}N^2} 
  \int_{-\infty}^{\infty}  
  \frac{\C(k,0) - \C(k,t)}{k^2} \D{k}
  \label{d8k}
\end{equation}
  where~\cite{note.factor-C} 
\begin{equation}
 \C(k,t) 
 = \frac{N}{L^2} \Av{\check\psi(k,t)\check\psi(-k,0)}
 \label{Corr=}.
\end{equation}
In Sec.~\ref{sec:4p},
  this formula will be re-derived 
  as a special case of Eq.~(\ref{AP+}).

Once the formula (\ref{d8k}) is derived, 
  all depends on the knowledge of $\C$.
In particular, 
  the long-time behavior of MSD
  is found by the linear analysis of Eq.~(\ref{*3+}).
From the linearized equation,
  the correlation $\C$ 
  is calculated as 
\begin{equation} 
 \C(k,t) 
 = \frac{S}{L^2} e^{-(D_*/S) k^2 t}, \quad
 S = S(k) \simeq S(0)  
 \label{d4psi}
\end{equation}
  with the aid of Eq.~(\ref{f4}).
On the other hand,
  linear analysis implies 
  that the difference 
  between the Eulerian and Lagrangian descriptions vanishes
  up to a trivial rescaling of the variables.
Thus
  the linearized version of Eq.~(\ref{d8k}) 
  corresponds to the approximate formula 
  by Alexander and Pincus~\cite{Alexander.PRB18},
\begin{equation}
  \Av{R^2} 
  \simeq \text{const.}\times 
  \int_{-\infty}^{\infty} \frac{F(q,0) - F(q,t)}{q^2} \D{q}
  \label{AP},
\end{equation}
which evidently reproduces 
  the sub\-diffusion law in Eq.~(\ref{Kollmann}).
For the special case of point\-like particles ($\sigma=0$), 
  $S$ in Eq.~(\ref{d4psi}) is replaced with unity;
  more generally,
  the linear formulation is also readily applicable 
  to systems with arbitrary interaction potential
  and, in this sense, turns out to be equivalent 
  to the theory of Kollmann \cite{Kollmann.PRL90}.
A refined treatment, 
  including the effect of the nonlinear term on $\C$,
  requires Eq.~(\ref{d8k}) instead of Eq.~(\ref{AP}).
We refer to Eq.~(\ref{d8k})
  as the modified Alexander--Pincus formula:
  the modification consists 
  in the adoption of the Lagrangian description.

In the linear (Edwards--Wilkinson) case,
  a $\nd$-dimensional version of Eq.~(\ref{AP}) has appeared 
  in the literature~\cite{Honda.PRE55,Toninelli.PRE71,Ikeda.JCP138};
  we will discuss it later in Sec.~\ref{sec:discuss},
  calling attention to some delicate points 
  about the extension to the $\nd$-dimensional liquid dynamics. 

\section{Four-point correlation}
\label{sec:4p}

\begin{figure*}
  \smash{(a)\quad}\par
  \includegraphics[width=13.5cm]{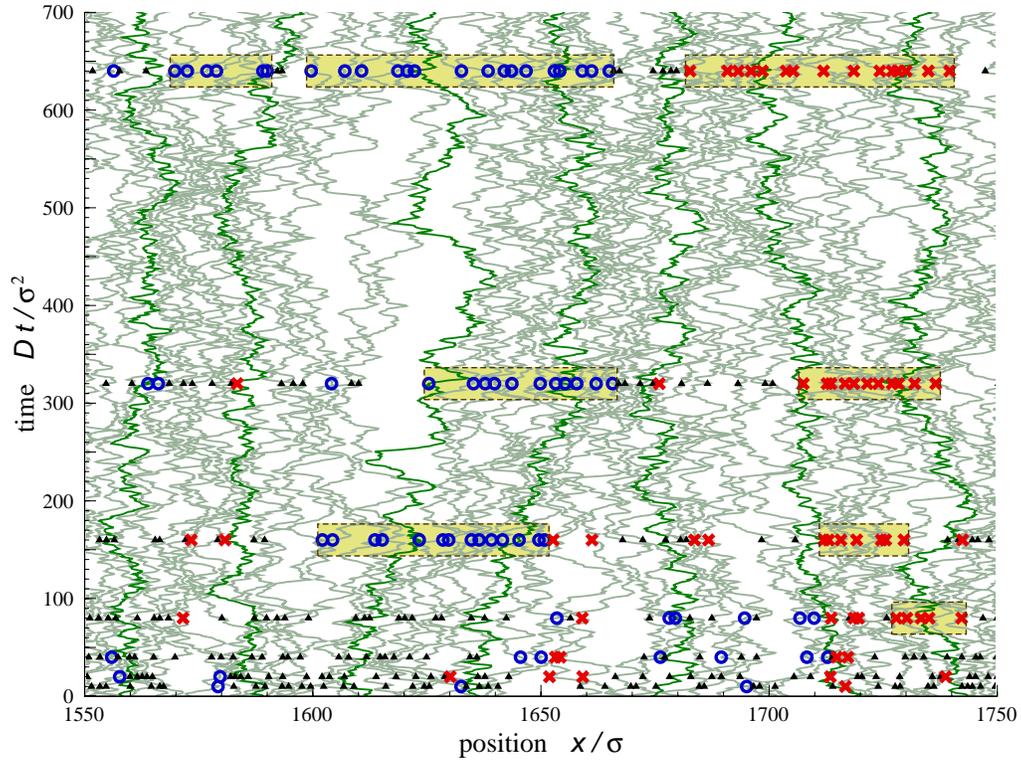}
  \par
  \smash{(b)\quad}\par
  \includegraphics[width=15.0cm]{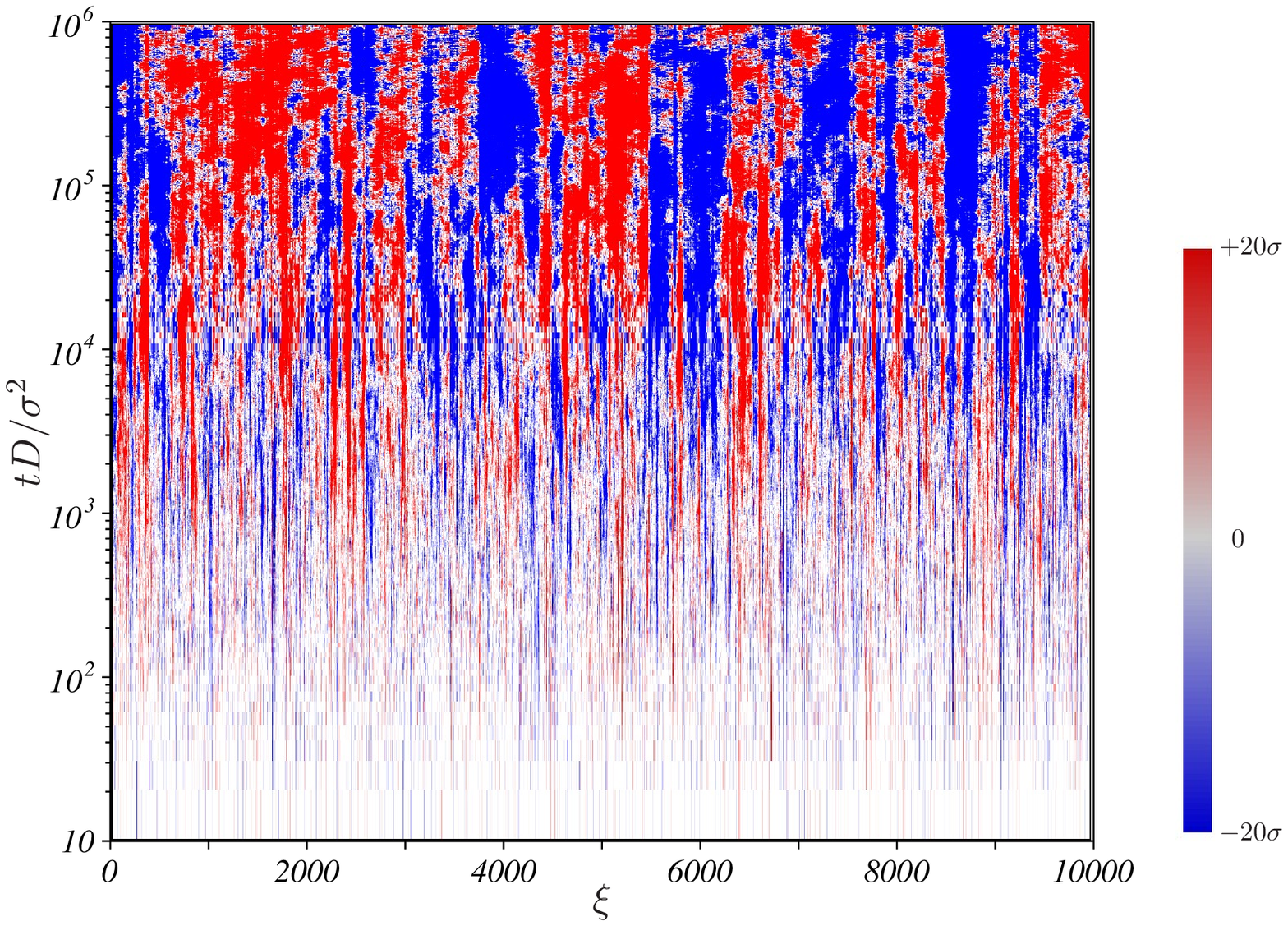}
  \caption{\label{Fig:SFD.cluster}%
    (Color online)
    Clusters in cooperative motion
    visualized in space-time diagrams of particles in SFD,
    calculated for $\rho_0 = N/L = 0.25\,\sigma^{-1}$. 
    (a) World\-lines of particles in the $(x,t)$-plane.
    The symbols $\bigcirc$ and $\times$
    mark particles displaced (by more than $5\sigma$) 
    rightward and leftward, respectively.
    The unmoving particles 
    are indicated with small triangles.
    Each cluster in cooperative motion,
    involving 5 particles at least,
    is highlighted in a box.
    (b) The displacement $R$ 
    depicted in gray scale (blue-to-red scale online) 
    as a function of $\xi$ and $t$.
    Unmoving particles are shown in white.
    }
\end{figure*}

\subsection{Cooperative motion in SFD}

The slow dynamics in SFD 
  is associated with collective motion of the particles.
This collective motion,
  obtained by numerical integration of Eq.~(\ref{Langevin.X}),
  is depicted in Fig.~\ref{Fig:SFD.cluster}.
In the numerical calculations 
  (see Appendix~\ref{app:num} for details),
  after the system has reached the thermal equilibrium,
  we choose some instance as $t=0$
  and record the ``initial'' position of each particle,
  say, $X_i(0)$.
To produce Fig.~\ref{Fig:SFD.cluster}(a), 
  at $t = 2^n\times 10\,\sigma^2/D$
  (with $n = 1, 2, \ldots$)
  we measured the displacement 
  $R_i(t) = X_i(t) - X_i(0)$ for each $i$.
If $R_i(t) > 5\sigma$, 
  we mark the position of the particle 
  with a circle ($\bigcirc$);
  if $R_i(t) < -5\sigma$, 
  we mark it with a cross ($\times$). 
As the time difference $t$ increases, 
  a string of the same kind of symbol is formed, 
  expressing a cluster of particles in cooperative motion.
While Fig.~\ref{Fig:SFD.cluster}(a) presents 
  a close-up for a relatively limited time span, 
  a long shot up to $t = 10^6 \sigma^2/D$ 
  is shown as Fig.~\ref{Fig:SFD.cluster}(b).
Formation of large clusters in cooperative motion
  is visible.
By the time difference $t = 10^6 \sigma^2/D$,
  a typical cluster
  reaches the size of several hundred particles,
  occupying a length on the order of $10^3\sigma$ (as 
  each particle is assigned a space of $1/\rho_0 = 4\sigma$).

Quantitative description of this collective motion 
  requires some four-point correlation functions,
  such as those used 
  for the analysis of dynamical heterogeneity
  in glassy systems.
In regard to SFD, 
  it seems especially natural 
  to consider the four-point correlation
  to deal with the topological constraint 
  of the ``no passing'' rule,
  as the requirement 
  that two world lines should not intersect 
  involves four points in the space-time,
  namely $X_i(0)$, $X_j(0)$, $X_i(t)$, and $X_j(t)$.
This is 
  why Miyazaki and Yethiraj \cite{Miyazaki.JCP117} 
  needed to calculate a four-point correlation 
  [$G(1,2;3,4)$ in their notation]
  to study the entanglements of rod polymers 
  within the framework of liquid state theory,
  and why 
  Abel \textit{et al.}~\cite{Abel.PNAS106},
  to improve on the conventional MCT 
  which fails in reproducing the sub\-diffusive law for SFD,
  had to examine the irreducible four-point propagator
  (denoted with $\xirr$).

Here we demonstrate 
  that the label-variable method 
  is also capable 
  of calculating a kind of four-point correlation,
  in the form of 
  two-particle displacement correlation.
Though it was originally presented 
  as $\Av{{R_i}{R_j}}$ in Eq.~(\ref{RR=}),
  there is no trouble 
  in replacing the particle numbering with the label variable.
Then, our task 
  is derivation of a formula 
  to calculate $\Av{R(\xi,t) R(\xi',t)}$
  with $R(\xi,t) = x(\xi,t) - x(\xi,0)$,
  which will be shown in the next subsection.

\subsection{Two-Particle Displacement Correlation}

Aiming for an analytical expression 
  for $\Av{ R(\xi,t) R(\xi',t) }$ in SFD,
  let us extend Eq.~(\ref{d8k})
  so that we can calculate $\Av{R(\xi,t)R(\xi',t)}$ 
  from the correlation $\C$ defined by Eq.~(\ref{Corr=}).
We start with noticing 
  that $Q/\rho$ stands for the velocity, 
  whose integral in regard to $t$ gives 
  the displacement of the particle labelled with $\xi$:
\begin{equation}
 R(\xi,t)  
 = 
 \int_0^t \D{t}' \frac{Q(\xi,t')}{\rho(\xi,t')} 
 \notag. 
\end{equation}
Into this equation 
  we substitute 
  $Q/\rho = \dXi^{-1}\dt(1/\rho)$,
  obtained from Eq.~(\ref{L7})
  upon integration over $\xi$,
  to find 
\begin{align}
 R(\xi,t) 
 &=  \dXi^{-1}\! 
 \left. \left( \frac{1+\psi}{\rho_0} \right) \right|_0^t
 \notag \\ 
 & = \frac{1}{\rho_0} 
 \sum_k   \frac{e^{-{\II}k\xi}}{-{\II}k}
 \left[  \check\psi(k,t) - \check\psi(k,0) \right]
 \label{R.psi+}.
\end{align}

Subsequently,
  we multiply Eq.~(\ref{R.psi+}) 
  by its duplicate with $(\xi,k)$ changed to $(\xi', -k')$,
  and take the statistical average.
The double summation on the right-hand side
  reduces to the single one, 
  if we assume 
  that the contribution from the terms with $k\ne k'$ 
  vanishes.
This is true in the linear case, 
  and also seems to be justifiable for nonlinear cases
  within the framework 
  of the direct-interaction approximation (explained later).
Thus we obtain a formula 
  allowing us to calculate 2pDC
  from $\Av{\psi\psi}$:
\begin{multline}
 \Av{R(\xi,t)R(\xi',t)}  \\ {}
 = \frac{L^4}{\pi N^2}  \int_{-\infty}^{\infty} \D{k}\,
  e^{-{\II}k(\xi-\xi')}  \frac{\C(k,0) - \C(k,t)}{k^2}
 \label{AP+}, 
\end{multline}
  with $\C(k,t)$ defined by Eq.~(\ref{Corr=}).
Note that Eq.~(\ref{AP+}) includes 
  the modified Alexander--Pincus formula (\ref{d8k})
  as a special case with $\xi=\xi'$,
  as it ought to be.
In this sense,
  the formula (\ref{AP+}) could be referred to 
  as the \emph{extended Alexander--Pincus formula}.
We emphasize 
  that Eq.~(\ref{AP+}) does not 
  rely on smallness of deformation,
  nor it requires such kind of approximation at all,
  as long as the Lagrangian description is strictly followed.

\subsection{Calculation of 2pDC: linear approximation}

For $\C$ in Eq.~(\ref{d4psi}) 
  calculated from the linear approximation
  of Eq.~(\ref{*3+}),
  the extended Alexander--Pincus formula (\ref{AP+}) 
  gives  
\begin{multline}
   \Av{ R(\xi,t) R(\xi',t) }  \\ {}
   = 
   \frac{2S}{\rho_0} \sqrt{\frac{{\Dc}t}{\pi}} 
   \exp\left[  - \frac{(\xi-\xi')^2}{4\rho_0^2 {\Dc}t} \right]  
    - \frac{S}{\rho_0^2} |\xi-\xi'|
    \erfc \frac{|\xi-\xi'|}{2\rho_0\sqrt{{\Dc}t}}
    \label{R1*R2} ,
\end{multline}
reproducing Eq.~(B3) in Ref.~\cite{Ooshida.JPSJ80}; 
  the same result was reported  
  in regard to the generalized elastic model \cite{Taloni.EPL97}.
This is expressible 
  in terms of a similarity variable 
\begin{equation}
  \theta = \dfrac{\xi-\xi'}{2\rho_0 \sqrt{{\Dc}t}}
  \label{th=}
\end{equation}
  as  
\begin{equation}
 \frac{\Av{ R(\xi,t) R(\xi',t) }}{\sigma \sqrt{{\Dc}t}} 
 = \frac{2S}{{\rho_0}\sigma}
 \left( 
   \frac{{~}e^{-\theta^2}\!}{\sqrt{\pi}} 
   - |\theta| \erfc|\theta| 
 \right)
 =
 \varphi(\theta)
 \label{R1*R2.sim} \tag{$\ref{R1*R2}'$}.
\end{equation}
From this similarity solution
  we can read the dynamical correlation length
\begin{equation}
  \lambda = \lambda(t) = 2\sqrt{{\Dc}t}
  \label{ld},
\end{equation}
  indicating the size $\lambda$ 
  of a cluster in a cooperative motion.
We have already seen such clusters
  in Fig.~\ref{Fig:SFD.cluster},
  though care should be taken in regard to the difference 
  that Eq.~(\ref{ld}) is a statistical law
  while Figs.~\ref{Fig:SFD.cluster}(a) and (b) 
  present a single run,
  and only a small portion of it 
  is shown in Fig.~\ref{Fig:SFD.cluster}(a). 
The dynamical correlation length $\ld$ in Eq.~(\ref{ld})
  is the diffusive one (with the exponent $1/2$),
  which may occur 
  also in different contexts,
  such as roughening of growing surfaces \cite{Krug.AdvPhys46}
  and kinetically constrained models 
  of defect-mediated glassy dynamics \cite{Toninelli.PRE71}.

\begin{figure}
  \includegraphics[clip,width=0.75\linewidth]{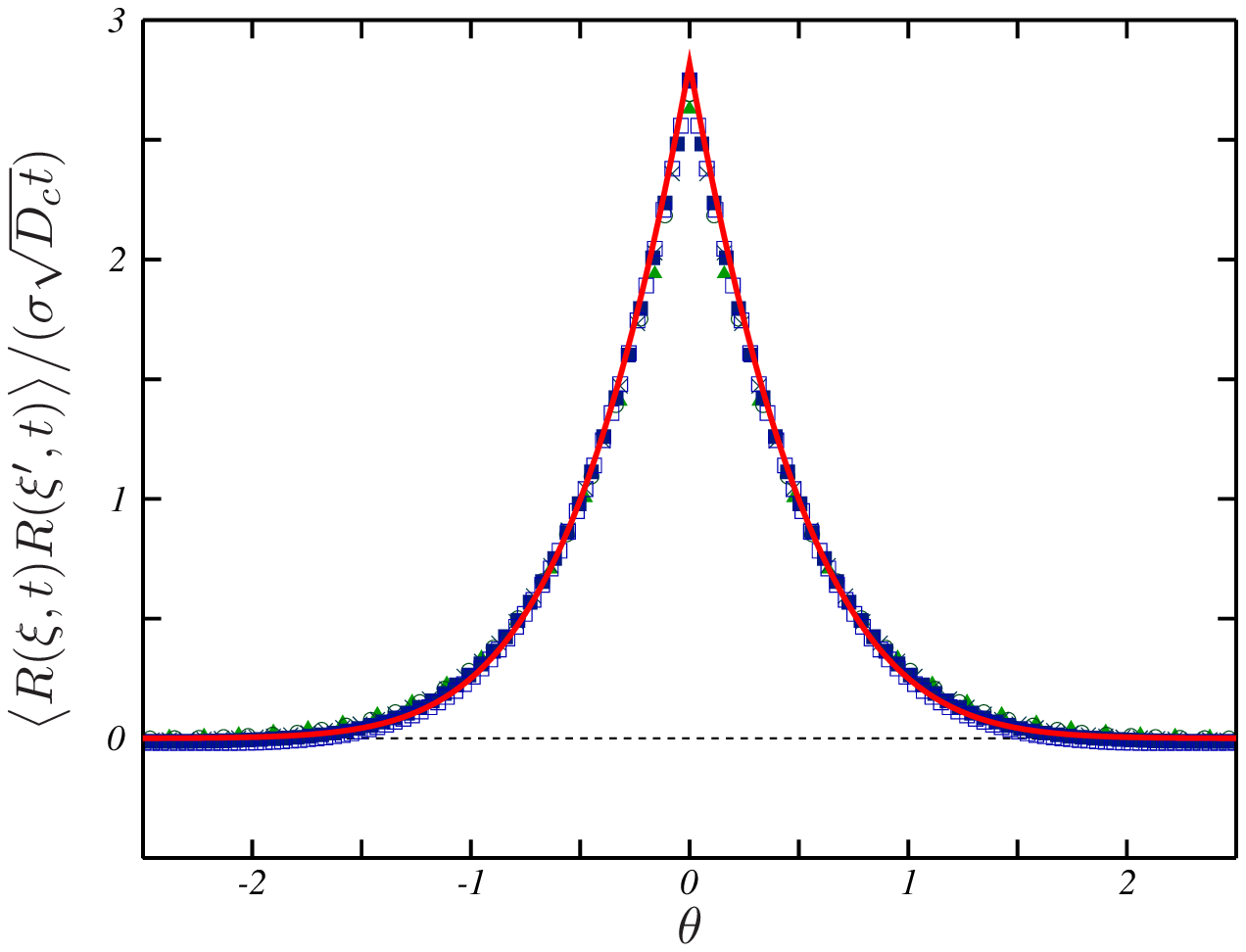} 
  \caption{\label{Fig:RR.sim}%
    (Color online)
    Comparison of Eq.~(\ref{R1*R2}) 
    with the computed 2pDC 
    in the same system as in Fig.~\ref{Fig:MSD}
    ($N = 2^{15} = 32768$, $\rho_0 = N/L = 0.25\,\sigma^{-1}$).
    The solid (red) line 
    shows the self-similar solution 
    in Eq.~(\protect\ref{R1*R2.sim}),
    while the simulation data 
    are plotted with symbols: 
    solid triangles 
    ($\blacktriangle$)         for $t = 100\,\sigma^2/D$, 
    open circles ($\bigcirc$)  for $t = 200\,\sigma^2/D$, 
    crosses        ($\times$)  for $t = 400\,\sigma^2/D$, 
    open squares   ($\square$) for $t = 800\,\sigma^2/D$, and 
    solid squares  ($\blacksquare$) for $t = 1600\,\sigma^2/D$.%
  }
\end{figure}

Equation~(\ref{R1*R2}) is compared 
  with statistical results of particle simulations 
  in Fig.~\ref{Fig:RR.sim}.
Except for the transient behavior
  slightly visible for $t = 100\,\sigma^2/D$, 
  all of the simulation results 
  are consistent with Eq.~(\ref{R1*R2}).
The transient behavior 
  can be studied by taking 
  the nonlinear terms in Eq.~(\ref{*3+}) into account,
  which will be discussed in the next section.

The variable for the horizontal axis of Fig.~\ref{Fig:RR.sim}
  requires some consideration.
As a quantity corresponding to $\xi-\xi'$,
  here we have taken 
  the distance in the label numbering, 
  say, $|i-j|$ between the $i$-th and $j$-th particle.
Though adequate in the present case,
  this is not convenient for extensions
  to multidimensional cases
  \cite{Donati.PRL82,Doliwa.PRE61},
  because the result may depend on the way of numbering.
A reasonable alternative
  in the particle simulation 
  is the initial distance such as $X_i(0) - X_j(0)$,
  which can be compared with the theoretical prediction 
  by assuming 
  that $(\xi-\xi')/\rho_0$ corresponds to $X_i(0) - X_j(0)$
  after statistical averaging.
Numerical calculations show 
  that this is indeed valid 
  as far as the long-time behavior is concerned, 
  but transiently there are additional modifications 
  due to triple correlations 
  such as 
  $\Av{{\check\psi(-p,0)}{\check\psi(-q,0)}{\check\psi(-k,t)}}$.
Before discussing these modifications, 
  let us develop a nonlinear closure theory for $\C$,
  which %
  introduces the entropic nonlinearity 
  into the theory of SFD,
  making the triple correlation 
  available as a byproduct.

\section{Nonlinear theory for finite-time effects}
\label{sec:NL}

\subsection{Inclusion of nonlinearity: DIA for SFD}

The expression for 2pDC in Eq.~(\ref{R1*R2})
  is based on the linear approximation to Eq.~(\ref{*3+}),
  which is only asymptotically valid 
  for sufficiently long time.
For finite values of $t$,
  there should be a correction to Eq.~(\ref{R1*R2})
  describing the transient behaviors of 2pDC and MSD;
  to find this correction,
  now we develop a nonlinear closure theory 
  for the correlation $\C$.

In short, what we present here and in the next subsection
  is a systematic derivation of MCT equation for $\C$.
Except for some minor (but important) differences, 
  this is analogous to the attempt of a number of authors 
  \cite{Das.PRA34,Schmitz.PRL71,Miyazaki.JPA38}
  who tried to re-derive MCT equation 
  with the Martin--Siggia--Rose (MSR) formalism 
  \cite{Martin.PRA8}.
They were obstructed by the problem 
  of inconsistency with FDT:
  this difficulty is inevitable  
  for dense colloidal suspensions or supercooled liquids,
  as long as one uses the standard MSR formalism 
  with the plain $\hat\rho(\mb{k},t)$ as the field variable
  \cite{Andreanov.JStat2006,Kim.JPhA40,Kim.JStat2008,Nishino.PRE78}.
Only some special classes of Langevin equations
  are free from this difficulty.
Two such classes are known: 
  one is the class of models 
  whose nonlinearity comes from the gradient 
  of the thermodynamic potential (entropy) alone,
  such as the $p$-spin model,
  referred to as ``Class I'' 
  by Miyazaki and Reichman \cite{Miyazaki.JPA38};
  and the other one (called ``Class II'')
  is exemplified 
  by liquid models with Gaussian approximation to the entropy
  \cite{Schmitz.PRL71,Miyazaki.JPA38}.
Fortunately, 
  our equation for $\check\psi$ belongs to Class I 
  and therefore consistency with FDT is expected. 

Let us return to the nonlinear Langevin equation (\ref{*3+})
  governing $\check\psi$, 
  with the random force statistics in Eq.~(\ref{f4}).
The correlation $\C$
  is then subject to an equation 
  containing triple correlations: 
\begin{multline}  
 \left( \dt + \frac{D_*}{S} k^2 \right) \C(k,t)  \\{}
 = 
 \frac{N}{L^2} \sum_{p+q+k=0} 
 \Vertex_k^{pq} 
 \Av{\check\psi(-p,t)\check\psi(-q,t)\check\psi(-k,0)}
 \notag,  
\end{multline}
  with the $O(\check\psi^3)$ term in Eq.~(\ref{*3+})
  discarded,
  and with $\sum_{p+q+k=0}$ denoting 
  the same summation over $(p,q)$ as in Eq.~(\ref{*3+}).
Note the absence of $\Av{\CheckFL{\check\psi}}$
  on the right-hand side;
  this term vanishes 
  because $\fL$ is not multiplicative,
  which means 
  that $\Av{\CheckFL(k,t) \CheckFL(-k',t')}$ 
  is independent of the $\check\psi$'s
  according to Eq.~(\ref{f4}).
To provide a closure to this equation, 
  we apply the formalism of 
  \emph{direct-interaction approximation} (DIA)
  \cite{Kraichnan.JFM5,Kraichnan.PhF8,Kaneda.JFM107,%
  Kida.JFM345,Goto.PhysicaD117}.
The procedure of the calculation
  is essentially the same 
  as that in Ref.~\cite{Matsumoto.PTPS195}
  and is explained briefly in Appendix~\ref{app:DIA}.
As a result, 
  we obtain a set of equations:
\begin{align}
 \left( \dt + \frac{D_*}{S} k^2 \right) \C(k,t)
 &  = \int_{t_0}^t \D{t'} M_G(k,t-t') \C(k,|t'|)  \notag \\ & \quad {}
 + \int_{t_0}^0 \D{t'} M_C(k,t-t') \bar{G}(-k,-t')
 \label{DIA.C}
 \\
 \left( \dt + \frac{D_*}{S} k^2 \right) \bar{G}(k,t)
 & = \int_0^t \D{t'} M_G(k,t-t')  \bar{G}(k,t')
 \label{DIA.G}
\end{align}
  for the correlation $\C$ and
\[
  \bar{G}(k, t-t') = \Av{G(k,t;k,t')} 
  = \Av{%
  \frac{\delta{\check\psi(k,t)}}{\delta{\check\psi(k,t')}}
  }
  \notag,
\]
  with $t_0\,(< 0)$ denoting the time 
  at which the ``direct interactions'' are switched off,
  and $M_G$ and $M_C$ are memory kernels 
  given by
\begin{align}
 & M_G(k,s) 
 = \frac{4L^2}{N} 
 \sum_{p+q+k=0} 
 \Vertex_k^{pq} \Vertex_q^{pk} \C(p,s) \bar{G}(-q,s) 
 \label{MG},
 \\
 & M_C(k,s)
 = \frac{2L^2}{N} 
 \sum_{p+q+k=0} 
 \left( \Vertex_k^{pq} \right)^2  \C(p,s) \C(q,s) 
 \label{MC}.
\end{align}
Note that, in the present case, 
  the propagator (Kraichnan's response function) $\bar{G}$
  is essentially equivalent 
  to the response function to an externally applied probe force,
  because the random forcing term 
  $\rho_0\CheckFL$ in Eq.~(\ref{*3+})
  is not multiplicative but additive.

In regard to the choice of $t_0$,
  we consider two possibilities.
Choosing $t_0\to -0$ would admit  
  a solution of the form $\C(t) = \bar{G}(t)\C(0)$, 
  which corresponds to the ``Class II'' approximation;
  we do not take this choice, 
  as this would require the Langevin equation 
  to belong to Class I and II simultaneously,
  leading to a result 
  that is either trivial or inconsistent with FDT.
Instead, 
  we take $t_0 \to -\infty$,
  so that Eqs.~(\ref{DIA.C}) and (\ref{DIA.G})
  become identical to the standard one-loop result 
  of the MSR formalism \cite{Martin.PRA8}.

\subsection{Label-based MCT equation for SFD}

In principle,
  Eqs.~(\ref{DIA.C}) and (\ref{DIA.G}) 
  with the initial conditions
\begin{equation}
 \C(k,0) = \frac{S(k)}{L^2}, \quad
 \bar{G}(k,0) = 1
 \notag 
\end{equation}
  should suffice for determination of $\C$ and $\bar{G}$.
However, 
  as soon as we start to calculate them 
  in this straightforward way,
  we find us confronted with difficulties.
Since we have $t_0 \to -\infty$,
  the equations are acausal.
Besides, 
  the memory terms seem to suffer ultraviolet divergence.  
To elude these difficulties, 
  we demonstrate 
  that one of the two equations can be replaced 
  with a simpler relation between $\C$ and $\bar{G}$,
  later shown as Eq.~(\ref{FDT.G}).
We  start with
  differentiating Eq.~(\ref{DIA.C}) 
  in regard to $t$
  and adding it to Eq.~(\ref{DIA.G}) 
  multiplied by $\alpha_0 k^2$
  with some constant $\alpha_0$,
  to write an equation for $\dt\C + \alpha_0 k^2 G$:
\begin{widetext}%
\begin{align}
 \hspace*{-0.5em}
 \left( \dt + \frac{D_*}{S} k^2 \right) 
 \left[ \dt\C(k,t) + \alpha_0 k^2 G(k,t) \right]  
 &  = \int_0^t \D{t'} M_G(k,t-t')  
 \left[ \dt\C(k,t') + \alpha_0 k^2 G(k,t') \right] 
 \notag \\ & \quad {}
 + \int_{t_0}^0 \D{t'} 
 \left\{ M_G(k,t-t') \partial_{t'}\C(k,-t')
         - [\partial_{t'} M_C(k,t-t')] \bar{G}(-k,-t') \right\}
 \relax,
 \label{dC/dt+k2*G}
\end{align}
  with $t_0\to{-\infty}$ taken into account.
The second term on the right-hand side 
  includes $\Vertex$'s through the memory kernels,
  which we rewrite 
  by substituting Eqs.~(\ref{MG}) and (\ref{MC}).
Subsequently, 
  introducing $W$ by $\Vertex_k^{pq} = D_* k^2 W_{kpq}$
  and making use of the symmetry of $W$
  [see Eq.~(\ref{vertex})],
  after some algebraic manipulation,
  we find 
\begin{align}
 & {[\mbox{the integrand 
 in the 2nd term on RHS of Eq.~(\ref{dC/dt+k2*G})}]}
 \notag \\ 
 &= 
 \frac{4L^2}{N} D_*^2 k^2 \sum W_{kpq}^2
 \left[ 
   q^2 \bar{G}(-q, t-t') \partial_{t'} \C(k,-t') 
   - k^2 \bar{G}(k, -t') \partial_{t'} \C(q,t-t') 
 \right]  \notag \\
 &= 
 \frac{4L^2}{N} D_*^2 k^2 \sum W_{kpq}^2 \times \notag\\ & \qquad  
 \left\{
   q^2 \bar{G}(-q, t-t') 
   \left[ \partial_{t'} \C(k,-t') - \alpha_0 k^2 \bar{G}(k, -t') \right]
   - k^2 \bar{G}(k, -t') 
   \left[ \partial_{t'} \C(-q,t-t') - \alpha_0 k^2 \bar{G}(-q, t-t') \right] 
 \right\}  
 \label{term2@RHS}.
\end{align}
\end{widetext}
Then Eq.~(\ref{dC/dt+k2*G}) 
  can be replaced with a simpler relation 
\begin{equation}
 \dt\C(k,t) + \alpha_0 k^2 \bar{G}(k,t) = 0 
 \quad
 (\text{for}~\forall{k})
 \label{FDT.G},
\end{equation}
  in the sense 
  that both sides of Eq.~(\ref{dC/dt+k2*G}) vanishes
  if Eq.~(\ref{FDT.G}) holds, 
  with Eq.~(\ref{term2@RHS}) taken into account, of course.
The constant $\alpha_0$
  is determined to be
  $ \alpha_0 = {D_*}/{L^2}$
  by the initial condition.

Taking notice of the property of Eq.~(\ref{*3+})
  that the propagator $\bar{G}$ 
  is equivalent to the response to the probe force,
  we note 
  that Eq.~(\ref{FDT.G}) states 
  the fluctuation--dissipation theorem (FDT), 
  which can be derived 
  directly from the Langevin equation~(\ref{*3+}) 
  through the distribution function
  \cite{Marconi.FR461,Risken.Book1996,van-Kampen.Book2007}.
In other words, 
  FDT is \emph{already included} 
  in Eqs.~(\ref{DIA.C}) and (\ref{DIA.G}). 
This inclusion is a remarkable feature 
  of Eq.~(\ref{*3+}), or Eq.~(\ref{*2+}),
  if we compare it with an analogous calculation 
  starting from the Fourier representation
  of the ``Eulerian'' equation (\ref{*rho}),
  as opposed to the ``Lagrangian'' equation~(\ref{*2+}).
In the ``Eulerian'' case, 
  the step corresponding to the rearrangement 
  of Eq.~(\ref{dC/dt+k2*G}) 
  turns out to be inconsistent with FDT 
  \cite{Matsumoto.PTPS195}. 
This inconsistency 
  is due to the hidden dependence of $\fRho(x,t)$ on $\rho$
  in its statistics in Eq.~(\ref{f2}),
  known as the multiplicative noise 
  \cite{Miyazaki.JPA38},  
  which makes the ``Eulerian'' equation 
  intractable with a DIA-like expansion. 
Contrastively, 
  the statistics of $\fL$ 
  on the right-hand side 
  of the ``Lagrangian'' equation (\ref{*2+})
  is given by Eq.~(\ref{f3})
  which is \emph{independent} of $\psi$.
This is why the DIA equations (\ref{DIA.C}) and (\ref{DIA.G}) 
  successfully reproduce FDT.

Equation~(\ref{FDT.G}) allows us 
  to eliminate $\bar{G}$ from Eq.~(\ref{DIA.C})
  and thereby elude the difficulties 
  mentioned at the beginning of this subsection,
  as it implies
\begin{equation}
  \bar{G}(k,t) = -\frac{1}{\alpha_0 k^2} \dt\C(k,t)
  \tag{$\ref{FDT.G}'$},
\end{equation}
  from which we can show
\[
 M_G(k,s) = -\frac{1}{\alpha_0 k^2} \ds M_C(k,s).
\]
Then we substitute it into Eq.~(\ref{DIA.C}),
  and the result reads 
\begin{align*}
 & \left( \dt + \frac{D_*}{S} k^2 \right) \C \notag \\
 &= 
 \frac{1}{\alpha_0 k^2} \left[
  M_C(k,0) \C
  - \int_0^t \D{t'} M_C(k,t-t') \partial_{t'} \C(k,t')
 \right]
 \notag.
\end{align*}
The source of the ultraviolet divergence 
  is now isolated in $M_C(k,0)$,
  which we should discard,
  as this term seems to have originated 
  from an inappropriate treatment 
  of the self-interaction in DIA~\cite{note.self}.
Thus we arrive at the MCT equation:
\begin{equation}
 \left( \dt + \frac{D_*}{S} k^2 \right) \C(k,t)
 =  - \int_0^t \D{t'} M(k,t-t') \partial_{t'} \C(k,t')
 \label{MCT}
\end{equation}
  where
\begin{align}
 M(k, s) 
 &= \frac{M_C(k,s)}{\alpha_0 k^2}  \notag \\  
 &= \frac{2L^4}{N} D_* k^2 
 \sum_{p+q+k=0}  W_{pqk}^2 \C(p,s) \C(q,s)
 \label{M}
\end{align}
  with the summation taken over $(p,q)$
  satisfying the triad condition $p+q+k=0$.

\subsection{Solution to MCT equation}
\label{subsec:sol}

Now the finite-time correction 
  to Eq.~(\ref{R1*R2}) for 2pDC
  is within our reach:
  all we need to do is to solve Eq.~(\ref{MCT})
  and substitute the solution $\C$ 
  into the extended Alexander--Pincus formula (\ref{AP+}).
Although
  one may switch to numerical remedy, 
  here we prefer 
  to stick to the fully analytical calculation,
  which is possible 
  by assuming the dilute limit  
  ($\rho_0 \sigma \to +0$; $S=1$, $\Dc = D$).
This does not trivialize the problem,
  because nonlinearity still exists  
  due to 
\begin{equation}
  \frac{1}{1+\psi} = 1 - \psi + \psi^2 - \cdots 
  \label{1/rho}
\end{equation}
  and therefore the right-hand side 
  of the MCT equation (\ref{MCT})
  does not vanish.
Let us evaluate it 
  using the linear solution Eq.~(\ref{d4psi})
  as the zeroth approximation valid for ${t \to +\infty}$,
  which now reads 
\begin{equation}
  \C(k,t) \simeq  \frac{1}{L^2} e^{-D_* k^2 t}  
  \label{d4psi.S=1}
\end{equation}
  as $S = 1$.

To start with,
  we calculate $M(k,s)$ 
  by substituting the approximate solution into Eq.~(\ref{M}).
Parametrizing the variables in the summation
  as $(p,q) = (-k/2+m, -k/2-m)$
  and denoting the wave number interval  
  with $\Delta{m} = 2\pi/N$,
  we find 
\begin{equation}
 M(k,s) = \frac{D_* k^2}{\pi} 
 \sum_m \exp\left[ -D_* \left( {\frac12}k^2 + 2 m^2 \right)s \right]
 \Delta{m}
 \notag.
\end{equation}
The summation is then replaced with an integral, 
  which readily yields 
\begin{equation}
 M(k,s) 
   = \frac{D_* k^2}{\sqrt{{2\pi}D_* s}}  e^{ -{\frac12}D_* k^2 s }
   \label{M1}.
\end{equation}
At this point, 
  the nonlinear integro-differential equation,
  consisting of Eqs.~(\ref{MCT}) and (\ref{M}), 
  is approximated with a linear integro-differential equation 
  that can be obtained 
  by substituting Eq.~(\ref{M1}) into Eq~(\ref{MCT}).
The equation is then formally solved 
  in terms of Laplace transform,
  but its inversion is difficult to perform analytically.
Thus we need a further approximation:
  using both Eq.~(\ref{d4psi.S=1}) and Eq.~(\ref{M1}),
  we have
\begin{multline}
 {[\mbox{RHS of Eq.~(\ref{MCT})}]} \\ {}
 = \frac{D_*^2 k^4}{L^2}
 e^{ -D_* k^2 t }
 \int_0^t 
 \frac{\D{t'}}{\sqrt{{2\pi}D_* (t-t')}}  
 e^{+{\frac12}D_* k^2 (t-t')}
 \label{M.RHS.approx},
\end{multline}
  so that Eq.~(\ref{MCT}) 
  is now approximated 
  by a linear inhomogeneous differential equation.

Though the integral in Eq.~(\ref{M.RHS.approx}) 
  can be evaluated rigorously 
  in terms of the error function with an imaginary argument, 
  it is more convenient to evaluate it 
  by expanding the integrand in powers of $t' - t$,
  as the main contribution to the integral 
  comes from the vicinity of $t' = t$.
Thus we find 
\begin{multline}
 {[\mbox{RHS of Eq.~(\ref{MCT})}]} \\ {}
 = \frac{D_* k^4}{L^2}
 e^{ -D_* k^2 t }
 \left[
 \sqrt{{\frac2\pi}D_* t} 
 + \frac{k^2}{3} \sqrt{\frac{(D_* t)^3}{2\pi}} 
 + \cdots  
 \right]
 \notag, 
\end{multline}
  which allows us to integrate Eq.~(\ref{MCT}) 
  as 
\begin{multline}
 \C = \frac{1}{L^2}  e^{ -D_* k^2 t }  \times \\ {}
 \left[
  1 + {\frac23}\sqrt{\frac2\pi} k^4 (D_*t)^{3/2}
  + \frac{2}{15\sqrt{2\pi}} k^6 (D_*t)^{5/2} + \cdots 
 \right] 
 \label{C.approx}.
\end{multline}
It should be possible, at least in principle, 
  to substitute Eq.~(\ref{C.approx}) 
  into Eq.~(\ref{M}) 
  and the right-hand side of Eq.~(\ref{MCT})
  for the second approximation,
  but for the present 
  let us content ourselves with this first approximation
  and go ahead.

However, 
  some remarks on the properties of $W_{kpq}$
  with finite $\rho_0\sigma$
  may be in order here.
At first glance, 
\[
  W_{kpq}
  =  1
  + \frac{k}{pq} \sin{\rho_0\sigma k}
  + \frac{p}{kq} \sin{\rho_0\sigma p} 
  + \frac{q}{kp} \sin{\rho_0\sigma q}  
\]
  might be reminiscent of the MCT vertex 
  for fluids in disordered porous media  
  \cite{Krakoviack.PRE75,Krakoviack.PRE79}
  and give an impression 
  that it exhibits some singularity
  for $k\to0$,
  but in actuality it does not.
Under the condition that $k+p+q=0$,
  we have 
\[
  W_{kpq} \simeq 
  1 + \rho_0\sigma\frac{k^3+p^3+q^3}{kpq}
  = 1 + 3\rho_0\sigma,
\]
  which is evidently finite;
  the full treatment 
  of the trigonometric functions in $W_{kpq}$
  does not change the result.
This behavior is parallel to that of the vertex 
  for the corresponding Eulerian MCT.
The one-dimensional Eulerian MCT 
  for rigid particles with diameter $\sigma$
  is given by Eq.~(\ref{MCT.F}) and  
\begin{equation}
  M(k,s) \propto 
  D k^2 \sum 
  \left(
  \frac{\sin{{\sigma}p}+\sin{{\sigma}q}}{k} 
  \right)^2 
  F(p,s) F(q,s)
  \notag;  
\end{equation}
  using $p+q+k=0$,
  we have $(\sin{{\sigma}p}+\sin{{\sigma}q})/k 
  \simeq -\sigma$ 
  for the long-wave behavior of the vertex,
  which exhibits no singularity.
Thus it is found
  that both the Lagrangian MCT vertex for $\C$
  and the Eulerian MCT vertex for $F$
  are regular for long waves.
The long-wave singularity
  responsible to the anomalous diffusion 
  resides not in the memory kernel 
  but in the modified Alexander--Pincus formula (\ref{d8k}).

One may also wonder 
  whether the Lagrangian MCT equation (\ref{MCT})
  exhibits an MCT transition 
  and, if it occurs, what would be its consequence.
A full study of the possible MCT transition,
  which means emergence of a non-trivial fixed point 
  in the MCT dynamics,   
  requires numerical evaluation 
  of the wavenumber integral in Eq.~(\ref{M})
  and therefore out of the scope of the present study.
However, we may conjecture 
  that the MCT transition would not affect 
  the results of the present analysis seriously.
Since the MSD 
  given by the modified Alexander--Pincus formula (\ref{d8k})
  is dominated by the long-wave components of $\C$,
  which is supposed to evolve very slowly,
  the behavior of $\Av{R^2}$ may remain essentially unchanged,
  at least within some limited time scale, 
  even if an MCT transition occurs 
  and $\C(k,t)$ is destined to have some non-zero value
  for $t\to+\infty$.
Numerical studies of Eq.~(\ref{MCT})
  may clarify the validity range of this conjecture 
  and will be reported elsewhere.

\subsection{Effects of the nonlinear term
            on transient behaviors of MSD and 2pDC}

Since $\C$ is now available in Eq.~(\ref{C.approx})
  as a result of nonlinear closure theory,
  we can evaluate $\Av{R(\xi,t) R(\xi',t)}$ 
  using the formula (\ref{AP+}).
The procedure is analogous 
  to that for the derivation of Eq.~(\ref{R1*R2})
  from the linear solution in Eq.~(\ref{d4psi}).

If we take into account the term of order $(D_*t)^{3/2}$
  and ignore that of order $(D_*t)^{5/2}$ in Eq.~(\ref{C.approx}),
  by substituting Eq.~(\ref{C.approx}) into 
  the formula~(\ref{AP+})
  we obtain   
\begin{widetext}%
\begin{align}
  \Av{R(\xi,t)R(\xi',t)}  
  &=  
   \frac{2}{\rho_0} \sqrt{\frac{Dt}{\pi}} 
   \exp\left[  - \frac{(\xi-\xi')^2}{4\rho_0^2 Dt} \right]  
    - 
    \frac{|\xi-\xi'|}{\rho_0^2} 
    \erfc \frac{|\xi-\xi'|}{2\rho_0\sqrt{Dt}} 
    \vphantom{\frac{\int^0_0}{0}}
    -  
    \frac{\sqrt2}{3\pi}\rho_0^{-2}
    \left[ 1 - \frac{(\xi-\xi')^2}{2\rho_0^2 Dt} \right]
   \exp\left[  - \frac{(\xi-\xi')^2}{4\rho_0^2 Dt} \right]  
   \notag\\ 
  &=
   \sigma\sqrt{Dt}\,\varphi(\theta)
   - 
   \frac{\sqrt2}{3\pi}\rho_0^{-2}
   \left( 1 - 2\theta^2 \right) e^{-\theta^2}
   \label{R1*R2.NL}
\end{align}%
\end{widetext}
  where 
  $\theta = (\xi-\xi')/(2\rho_0\sqrt{Dt})$
  and the function $\varphi$ is defined 
  in Eq.~(\ref{R1*R2.sim}) with $S=1$.
As a special case for $\xi = \xi'$,
  Eq.~(\ref{R1*R2.NL}) 
  gives correction to $\Av{R^2} \propto \sqrt{t}$:
\begin{equation}
 \Av{R^2}
 = \frac{2}{\rho_0} \sqrt{\frac{D t}{\pi}} 
 - \frac{\sqrt2}{3\pi} \rho_0^{-2}
 \label{MSD.NL}.
\end{equation}
The first term 
  reproduces Eq.~(\ref{Kollmann}) with $S = 1$,
  while the second term gives a correction to it.
The contribution 
  from the higher-order terms in Eq.~(\ref{C.approx})
  slightly enlarges 
  the coefficient of the correction term,
  but the form of Eq.~(\ref{MSD.NL}) itself is not affected.
It is interesting to note 
  that Eq.~(\ref{MSD.NL}), 
  if combined with the relation 
  \cite{Percus.PRA9,van-Beijeren.PRB28}
\begin{equation}
  \frac{\D^2\Av{R^2}}{\D{t}^2} 
  = 2 \Av{u(t)u(0)}
  \label{u*u}
\end{equation}
  with $u = \D{R}/\D{t}$, 
  gives the same expression 
  as the asymptotic one without the correction term.

\begin{figure}
 \includegraphics[clip,width=0.9\linewidth]{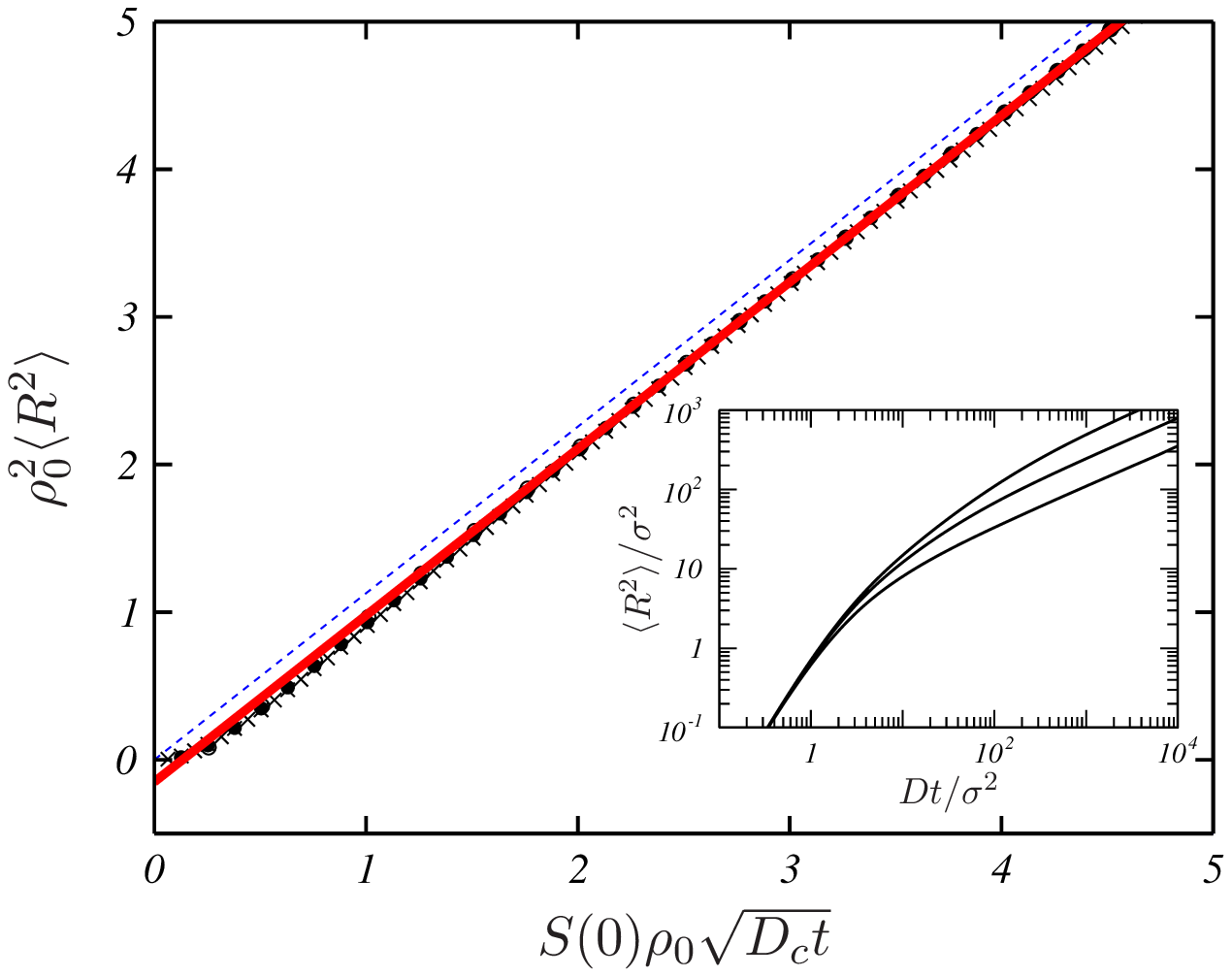}
 \caption{\label{Fig:MSD.NL}%
   (Color online)
   Comparison of Eq.~(\ref{MSD.NL})
   with numerical data,
   by means of rescaled plotting of MSD versus time
   for three different values of the density:
   $\rho_0 = N/L = (1/4)\sigma^{-1}$, $(1/8)\sigma^{-1}$,
   and $(1/16)\sigma^{-1}$.
   The number of the particles 
   is fixed at $N = 2^{15} = 32768$.
   The solid (red) line 
   represents the prediction of the nonlinear theory 
   in Eq.~(\ref{MSD.NL}),
   which is compared with Kollmann's law (\ref{Kollmann}) 
   in the dotted line.
   See the text for the reason 
   why $S(0)$ is included in the rescaling.
   The inset shows the same data 
   ($\rho_0\sigma = 1/4$, $1/8$, and $1/16$, from bottom to top)
   without rescaling, 
   using the simple nondimensionalization 
   with $\sigma^2$ and $\sigma^2/D$. 
 }
\end{figure}

While the first term in Eq.~(\ref{MSD.NL})
  is a mere reconfirmation of the classical result,
  the second term is something non-trivial
  which should be verified numerically. %
Without the second term in Eq.~(\ref{MSD.NL}),
  plotting $\Av{R^2}$ against $\sqrt{t}$
  would yield a graph of a straight line 
  passing through the origin.
In actuality,
  the second term in Eq.~(\ref{MSD.NL})
  shifts the asymptotic straight line,
  making a positive intercept on the $\sqrt{t}$-axis
  and a negative intercept on the $\Av{R^2}$-axis.
These intercepts 
  are already visible 
  in the inset of Fig.~\ref{Fig:MSD}, 
  and also in Fig.~\ref{Fig:MSD.NL} 
  (a plot analogous to Fig.~\ref{Fig:MSD} 
  but with the axes rescaled, as we explain below). 
The presence of the correction itself  
  is probably not surprising,
  because the short-time diffusion should behave 
  as $\Av{R^2} \propto Dt$ 
  before the collisions begin to take effect;
  it is more noteworthy  
  that, since this single-particle behavior 
  plays the role of the ``mode-coupling'' 
  in the Fourier representation,
  the description of the transient behavior 
  requires a nonlinear theory such as MCT.    
Viewed in the light of the thermodynamic potential,
  the nonlinearity in Eq.~(\ref{1/rho})
  originates from that of the configurational entropy.
This nonlinearity is captured 
  by adopting $\psi(\xi,t)$ as the field variable,
  which could be interpreted 
  as a kind of kinetic free-volume theory. 

The prediction of the nonlinear theory 
  in Eq.~(\ref{MSD.NL}), 
  including its dependence on the density $\rho_0$, 
  is compared with the computed MSD 
  in Fig.~\ref{Fig:MSD.NL}.
Having noticed 
  that the quantitative comparison
  requires us 
  to take into account
  the effects of the finite density,
  we revived $S = S(0)$ 
  in the first term of Eq.~(\ref{MSD.NL}),
  and plotted $\rho_0^2\Av{R^2}$ 
  against $S\rho_0\sqrt{{\Dc}t}$.
The MSD
  computed for three different values of density
  ($\rho_0\sigma = 1/4$, $1/8$, and $1/16$; see the inset)
  seem to collapse into a single curve
  whose asymptote 
  is the straight line given by Eq.~(\ref{MSD.NL}).
Improvement of Eq.~(\ref{MSD.NL}) 
  accounting for the small deviation from the straight line,
  as well as justification 
  for the revival of $S$ in Eq.~(\ref{MSD.NL}), 
  will be performable 
  with a careful numerical calculation 
  of the MCT equation, 
  which will be reported elsewhere.

\section{Other forms of four-point correlation derived from 2pDC} 
\label{sec:chi}

\subsection{Behavior of 2pDC as a function of the initial distance}
\label{subsec:chiR}

\begin{figure}
  \includegraphics[clip,width=0.75\linewidth]{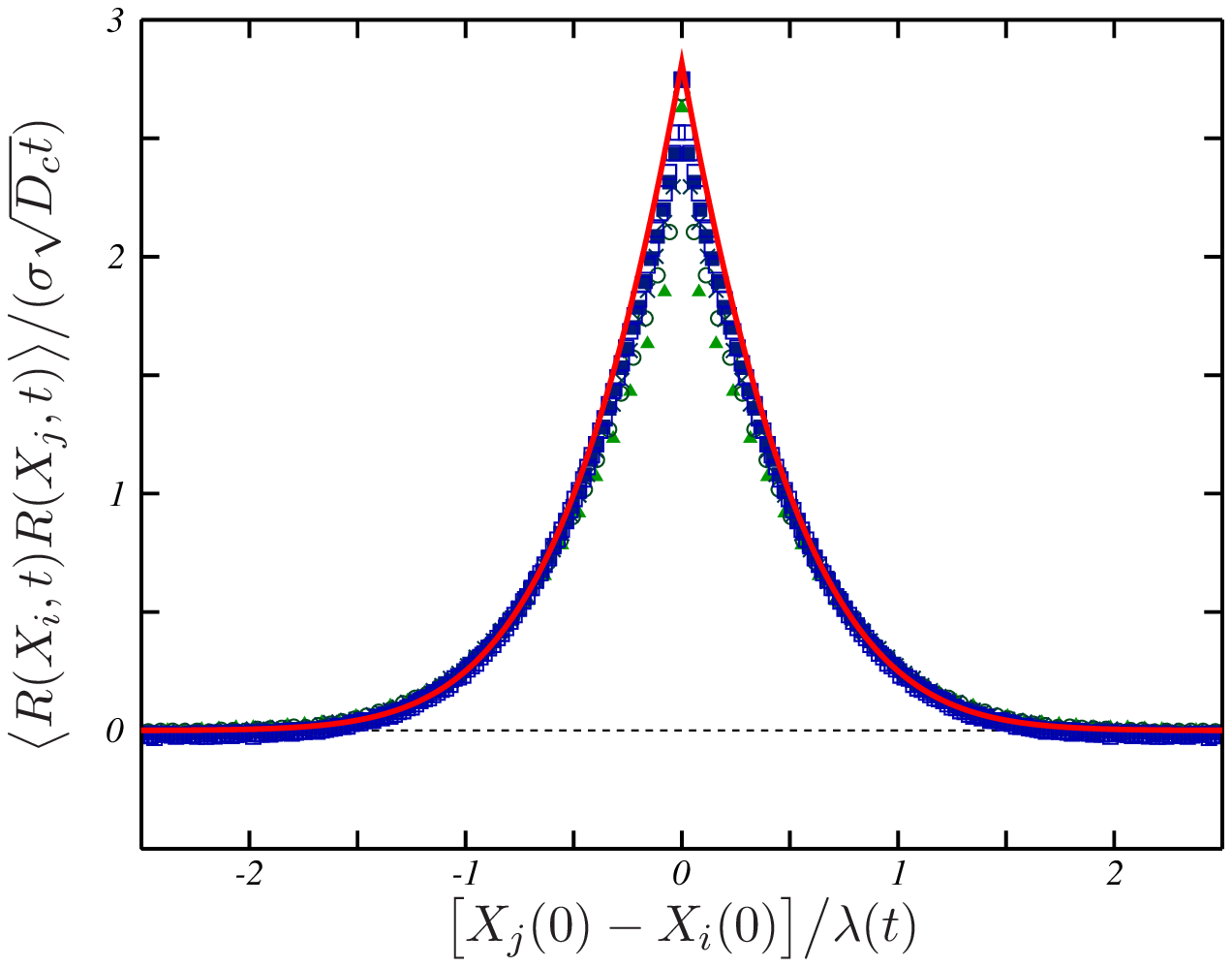} 
  \caption{\label{Fig:RR-vs-init}%
    (Color online)
    Eq.~(\ref{R1*R2}) 
    compared with simulation data 
    (with the same $N$ and $\rho_0$ 
    as in Figs.~\ref{Fig:MSD} and \ref{Fig:RR.sim}),
    on the basis of the initial distance $X_j(0) - X_i(0)$
    instead of the particle numbering.
    The solid (red) line shows the theoretical prediction 
    for $t\to+\infty$ by Eq.~(\ref{R1*R2}),
    while the simulation data 
    are plotted with the same symbols
    as in Fig.~\ref{Fig:RR.sim}.
  }
\end{figure}

As we have already shown in Fig.~\ref{Fig:RR.sim}, 
  the theoretical prediction for $\Av{R(\xi,t)R(\xi',t)}$ 
  is almost perfectly consistent with numerical calculation.
However, taking the label distance $\xi-\xi'$
  or $\theta = (\xi-\xi')/[\rho_0\ld(t)]$
  for the horizontal axis of the graph 
  is sometimes inconvenient,
  for example, when we try to compare the result 
  with two-dimensional or three-dimensional 
  numerical calculations.
For this purpose,
  it may be more convenient 
  to re-express the result 
  as a function of the initial distance, 
  $X_j(0) - X_i(0)$,
  and plot the two-particle displacement correlation 
  against $[X_j(0) - X_i(0)]/\ld(t)$.
Such a graph is shown 
  in Fig.~\ref{Fig:RR-vs-init}. 
The analytic curve in Fig.~\ref{Fig:RR-vs-init} 
  is drawn   
  by simply equating $[X_j(0) - X_i(0)]/\ld(t)$ 
  with $\theta$ in Eq.~(\ref{R1*R2.sim}).
This seems to be valid for $t\to+\infty$, 
  but a considerable discrepancy is seen 
  for shorter times. 

Although $X_j(0) - X_i(0)$ and $(\Xi_j - \Xi_i)/\rho_0$
  are equal on the average, 
  generally they are different, 
  as is evident from the relation 
\begin{equation}
  X_j(0) - X_i(0)
  = 
  \int_{\Xi_i}^{\Xi_j}
  \frac{1+\psi(\xi,0)}{\rho_0}\D\xi
  \label{*X}.
\end{equation}  
This difference is responsible 
  for the discrepancy in Fig.~\ref{Fig:RR-vs-init}
  for finite $t$.
Taking this difference into account, 
  we can evaluate 2pDC theoretically 
  as a function of the initial distance.
Although its complete evaluation 
  is out of the scope of the present paper, 
  as it seems to involve wavenumber integrals 
  that are difficult to perform analytically, 
  we can outline the procedure of the calculation
  at least.

With the value of the initial distance 
  denoted with $\tilde{d}$,  
  the function that gives 2pDC, 
  which we denote with  $\chiR(\tilde{d},t)$,
  is formally written as 
\begin{multline}
  \chiR(\tilde{d},t)  \\ {} 
  = \Av{%
  \frac{1}{L}\iint
  \delta({x_1}-{x_2}-{\tilde{d}}\,) R(\xi_1,t)R(\xi_2,t)
  \D{x_1}\D{x_2}
  }
  \label{RR.i1}
\end{multline}
  where 
\[
  x_1 = x_1(\xi_1) = x(\xi_1,0), \quad  
  x_2 = x_2(\xi_2) = x(\xi_2,0), 
\]
  and therefore 
\[
  x_2 - x_1 
  = \frac{1}{\rho_0}
  \left[
    \xi_2 - \xi_1 + \int_{\xi_1}^{\xi_2}\psi(\xi,0)\D\xi
  \right]
  \relax.
\]
Changing 
  the variables of the integral in Eq.~(\ref{RR.i1}) 
  from $(x_1,x_2)$ to $(\xi_1,\xi_2)$
  with
\[
  \D{x_1} 
  = \frac{1+\psi(\xi_1,0)}{\rho_0}\D\xi_1, 
  \quad
  \D{x_2} 
  = \frac{1+\psi(\xi_2,0)}{\rho_0}\D\xi_2, 
\]
  and introducing the Fourier representation 
  of the delta function, 
\[
  \delta(\tilde{x})
  = \frac{1}{L} \sum_q e^{-{\II}q\tilde{x}}
  \quad
  \left(\frac{q}{2\pi/L} \in\Z \right),
\]
  we rewrite Eq.~(\ref{RR.i1}) as 
\begin{equation}
  \chiR(\tilde{d},t) 
  = \sum_k e^{{\II}k{\rho_0}\tilde{d}}
  \Av{{\tilde{R}(k,t)}{\tilde{R}(-k,t)}}
  \label{RR.i2}
\end{equation}
  where $k=q/\rho_0$ and 
\begin{align}
  \tilde{R}(k,t)  
  &= {\frac1L}
  \int_0^L e^{{\II}qx} R(\xi(x,0),t)\D{x}
  \notag \\ 
  &= {\frac1N}
  \int_0^N \! \exp\left(
  {\II}k\left[ \xi + \int_0 \psi \,\D\xi \right] 
  \right) 
  R(\xi,t)\, (1 + \psi)\D\xi
  \label{RR.i3}, 
\end{align}
  with 
\[
  \psi = \psi(\xi,0), \quad
  \int_0 \psi\,\D\xi
  = \int_0^\xi \psi(\tilde\xi,0)\D\tilde\xi
  \relax.
\]
Then we express $\psi$ in Eq.~(\ref{RR.i3}) 
  with $\check\psi$ in Eq.~(\ref{m1})
  and also substitute Eq.~(\ref{R.psi+}).
After some rearrangement, 
  we obtain 
\begin{multline}
  \tilde{R}(k,t) 
  = \frac{L}{N}\times 
  \frac{\check\psi(k,t)-\check\psi(k,0)}{-{\II}k} \\{}
  + \frac{L}{N}
  \sum_{p+{p'}=k}
  \frac{\check\psi(p,0)\left[ 
    \check\psi(p',t) - \check\psi(p',0)
  \right]}{{\II}p}
  + O(\check\psi^3)
  \label{RR.i4}.
\end{multline}
Substituting Eq.~(\ref{RR.i4}) into Eq.~(\ref{RR.i2})
  yields
  an expression of $\chiR(\tilde{d},t)$
  that consists of two parts:
  the first part simply reproduces Eq.~(\ref{AP+})
  with $\xi-\xi'$ replaced with $\rho_0\tilde{d}$, 
  and the second part involves 
  triple correlations such as 
  $\Av{\check\psi(-p,0)\check\psi(-q,t)\check\psi(-k,t)}$
  with $p+q+k=0$.
These triple correlations 
  can be calculated with DIA, 
  and as a result,
  we obtain a correction term 
  whose magnitude relative to the leading term 
  decreases in proportion to $t^{-1/2}$
  for $t\to+\infty$.
Detailed results of the calculation 
  will be reported elsewhere.

We note that the definition of $\chiR$ in Eq.~(\ref{RR.i1}) 
  is readily generalized to three-dimensional cases,
  as 
\begin{align}
  \ChiR(\tilde{\mb{d}},t) 
  &= \Av{\frac{1}{L^3}\iint
  \delta^3( \mb{r}_{12}-\tilde{\mb{d}} )
  \mb{R}_1\otimes\mb{R}_2\,
  \D^3\mb{r}_1
  \D^3\mb{r}_2
  }\notag \\
  &=
  \Av{%
  \frac{L^3}{N^2}
  \sum_i \sum_j
  \frac{%
  \delta^3( \mb{r}_{ij} - \tilde{\mb{d}} )
  }{g_2(\mb{r}_{ij})}
  \mb{R}_i\otimes\mb{R}_j
  }
  \relax,
  \label{RR.3D}
\end{align}
  where 
  $\mb{r}_{ij} = \mb{r}_j - \mb{r}_i$,   
  $\mb{R}_i = \mb{R}(\BoldXi_i,t) 
  = \mb{r}_i(t) - \mb{r}_i(0)$,
  and 
\[
  g_2(\mb{r}) 
  = \frac{L^3}{N^2}\sum_{i'}\sum_{j'} 
  \delta^3(\mb{r}_{j'}-\mb{r}_{i'}-\mb{r})
  \relax.
\]
This $\ChiR$ is similar to the quantity 
  calculated by Donati \textit{et al.} \cite{Donati.PRL82}
  ($g_u$ in their notation),
  except for two main differences:
  in their $g_u$, 
  the two-body density $g_2$ is 
  absent in the denominator,
  and a product of scalar displacements,
  ${|\mb{R}_i|}\,{|\mb{R}_j|}$, 
  is used instead of the tensor product 
  $\mb{R}_i\otimes\mb{R}_j$.
The presence or absence of $g_2$ 
  is not essential,
  though it indeed makes it difficult 
  to define $\ChiR$ 
  for small values of the initial distance, 
  $|\di|<\sigma$,
  in which we are not interested.
The other difference is crucial:
  the absolute value signs 
  obstruct analytical evaluation of $g_u$
  even in the one-dimensional cases.
Besides, 
  the tensorial character of $\ChiR$ can provide useful information 
  on the geometry of the collective motion
  in the three-dimensional glassy systems.
We will return to this point in Sec.~\ref{sec:discuss},
  but before that,
  let us relate the one-dimensional 2pDC to $\chi_4$.

\subsection{Calculation of $\chi_4$ from 2pDC}
\label{subsec:chiS}

With the knowledge 
  of the displacement correlation $\Av{R(\xi,t)R(\xi',t)}$
  in Eq.~(\ref{R1*R2.NL}),
  we can also calculate a one-dimensional version 
  of a quantity 
  which is commonly referred to as $\chi_4(t)$.
To be precise,
  we consider the $\Q$-based $\chi_4$ 
  \cite{Glotzer.JCP112,Lacevic.JCP119},
  as opposed to other variants of $\chi_4$ 
  such as the $F$-based $\chi_4$ 
  \cite{Berthier.PRE69,Toninelli.PRE71}
  defined through the fluctuation 
  of the intermediate scattering function $F$ 
  or its self part.
If we consult Glotzer \textit{et al.} \cite{Glotzer.JCP112}
  and adapt their equations~(4) and (5) 
  for one-dimensional cases, 
  we have 
\begin{gather}
  \Q
  = \sum_i \sum_j \Da(X_j(t)-X_i(0))
  \label{chi4.Q}, 
  \\
  \chi_4(t) = \frac{L}{\kT}
  \frac{\Av{\Q^2} - \Av{\Q}^2}{N^2}
  \label{chi4=},
\end{gather}%
with some radius $a$ as a criterion of overlapping;
  $\Da$ denotes the overlapping function,
  which has a finite value around $r=0$ 
  and vanishes for $r \gg a$.

This type of four-point correlation function
  has been studied by many authors
  \cite{Dasgupta.EPL15,Glotzer.JCP112,Lacevic.JCP119,%
  Dauchot.PRL95,Lechenault.EPL83a,Shiba.arXiv1205}
  as an indicator of cooperative motion in glassy systems.
To our knowledge, 
  most of these studies 
  are based on direct numerical simulations 
  of particle systems
  and there are also experiments 
  grounded on observation of particles,
  but analytical calculations are quite rare.
What makes it difficult to calculate $\chi_4$ analytically  
  is that, in the usual formulation,  
  the \emph{four-point} correlation
  implies a \emph{four-body} correlation.
More concretely, 
  as $\Q$ in Eq.~(\ref{chi4.Q})
  already contains double summation, 
  calculation of $\chi_4$ requires 
  dealing with quadruple summation
  whose summand involves four particles simultaneously;
  this would be a hopeless task.

To facilitate calculation of four-point correlation, 
  here we introduce two modifications 
  to Eqs.~(\ref{chi4.Q}) and (\ref{chi4=}).
Firstly, 
  we target on the ``self part'' ($i=j$) of $\Q$
  and its contribution to $\chi_4$,
  denoting them as 
  \cite{note.chiS}
\begin{equation}
 \Qs = \sum_i \Da(R_i(t)),
 \quad
 \chiS(t) = \frac{L}{\kT}
 \frac{\Av{\Qs^2} - \Av{\Qs}^2}{N^2}
 \label{chi4s=}.
\end{equation}
Since Glotzer \textit{et al.}~\cite{Glotzer.JCP112} reported 
  that the contribution of the self part ($i=j$) 
  is dominant over that of the distinct part (${i}\ne{j}$)
  in the three-dimensional case,
  it is justifiable to calculate $\chiS$ instead of $\chi_4$.
Secondly, 
  as the overlap function $\Da$,
  we adopt a Gaussian function
\begin{equation}
  \Da(r) = e^{-r^2/a^2}
  \label{overlap.Gauss}
\end{equation}
  instead of the step function 
  used by Glotzer \textit{et al.}~\cite{Glotzer.JCP112}.
We note that, although there exists a variant of $\chi_4$
  from whose definition 
  the probe length can be totally expelled
  \cite{Ikeda.JCP138},
  the probe length $a$ is indispensable to $\Qs$.


From Eq.~(\ref{R1*R2.NL})
  we already know 
  the covariance $\Av{{R_i}{R_j}}$ for all $(i,j)$ 
  and for arbitrary $t$ (within a certain limitation, of course).
The problem is 
  how to evaluate $\Av{\Qs}$ and $\Av{\Qs^2}$
  in Eq.~(\ref{chi4s=}) using this information.
This is possible, 
  if we assume that $(R_1,R_2,\ldots,R_N)$
  is subject to a joint (multivariate) Gaussian distribution,
  which is determined uniquely
  as the covariance is given  
  and the mean is known to vanish. 
For the purpose of calculating $\Av{\Qs^2}$,
  it suffices to determine 
  the two-body distribution function 
  for $(R_i,R_j)$,
  which we denote with 
\begin{multline}
 P(R_i,R_j)  \\ {}
 = \frac{1}{2\pi\sqrt{\Delta_{ij}}} 
 \exp\left[ 
 -\frac{\Av{R^2}(R_i^2 + R_j^2) 
 - 2\Av{{R_i}{R_j}} {R_i R_j}}
 {2\Delta_{ij}} 
 \right]
 \notag 
\end{multline} 
  where 
\[
  \Delta_{ij} = \Av{R^2}^2 - \Av{{R_i}{R_j}}^2, \quad
  \Av{R^2} = \Av{R_i^2} = \Av{R_j^2}
  \relax.
\]
Using this joint distribution function $P(R_i,R_j)$
  and adopting Eq.~(\ref{overlap.Gauss}) 
  for the overlapping function,
  we obtain 
\begin{align}
 \Av{\Da(R_i)}
 &= \int \Da(R_i) P(R_i,R_j)\D{R_i}\D{R_j} \notag\\
 &= \frac{1}{\sqrt{1 + \dfrac{2\Av{R^2}}{a^2}}}
 \label{Ex.d}
\end{align}
  and
\begin{align}
 \Av{\Da(R_i)\Da(R_j)} 
 &= \int \Da(R_i)\Da(R_j) 
 P(R_i,R_j)\D{R_i}\D{R_j} \notag \\ 
 &= \frac{1}{\sqrt{%
 \left( 1 + \dfrac{2\Av{R^2}}{a^2} \right)^2 
 - \dfrac{4\Av{{R_i}{R_j}}^2}{a^4}}
 }
 \label{Ex.d2};
\end{align}
  note that Eq.~(\ref{Ex.d2}) 
  is confirmed separately for $i\ne j$ and $i=j$.
With Eq.~(\ref{Ex.d}) and (\ref{Ex.d2}),
  now we can evaluate $\chiS$ 
  in Eq.~(\ref{chi4s=}),
  taking the uniformity of the system into account.
As a result, 
  we obtain 
\begin{widetext}
\begin{align}
 \chiS 
 &= 
 \frac{L}{{N^2}\kT}
 \left\{
   \sum_i \sum_j \Av{\Da(R_i)\Da(R_j)} 
   - 
   \left[ \sum_i \Av{\Da(R_i)} \right]^2
 \right\}
 \notag \\ 
 &=
 \frac{L}{N\kT}
 \sum_l 
 \left[
   \frac{1}{\sqrt{%
   \left( 1 + \dfrac{2\Av{R^2}}{a^2} \right)^2 
   - \dfrac{4\Av{{R_i}{R_{i+l}}}^2}{a^4}}
   }
   -
   \frac{1}{1 + \dfrac{2\Av{R^2}}{a^2}} 
 \right]
 \label{chi4s.sum}.
\end{align}%
\end{widetext}
Note that 
  the double summation 
  $\sum_i \sum_j \Av{\Da(R_i)\Da(R_j)}$
  in Eq.~(\ref{chi4s.sum})
  is a result of the simplification 
  by the replacement of $\chi_4$ 
  with $\chiS$ (retaining only the self part):
  if this simplification were not introduced,
  we would have to struggle with a quadruple summation 
  such as
\[
  \sum_i \sum_j \sum_k \sum_l 
  \Av{\Da(X_j(t)-X_i(0))\Da(X_l(t)-X_k(0))},
\]
  whose evaluation would be much less workable 
  than $\Av{\Da(R_i)\Da(R_j)}$.

\begin{figure*}
  \raisebox{0pt}{\raisebox{6cm}{(a)}\ 
  \includegraphics[clip,width=7.5cm]{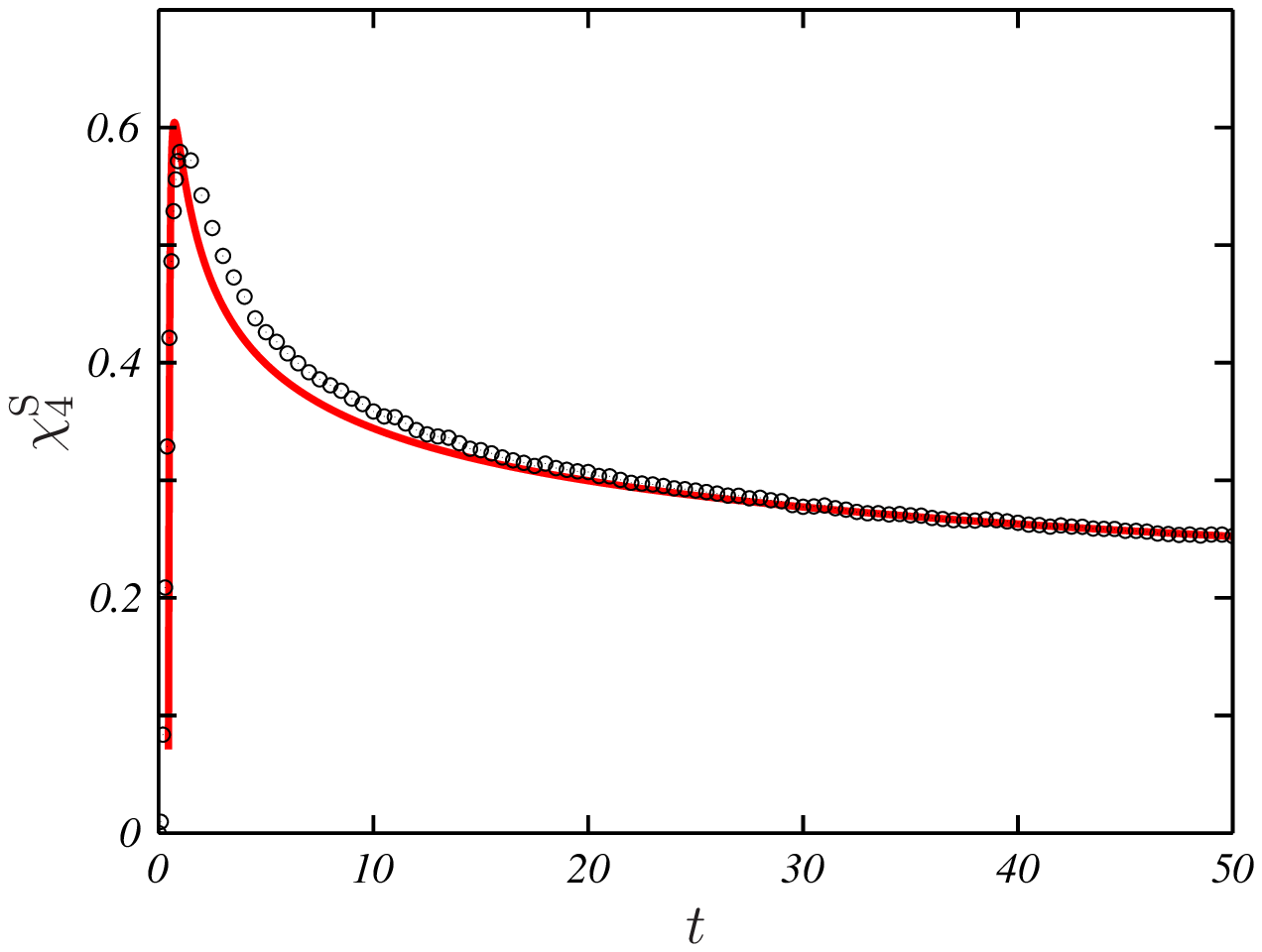}}%
  \quad
  \raisebox{0pt}{\raisebox{6cm}{(b)}\ 
  \includegraphics[clip,width=7.5cm]{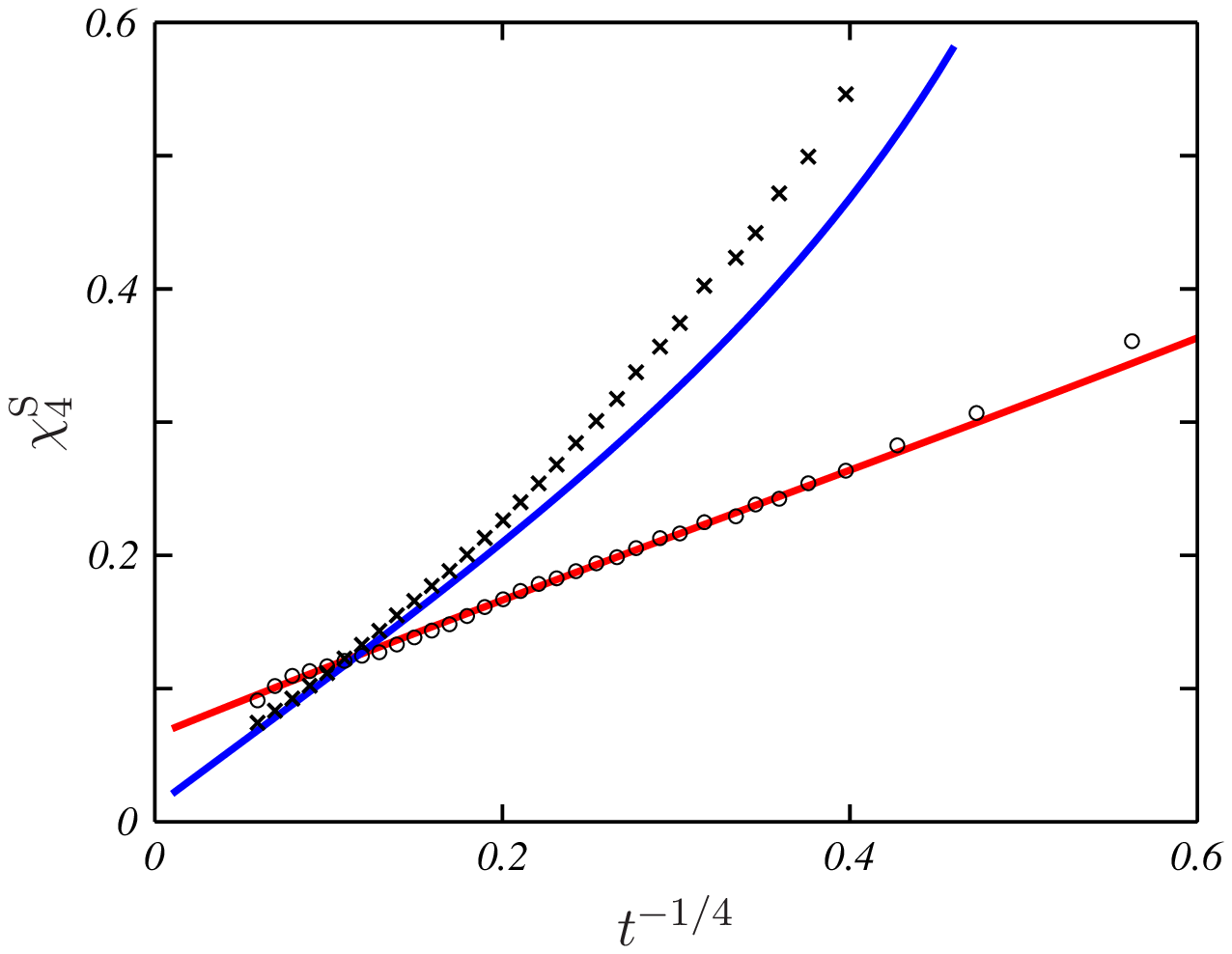}}%
  \caption{\label{Fig:chi4s.SFD}%
    (Color online) 
    Short-time and long-time behavior of $\chiS$ for SFD,
    compared with the analytical prediction
    given by the sum 
    of  $(\chiS)_{\text{solo}}$ in Eq.~(\ref{chi4s.solo})
    and $(\chiS)_{\text{coll}}$ in Eq.~(\ref{chi4s.coll}),
    with the effect of $S\ne1$ and $\Dc/D \ne 1$
    taken into account.
    The probe length (radius of the overlapping function)
    was chosen as $a = 0.5\,\sigma$\protect.
    The time is non-dimensionalized with $\sigma^2/D$.  
    (a) Short-time behavior for the case 
    with $N = 256$ and $\rho_0 = N/L = (1/4)\,\sigma^{-1}$.
    The computed data are plotted with circles
    and the analytical prediction is shown  
    with a solid (red) line.
    (b) Long-time behavior 
    for $\rho_0 = (1/4)\sigma^{-1}$ (plotted with circles)
    and for $\rho_0 = (1/16)\sigma^{-1}$ 
    (plotted with crosses).
    The solid lines show analytical prediction 
    for the two cases.
    Note the agreement 
    between the computed results and the theoretical curves
    except for the transient discrepancy,
    as well as the asymptotic behavior of the curves 
    that takes the form of a straight line in the graph,
    exhibiting the decay of $(\chiS)_{\text{solo}}$
    in proportion to $t^{-1/4}$
    and the finite value of $(\chiS)_{\text{coll}}$
    that remains for $t \to +\infty$.
  }
\end{figure*}

Before applying Eq.~(\ref{chi4s.sum}) to SFD,
  we can test it with free Brownian particles.
From the Langevin equation
  obtained by setting $V=0$ in Eq.~(\ref{Langevin.X}),
  we have 
\begin{equation}
 \Av{{R_i}{R_j}}
 = 
 \begin{cases}
   \Av{R^2} 
    = 2D\left[ 
     t - \tauB \left( 1 - e^{-t/\tauB} \right) 
   \right]  & (i =   j) \\
   0        & (i \ne j) \relax
 \end{cases}
 \label{RR.free}
\end{equation}
  where $\tauB = m/\mu$.
This is substituted into Eq.~(\ref{chi4s.sum}),
  which yields 
\begin{align}
 \chiS 
 &= \frac{1}{\rho_0\kT}
 \left(
   \frac{1}{\sqrt{1 + 4\Av{R^2}/a^2}} 
   - \frac{1}{1 + 2\Av{R^2}/a^2}
 \right)  \notag \\
 &= (\chiS)_{\text{solo}} 
 \label{chi4s.solo}
\end{align}
  for free Brownian particles;
  note that all the contribution 
  comes from the term with $l=0$
  in Eq.~(\ref{chi4s.sum}), 
  which we refer to as the ``solo'' part.
Taking notice of the $t$-dependence of $\Av{R^2}$
  in Eq.~(\ref{chi4s.solo})
  and making some calculation,  
  we find $(\chiS)_{\text{solo}}$ to have a peak
  at the instant when $\Av{R^2} = 2.6\,a^2$ approximately;
  see the short-time side 
  of Fig.~\ref{Fig:chi4s.SFD}(a).
Obviously, this short-time peak 
  is irrelevant to particle interaction.
After this peak, 
  $(\chiS)_{\text{solo}}$
  decreases monotonically toward zero,
  in proportion to $t^{-1/2}$ for $t\to +\infty$. 


Now let us calculate $\chiS$ for SFD,
  combining Eq.~(\ref{chi4s.sum}) 
  with the result for $\Av{RR}$ in Eq.~(\ref{R1*R2.NL}).
We evaluate $\chiS$ in Eq.~(\ref{chi4s.sum}) 
  as a sum of the ``solo'' part ($l=0$)
  and the collective part (contribution 
  from the terms with $l\ne0$).
The solo part 
  is given by Eq.~(\ref{chi4s.solo})
  and depends on $\Av{R^2}$ alone,
  for which we use Eq.~(\ref{MSD.NL})
  that was obtained by setting $\xi = \xi'$ 
  in Eq.~(\ref{R1*R2.NL}).
As a matter of course, 
  we must exclude cases of very short time, 
  for which Eq.~(\ref{MSD.NL}) predicts 
  $\Av{R^2}$ to be negative;
  this is out of the validity range of Eq.~(\ref{MSD.NL}).
Subsequently, 
  to evaluate the contribution 
  from the terms with $l\ne0$,
  we use the asymptotic form of Eq.~(\ref{R1*R2.NL}) 
  for $t\to+\infty$,
  expressed as a self-similar solution 
  in Eq.~(\ref{R1*R2.sim}):
\[
 \Av{{R_i}{R_{i+l}}} 
 \simeq \frac{2\sqrt{Dt}}{\rho_0}\varphi(\theta_l), \quad
 \theta_l 
 = \frac{l}{\rho_0\ld(t)}
 = l\Delta\theta.
\]
The collective part
  is thereby written as 
\begin{multline}
 (\chiS)_{\text{coll}} \\ {}
 \simeq 
 \frac{1}{\rho_0\kT}\times
 \frac{1}{1+\dfrac{2\Av{R^2}}{a^2}}
 \sum_{l\ne0} 
 \left\{
   \frac{1}{\sqrt{%
   1 - \left[\dfrac{\varphi(\theta_l)}{\varphi(0)}\right]^2
   }}
   - 1
 \right\}
 \notag.
\end{multline}
The number of particles %
  contributing to the sum 
  is estimated to be
  $ N_{\text{coll}} 
  \sim {1/{\Delta\theta}} = \rho_0\ld(t)$,
  which yields, approximately, 
\begin{subequations}%
  \begin{align}
    (\chiS)_{\text{coll}} 
    &\sim
    \frac{1}{\rho_0\kT}\times 
    \frac{\rho_0\ld(t)}{1 + {2\Av{R^2}/a^2}}
    \label{chi4s.1D.estimation} \\ 
    &\sim
    \frac{a^2}{\kT}\times \frac{\ld(t)}{\Av{R^2}}
    \sim \frac{a^2\rho_0}{\kT}
    \label{chi4s.1D.t=inf}
  \end{align}%
  \label{eqs:chi4s.1D.estimation}%
\end{subequations}%
  for $t \to +\infty$.
More precisely, 
  the sum can be evaluated 
  by rewriting it as 
\begin{align}
 \sum_{l\ne0}(\cdots) 
 &= 
 {2\rho_0\ld(t)}  
 \sum_{l=1}^{\infty}
 \left\{
   \frac{1}{\sqrt{%
   1 - \left[\dfrac{\varphi(\theta_l)}{\varphi(0)}\right]^2
   }}
   - 1
 \right\}\Delta\theta
 \notag \\
 &\to 
 {2\rho_0\ld(t)}  
 \int_0^{\infty}
 \left\{
   \frac{1}{%
   \sqrt{1 - 
   \pi\left[\varphi(\theta)\right]^2
   }} - 1 
 \right\}\D\theta
 \notag
\end{align}
  and using the numerical value 
  of the integral
\begin{equation}
 \int_0^{\infty}
 \left\{
   \frac{1}{%
   \sqrt{1 - 
   \pi\left[\varphi(\theta)\right]^2
   }} - 1 
 \right\}\D\theta
  = 0.364124      
  \notag;
\end{equation}
  thus we obtain 
\begin{align}
 (\chiS)_{\text{coll}} 
 &= \frac{1}{{\rho_0}\kT}
 \times \frac{\rho_0\ld(t)}{\Av{R^2}/a^2} 
 \times 0.364124  \notag\\
 &= \frac{1}{{\rho_0}\kT}
 \times \sqrt{\pi}\rho_0^2 a^2 \times 0.364124  \notag\\
 &= 0.6454 \times \frac{\rho_0 a^2}{\kT}
 \label{chi4s.coll}
\end{align}
  and 
  $\chiS = (\chiS)_{\text{solo}} + (\chiS)_{\text{coll}}$,
  with $(\chiS)_{\text{solo}}$
  given by substituting Eq.~(\ref{MSD.NL}) 
  into Eq.~(\ref{chi4s.solo}).
If we take $S$ into account
  according to Eq.~(\ref{R1*R2.sim}),
  with the reservation 
  that both Eq.~(\ref{R1*R2.sim}) and Eq.~(\ref{chi4s.coll}) 
  are valid only asymptotically, 
  the right-hand side of Eq.~(\ref{chi4s.coll})
  is multiplied by $S^{-1}$.

The analytically calculated $\chiS$
  and its numerical values 
  are compared in Fig.~\ref{Fig:chi4s.SFD}. 
The effect of finite $\rho_0$ 
  that makes $S$ and $\Dc/D$ different from unity 
  is taken into account
  (see Table~\ref{Tab:Dc} in Appendix~\ref{app:num}).
The peak in the short-time regime 
  has nothing to do with the slow dynamics,
  as it appears even for free Brownian particles.
After this peak, 
  $(\chiS)_{\text{solo}}$ decreases slowly,
  asymptotically in proportion to $t^{-1/4}$,
  while, reflecting the endless growth of the cluster size,
  $(\chiS)_{\text{coll}}$ remains finite for $t\to+\infty$.
The behavior of the numerical solution 
  is consistent with this analytical prediction.

The limiting value of $\chiS$ for $t\to+\infty$,
  given by $(\chiS)_{\text{coll}}$ 
  in Eq.~(\ref{chi4s.coll}),
  is an increasing function of the density $\rho_0$.
This conclusion remains unchanged 
  also if the effect of $S\ne1$ is included,
  because $1/S$ is also an increasing function 
  of $\rho_0$.
In conformity with the theoretical prediction, 
  Fig.~\ref{Fig:chi4s.SFD}(b) shows 
  that the value of $\chiS$ 
  is greater for $\rho_0 = (1/4)\sigma^{-1}$ (circles)
  than for $\rho_0 = (1/16)\sigma^{-1}$ (crosses)
  if $t$ is sufficiently long.

\section{Discussion}
\label{sec:discuss}
\subsection{Quantification of collective dynamics}

We have studied 
  a one-dimensional system of Brownian particles 
  with repulsive interaction, 
  regarding it as a simplified model of the cage effect.
On one hand, 
  the cage confines every particle in a narrow space.
On the other hand,
  since the particles are mutually caged 
  and therefore forbidden to move uncooperatively,
  they must either wait still or move together.
Thus the cage effect 
  involves correlated motion of numerous particles,
  which is visualized 
  as dynamical clusters in space-time diagrams 
  (Fig.~\ref{Fig:SFD.cluster}),
  with the diffusive correlation length $\ld(t) = 2\sqrt{{\Dc}t}$.
The substance that diffuses 
  is not the particles but the space between them,
  represented by $\psi$ in our theory.
The two-particle displacement correlation (2pDC)
  is shown to be a useful indicator of the correlated motion.
Asymptotically, 2pDC becomes self-similar:
  it suggests a matryoshka-like structure,
  such that the small cages are confined in larger cages,
  which, in turn, are caught in still larger and slower cages.

Using the analytical result for 2pDC 
  which is valid both transiently and asymptotically, 
  we demonstrated how to calculate 
  the $\Qs$-based $\chi_4$ (denoted with $\chiS$).
Despite the endless growth of $\ld(t)$
  and the absence of $\alpha$ relaxation, 
  the result in Eq.~(\ref{chi4s.coll})
  shows that $\chiS$ converges to some constant 
  for $t \to +\infty$.
An implication of Eq.~(\ref{chi4s.coll})
  is that $\chiS$, and probably $\chi_4$ in general, 
  does not give a straightforward representation 
  of the cluster size.
Indeed, $\ld(t)$ is in the numerator
  of Eq.~(\ref{eqs:chi4s.1D.estimation}) or (\ref{chi4s.coll}),
  but the result is modified by the denominator,
  or a pre\-factor $1/(1+2\Av{R^2}/{a^2})$
  originating from $\Av{\Qs}^2$,
  which cancels 
  the temporal growth of the cluster size.
In three-dimensional systems,
  a direct relation between $\chiS$ and the cluster size 
  is expected only for some limited time scales  
  corresponding to the plateau of the MSD.

\subsection{Comparison with different theories of SFD}
\label{subsec:cmp}

Among the problems of diffusion in confined geometries 
  \cite{Burada.CPhC10},
  SFD has occupied a special position 
  which has attracted researchers 
  from various fields of science
  and with a variety of mathematical approaches.
Let us review some of their works briefly, 
  with which, subsequently, 
  the present theory will be compared.
  
In the oldest type of theory \cite{Levitt.PRA8},
  the single-file system was modeled 
  with an array of free Brownian particles 
  interchanging their labels upon collision,
  and analyzed with a technique 
  that makes a full use of the one-dimensional geometry,
  such as the Jepsen line \cite{Barkai.PRE81}.
In this line of argument, 
  Hahn \& K{\"a}rger \cite{Hahn.JPhA28} 
  showed that MSD for SFD
  can be obtained  
  from the corresponding free-particle dynamics 
  with the relation $\Av{R^2} \propto \Av{|R_{\text{free}}|}$,
  taking it into account
  that the constant of proportionality 
  depends on the particle diameter.

Later, 
  theories in the Fourier space emerged.
The concise theory 
  by Alexander and Pincus \cite{Alexander.PRB18}
  is of this type.
Generally speaking, 
  if the spatiotemporal dynamics of the system 
  is described by a linear equation 
  and the system is translationally invariant,
  the Fourier or Fourier--Laplace approach 
  is a quite natural choice.
Thus, starting from the ``diffusion-noise'' equation,
  which is a linear Langevin equation 
  for the density $\rho(x,t)$,
  Taloni and Lomholt \cite{Taloni.PRE78} 
  calculated MSD 
  via the velocity correlation in Eq.~(\ref{u*u}),
  and Lizana \textit{et al.}~\cite{Lizana.PRE81}
  showed that the linear dynamics is reducible 
  to a fractional Langevin equation.
These works rely
  on the assumption of linear dynamics, 
  as is evident 
  in the case of Lizana \textit{et al.}~\cite{Lizana.PRE81}
  who denominate it the harmonization technique.
We should be cautious 
  with limitations of the harmonization, however, 
  especially when the interparticulate potential $V(r)$
  has a non-analytic point 
  as in Eq.~(\ref{V=}) or in the rigid-sphere potential,
  as was pointed out by Ikeda \textit{et al.}
  \cite{Ikeda.JCP138}.

The simplest description 
  based on the linear ``diffusion-noise'' equation
  cannot account for the finite-time correction
  which should certainly exist 
  according to our particle-based computations.
More refined linear theories 
  may succeed in interpolating two limiting cases
  of $Dt\ll 1/\rho_0^2$ and $Dt\gg 1/\rho_0^2$;
  let us leave them aside, however, 
  because interpolations are usually less informative 
  than phenomenologies.
Among phenomenologies on transient behavior of SFD, 
  the theory of 
  van Beijeren \textit{et al.} \cite{van-Beijeren.PRB28}
  deserves a particular attention.
This theory deals with SFD on a lattice, 
  on the basis of the picture of migrating vacancies
  and making use of Eq.~(\ref{u*u}).
The approximation of independently diffusing vacancies
  leads readily to the asymptotic law, 
  $\Av{R^2} \propto \sqrt{t}$.
For the cases of finite density of vacancies,
  van Beijeren \textit{et al.} \cite{van-Beijeren.PRB28}
  noticed that a cluster of vacancies may be formed
  and thereby a kind of memory effect may arise.
Instead of developing a systematic treatment of the memory,
  however, they assumed some phenomenological rules 
  about the dynamics of a vacancy cluster
  and thereby calculated $\Av{R^2}$ for all $t$.
We note that the picture of diffusing vacancies
  is both conceptually suggestive and practically useful.
An asymptotic theory based on the vacancy picture 
  can be very concise \cite{Kaerger.PRA45}.
The idea of migrating defect has been used 
  also in problems other than SFD, 
  such as dielectric relaxation
  of supercooled isoamyl bromide \cite{Glarum.JCP33}.
In the context of glassy dynamics, 
  this idea is incarnated 
  in the kinetically constrained models,
  which can be regarded 
  as a kinetic version of the free-volume theory of glasses
  \cite{Berthier.RMP83,Sellitto.PRE62}.

A systematic treatment of the memory effect 
  necessitates a term with time integral.
Though such a term arises 
  in the fractional Langevin equation \cite{Lizana.PRE81},
  its physical interpretation is not straightforward. 
Rallison \cite{Rallison.JFM186} proposed 
  another phenomenological theory,
  whose memory integral can be understood quite clearly.
Suppose that $n$ Brownian particles 
  are strongly interacting 
  and moving together.
Then it is easily shown from the Langevin equation 
  that the effective diffusion coefficient
  for their center of mass is $D/n$,
  in the sense that 
\[
  \frac{\D}{\D{t}}\Av{R^2} = \frac{2D}{n} \relax.
\]
By replacing $n$ in the denominator 
  with $\mathcal{N}(\ld) = 1 + \rho_0\ld$,
  which is the number of particles 
  within the dynamical correlation length $\ld = \ld(t)$,
  Rallison \cite{Rallison.JFM186} obtained 
\begin{equation}  
  \Av{R^2}
  = \int_0
  \frac{2D\D{t}}{\mathcal{N}(\lambda)},
  \quad
  \lambda = \lambda(t) = \sqrt{{4\pi}Dt}
  \label{Rallison}.
\end{equation}
Upon integration, 
  Eq.~(\ref{Rallison}) gives 
  normal diffusion for small $t$,
  and for large $t$, 
  it gives sub\-diffusion with a logarithmic correction term.

The MCT approach provides with a nonlinear theory
  in the form of Eq.~(\ref{MCT.F}) for $F$
  and Eq.~(\ref{MCT.Fs}) for $\Fs$,
  in which the memory kernels are approximated 
  with the products of $F$ and $\Fs$.
As was mentioned in Sec.~\ref{sec:cont}, 
  the mathematical properties of the MCT kernels
  are such that they decay exponentially 
  for the most part.
This implies that SFD cannot be described 
  by the conventional MCT.
A possible approach 
  consists in adopting Eq.~(\ref{MCT.F}) for $F$
  and replacing Eq.~(\ref{MCT.Fs}) for $\Fs$
  with another equation for tracers 
  in which the four-point correlation 
  is directly taken into account.
The theory of Miyazaki and Yethiraj \cite{Miyazaki.JCP117}
  for rod polymers,
  as well as Kollmann's theory \cite{Kollmann.PRL90},
  belongs to this category.
We note that Kollmann \cite{Kollmann.PRL90} focused 
  on the long-time behavior 
  and therefore considered 
  only the long-wave limit of Eq.~(\ref{MCT.F}),
  so that the nonlinear effect is ignored  
  except for the change from $D$ to $\Dc$.

The theories of Fedders~\cite{Fedders.PRB17}
  and Abel \textit{et al.}~\cite{Abel.PNAS106}
  could be termed as a modified MCT approach,
  in which both Eq.~(\ref{MCT.F}) and Eq.~(\ref{MCT.Fs})
  are essentially retained,
  but $\Ms$ is modified.
Fedders~\cite{Fedders.PRB17} noticed 
  that the summation of the diagrams 
  must be performed with the restriction 
  corresponding to the ``no-passing'' rule.
In the formulation 
  of Abel \textit{et al.}~\cite{Abel.PNAS106},
  this restriction was taken into account 
  by a kind of re-weighting in diagrammatic expansion.
As a result, a wavenumber integral 
  that appears in an expression related to $\Ms$
  (the scaled irreducible memory function $G^{\text{irr}}$)
  is changed in a delicate way.
Without the ``no-passing'' rule,
  the original integral reads 
\begin{equation}
  G^{\text{irr}}(k,t)
  \propto \int\left[ 1 - \cos(p-q) \right] 
  \Fs(p,t) F(q,t)\,\D{p}
  \label{Abel.0}
\end{equation}
  and gives normal diffusion asymptotically.
This is replaced by 
\begin{equation}
  G^{\text{irr}}_{\text{modified}}(k,t)
  \propto \int
  \Fs(p,t) F(q,t)\,\D{p}
  \label{Abel.1}
\end{equation}
  due to the re-weighting,
  and it gives the correct anomalous diffusion.

Having reviewed 
  main existing theories on SFD,
  now let us compare the present theory with them.
The present theory is a nonlinear one, 
  consisting of the Lagrangian MCT equation (\ref{MCT})
  and the modified Alexander--Pincus formula (\ref{d8k}).
The adoption of the Lagrangian description 
  enabled us 
  to reproduce the asymptotic law for MSD 
  and calculate a correction to it 
  within the liquid-theoretical framework.
Some four-point space-time correlations 
  are also calculated analytically.

One of the main differences between Eq.~(\ref{MCT}) for $\C$
  and the corresponding Eulerian MCT equation (\ref{MCT.F})
  is that the diffusing entity in Eq.~(\ref{MCT}) 
  is the ``free volume'' between the particles,
  which is quite analogous to the diffusing vacancies 
  considered by 
  van Beijeren \textit{et al.} \cite{van-Beijeren.PRB28}
  and also by other authors.
While van Beijeren \textit{et al.} \cite{van-Beijeren.PRB28}
  gave up a systematic treatment of the memory effects
  in the vacancy dynamics,
  the present theory treats it with a systematic approximation.
The modified Alexander--Pincus formula (\ref{d8k})
  seems to be exact in the limit of large system size.
The formula (\ref{d8k}) itself is linear
  in regard to $\C$,
  though there is a hidden nonlinearity 
  in the mapping from the label distance 
  to the Eulerian--Euclidean distance.

The present theory gives a finite-time correction
  to the long-time asymptotic result, 
  as is shown in Eq.~(\ref{MSD.NL}).
The correction slightly differs 
  from that of the phenomenological equation (\ref{Rallison})
  by Rallison \cite{Rallison.JFM186}:
  probably this is attributable 
  to the inaccuracy of $\ld(t)$ or $\mathcal{N}(\ld)$
  assumed in Eq.~(\ref{Rallison}).
Another issue 
  that requires further consideration 
  is the relation 
  between the modified Alexander--Pincus formula (\ref{d8k})
  and the modified MCT equation for $\Fs$,
  which should be understood somehow in the future.

We emphasize that there are two origins of nonlinearity,
  and the present MCT-based approach
  is capable of treating both of them in principle.
One is the nonlinearity of the configurational entropy,
  from which the nonlinearity of $1/(1+\psi)$ 
  in Eq.~(\ref{1/rho}) originates.
The other is the nonlinearity
  whose coefficient involves $\sin{\rho_0\sigma k}$,
  which can be traced back 
  to the term including $U$ in Eq.~(\ref{*Q})
  and represents the effect of direct contact 
  between the particles.
Though we have omitted the analysis of the latter 
  to limit all the results 
  within the range of analytical calculation,
  it would be straightforward to deal with the cases
  in which these two nonlinearities are present,
  once a numerical scheme is constructed.

\subsection{Methodological insight into memory-correlation approaches}
\label{subsec:method}

In our derivation of Eqs.~(\ref{MCT}) and (\ref{M})
  for the Lagrangian correlation $\C$ 
  and the memory kernel $M$ associated with it,
  we took the Langevin equation for the density field 
  as the starting point
  and adopted a field-theoretical method 
  akin to the MSR formalism.
While the derived equation itself 
  has a form parallel to the Eulerian MCT equation (\ref{MCT.F}),
  the derivation processes are quite dissimilar.
Practically, 
  Eq.~(\ref{MCT.F}) is derived 
  directly from the microscopic equation of motion  
  by way of the Mori--Zwanzig 
  projection operator formalism 
  \cite{Gotze.RPP55,Goetze.Book2009,Reichman.JStat2005,Mori.PTP33}.
This is usually considered to be more convenient 
  than the field-theoretical derivation,
  because the latter suffers 
  from the difficulties due to the multiplicative noise,
  such as violation of the FDT.
Langevin equations with multiplicative noise 
  may draw a general criticism 
  for the {It{\^o}}--Stratonovich dilemma 
  \cite{van-Kampen.Book2007}, 
  though it can be avoided 
  when the Onsager coefficient satisfies a certain condition 
  \cite{Miyazaki.JPA38}.
Besides, in regard to the treatment 
  of the noise correlation itself,
  there seems to be a subtle confusion in the literature:
  compare 
  Eq.~(14) in Ref.~\cite{Dean.JPhAMG29},
  Eq.~(4) in Ref.~\cite{Andreanov.JStat2006},
  Eq.~(2) in Ref.~\cite{Kim.JPhA40}
  and Eq.~(6) in Ref.~\cite{Taloni.PRE78}.
All these difficulties 
  have made the Langevin equation for the density field, 
  such as Eq.~(\ref{*rho}),
  inconvenient as a starting point.

Interestingly,
  when the field $\psi(\xi,t)$ is adopted 
  instead of the conventional density field,
  the positions of the two methods are reversed.
In contrast to the projection of the particle motion 
  onto $\rho(\mb{r},t) = \sum_j\delta_j(\mb{r}-\mb{r}_j(t))$
  which can be performed naturally, 
  it is not evident how to project the motion of the particles 
  onto $\psi = \rho_0/\rho - 1$.
Direct employment of the microscopic definition of $\rho$
  for the denominator
  would give rise to delicate issues 
  concerning the procedure of coarse-graining.
Alternatively, 
  a one-dimensional projection-operator formalism
  may be possible by using $X_{j+1} - X_j$ 
  as the microscopic definition of $\psi$,
  but this leads to another complication, 
  because this definition of $\psi$ 
  depends on the assumption about the ordering of the particles.
On the side of the field-theoretical formalism,
  the difficulty of the multiplicative noise disappears
  quite naturally,
  which has allowed us to derive the Lagrangian MCT equation 
  without violating the FDT.

Thus the combination of the Lagrangian vacancy field 
  with the field-theoretical formalism 
  is not less advantageous 
  than the conventional projection operator route 
  with the Eulerian density field.
The new route that leads to the Lagrangian MCT 
  deserves further exploration, 
  especially if it may guide us 
  to some improved theories of three-dimensional systems 
  in the future.
In closing the current section,
  let us discuss this possibility.

\subsection{Future directions: possible relevance to glassy dynamics}
\label{subsec:future}

We have implemented the Lagrangian description 
  by explicitly introducing the label variable $\xi$
  and thereby constructing a stretchable coordinate system
  that sticks to the cages everywhere.
Probably some aspects of glassy dynamics, 
  such as dynamical heterogeneity 
  characterizable by bond breaking 
  \cite{Yamamoto.PRE58,Shiba.arXiv1205,Kawasaki.arXiv1210},
  may require the Lagrangian description by nature
  when its continuum counterpart is sought.

The Lagrangian description in higher dimensions 
  may not be so simple as in one-dimensional cases,
  but it is possible.
In three-dimensional cases, 
  a triplet of label variables $(\xi,\eta,\zeta)$
  is expected to be related 
  with $\rho$ and $\mb{Q}$
  by equations analogous to Eq.~(\ref{L3+4});
  see Eqs.~(6.6) and (6.7) in Ref.~\cite{Ooshida.JPSJ80}.
Besides, 
  we could adopt some methods 
  from three-dimensional theories of turbulence
  in which Lagrangian correlations are used 
  \cite{Kraichnan.PhF8,Kaneda.JFM107,Kida.JFM345}.
Turbulence theoreticians have even considered 
  the Lagrangian dynamics of a tetrad (four material points)
  \cite{Chertkov.PhF11},
  whose two-time correlation  
  involves eight points in the space-time.

In contrast to the ``Eulerian'' (standard) MCT 
  in which cage effects are represented 
  by the memory kernel (not successful in SFD),
  the Lagrangian theory can dispense with the memory integral
  as far as the asymptotic behavior is concerned.
A pivotal role is played 
  by the modified and extended Alexander--Pincus formulae 
  in Eqs.~(\ref{d8k}) and (\ref{AP+}).
Its linear version,
  namely Eq.~(\ref{AP}) or its multidimensional extension,
  has been used in the context of glassy dynamics
  by several authors
  \cite{Toninelli.PRE71,Lefevre.PRE72},
  who limited themselves to the approximation 
  with linear elasticity.
Since the Eulerian and the Lagrangian variables 
  are approximately interchangeable
  in the description of small elastic deformation,
  they did not bother to distinguish the two descriptions.
Needless to say, this treatment fails 
  when the system is more liquid-like.
In an attempt to introduce the $\alpha$ relaxation
  into the calculation of $\chi_4$
  based on the ``elastic'' theory,
  Toninelli \textit{et al.} feared 
  that it would make the model inconsistent,
  because the underlying lattice,
  needed to define the deformation field,
  would be totally melted 
  \cite{Toninelli.PRE71}.
Probably this is too pessimistic:
  the ``melting'' of the lattice 
  does not make the theory totally inconsistent
  but requires more careful distinction 
  between the Eulerian and the Lagrangian coordinates.

There will be another modification 
  to the theory of Toninelli \textit{et al.},
  when departing from the linear elasticity
  and trying to consider liquid-like behavior.
Their formula corresponding to Eq.~(\ref{AP+}), 
  namely Eq.~(A3) in Ref.~\cite{Toninelli.PRE71},
  reads
\begin{equation}
  \Av{R(\di)R(\mb{0})} \propto 
  \int\frac{1-e^{-D{\mb{k}^2}t}}{\mb{k}^2}
  e^{-\II{\mb{k}\cdot\di}}\D^{\nd}\mb{k}
  \label{Toninelli.A3}
\end{equation}  
  in our notation,
  as they seem to have identified 
  the Langevin equation for the displacement field 
  with the Edwards--Wilkinson equation \cite{Edwards.PRSLA381}.
In regard to Eq.~(\ref{Toninelli.A3}),
  we suspect
  that the vectorial character of the displacement 
  is not adequately taken into account.
Probably one needs to decompose 
  $\mb{R}$ into the longitudinal and transverse components,
  and treat them more carefully.
Unlike the two sound modes in elastic solids,
  the two modes in the liquids can have 
  quite different nature:
  the liquid may resist compression strongly
  but the resistance to shear may be much weaker. 

\begin{figure}
  \includegraphics[clip,width=0.70\linewidth]{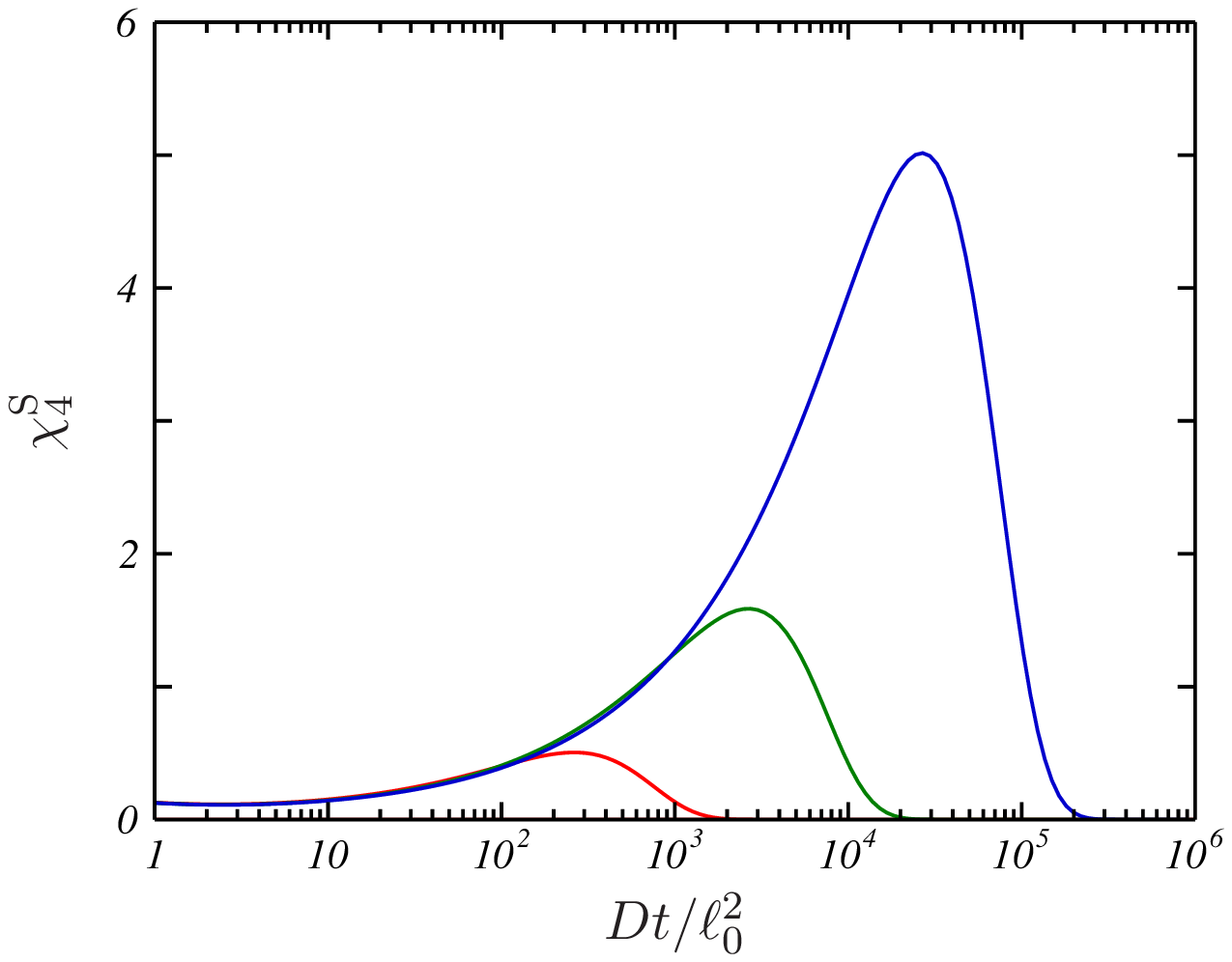}
  \caption{\label{Fig:chi4s.3D}%
  (Color online) 
  Behavior of $\chiS$ 
  given by Eq.~(\ref{chiS.cage+hop}),
  calculated through the three-dimensional 
  two-body displacement distribution function
  by phenomenologically extending the present theory 
  to the three-dimensional cases.
  The three curves 
  correspond to different values of $\tau$: 
  from left to right, $\tau = 10^3\ell_0^2/D$, 
  $10^4\ell_0^2/D$, and $10^5\ell_0^2/D$.
  }
\end{figure}

In Eq.~(\ref{RR.3D}), 
  we have proposed to define the three-dimensional 2pDC 
  as a tensorial quantity $\ChiR$.
Due to the isotropy 
  and the reflectional symmetry of the system,
  $\ChiR$ must be a sum 
  of the longitudinal and the transverse components:
\begin{equation}
  \ChiR
  = \Xl\frac{\di\otimes\di}{\di^2}
  + \Xtr\left( 
  \openone - \frac{\di\otimes\di}{\di^2} 
  \right)
  \notag.
\end{equation}
It is quite likely 
  that $\Xl$ and $\Xtr$
  will be characterized by different correlation lengths.
Taking the two different correlation lengths into account,
  we can extend the present theory phenomenologically 
  to the three-dimensional cases.
From the inferred distribution function
  for the displacements of two particles,
  shown in Appendix~\ref{app:3D},
  we can calculate $\chiS$
  in the same way as in Subsec.~\ref{subsec:chiS},
  as
\begin{equation}
  \chiS 
  \sim \frac{1}{\kT}\times 
  \frac{(1-\alpha)^2 \ld_\para \ld_\perp^2}%
  {\left( 1 + \dfrac{2X^0}{a^2} \right)^3}
  \label{chiS.cage+hop},
\end{equation}
  where $\ld_\para = \ld_\para(t)$ 
  and   $\ld_\perp = \ld_\perp(t)$ 
  denote the two correlation lengths, 
  $X^0 = X^0(t)$ is related to the MSD of the caged particles,
  and $\alpha = \alpha(t)$ stands for the relative number 
  of the particles that have hopped.
Assuming the $t$-dependence of these four quantities 
  phenomenologically as 
\begin{gather*}
  \ld_\para = \frac{\sqrt{Dt}}{1+\sqrt{t/\tau}},  \quad
  \ld_\perp = \left(1 + \sqrt{t/\tau}\right)\ell_0, \quad \\ 
  X_0 = \frac{Dt}{1 + Dt/\ell_0^2}, \quad
  \alpha = \tanh\frac{t}{\tau} \notag, 
\end{gather*}
  we can plot $\chiS$ against $t$
  for some different values of the time scale $\tau$.
The curves in Fig.~\ref{Fig:chi4s.3D} 
  reproduce some basic features of the $\Q$-based $\chi_4$
  calculated by La\v{c}evi\'c \textit{et al.} 
  \cite{Lacevic.JCP119},
  such as the shape of the uphill
  that looks steeper near the peak 
  in this semi-log plot.

To go beyond the linear theory,
  $\nd$-dimensional versions 
  of the extended Alexander--Pincus formula
  should be developed.
We suppose that 
  the $\nd$-dimensional formula 
  will contain,
  instead of $\C$, 
  correlations of \emph{deformation tensor}.
This is not so formidable 
  as it may appear,
  because many components of the correlation tensor 
  will turn out to vanish or 
  to have the same value as some other component.
The formula is now under development 
  and will be reported elsewhere.

It is also noteworthy  
  that, in the derivation of MCT, 
  the difficulty of FDT violation 
  disappeared quite naturally
  due to the Lagrangian description.
As is pointed out 
  by Miyazaki \& Reichman \cite{Miyazaki.JPA38},
  it has been difficult to construct 
  realistic models
  which does not violate FDT
  and can incorporate
  the effect of structural changes embodied in $S(k)$
  at the same time.
Since the ``Lagrangian MCT'' 
  is now shown to be consistent with FDT,
  study of its behavior for different $S(k)$ 
  may be quite intriguing.

For possible extensions in the future,
  we can mention several directions.
For example,
  one may include weak attractive interaction 
  and analyze the effect of the change in $S(k)$
  on the transient behavior of SFD.
One may also study non\-equilibrium behavior
  by driving the particles with an external force
  or changing the temperature suddenly.
Another interesting proposal
  is to permit overtaking as a rare event,
  which may play the role of the $\alpha$ relaxation.
We performed simulations 
  with some finite interaction potential,
  and a preliminary result \cite{Ooshida.Nagare31} shows 
  that normal diffusion is observed for $\Vmax = \kT$,
  while for $\Vmax = 5\kT$ 
  the behavior is essentially that of SFD
  in the time scale of the simulation.
The problem is to make a theory that can handle   
  the crossover between the two limiting cases.
The theory 
  allowing for rare overtaking events 
  may bridge the gap between the purely one-dimensional SFD
  and the behavior of three-dimensional rod polymers 
  \cite{Rallison.JFM186,Miyazaki.JCP117}.

We could also study 
  \emph{double-file diffusion},
  which would be analogous to two-lane models of traffic flows.
If a ``lane interaction'' is also introduced,
  the system would have also something common 
  with the Matsukawa--Fukuyama model of friction 
  formulated on a ladder lattice 
  \cite{Matsukawa.PRB49,Yoshino.PTPS184}.
The study of the double-file diffusion 
  may shed light to many related systems
  in which 
  frustration is dynamically created and annihilated, 
  such as template-assisted pattern formation 
  of colloid particles on a substrate with parallel channels
  \cite{Mondal.PRE85},
  frustrated Josephson-junction arrays
  in a magnetic field
  \cite{Yoshino.PTPS184,Yoshino.PRL105}, 
  and---hopefuly---also three-dimensional 
  dense colloidal suspensions.

\section{Conclusion}
\label{sec:conc}

We have developed a nonlinear theory of SFD 
  with a liquid-theoretical approach.
The theory 
  consists of the modified Alexander--Pincus formula (\ref{d8k})
  and the Lagrangian MCT equation (\ref{MCT}),
  which gives not only the established 
  Hahn--K{\"a}rger--Kollman law 
  on the long-time asymptotic behavior of $\Av{R^2}$,
  but also a finite-time correction to it,
  as is shown in Eq.~(\ref{MSD.NL}).
Though we have focused 
  on the nonlinearity of the configurational entropy,
  the proposed theoretical scheme makes it possible 
  to deal with the other nonlinear effects 
  in the thermodynamic potential as well,
  if the MCT equation is solved numerically.

Using this scheme,
  we have demonstrated 
  how to calculate four-point space-time correlations,
  such as the 2pDC and $\chiS$. 
These four-point space-time correlations
  quantify the collective dynamics in SFD
  as a simple model of ideal cages
  involving numerous particles.
The long-time behavior of $\chiS$ 
  exhibits convergence to a finite value,
  given by $(\chiS)_{\text{coll}}$ in Eq.~(\ref{chi4s.coll}),
  which is an increasing function of $\rho_0$.

The present work,
  in combination with the previous one~\cite{Ooshida.JPSJ80},
  is intended as several first steps toward a future theory 
  of three-dimensional glassy liquids,
  which will make it possible, for example, 
  to replace the semi-phenomenological curves 
  for $\chiS$ in Fig.~\ref{Fig:chi4s.3D}
  with a first-principle theoretical calculation.
Although the present theory is still embryonic, 
  it already suggests 
  that one of the important ingredients of the future theory 
  may be the displacement distribution function
  of two or more particles.
In the case of computational analysis,
  probably we should not insist 
  on some favorite statistical quantity alone,
  nor content ourselves 
  with the single-particle van Hove function,
  but try to deduce some suitable distribution function
  behind the computed statistical quantities.
The present analysis of one-dimensional cage dynamics
  and the concepts used for it 
  will provide a useful framework 
  both for analysis of numerical data 
  and for new development of theory of glassy liquids. 


\begin{acknowledgments}
  We express our cordial gratitude
    to Kunimasa Miyazaki, 
    whose lecture on SFD~\cite{Miyazaki.JPS08s} 
    motivated our study,
    and whose comments were helpful and enlightening.
  We also appreciate fruitful discussions
    with Shin-ichi Sasa, 
    Takeshi Kawasaki, Hajime Yoshino, and So Kitsunezaki.
  This work was supported 
    by Grants-in-Aid for Scientific Research 
    (\textsc{Kakenhi})  
    (C) No.~21540388 and (C) No.~24540404, JSPS (Japan).
\end{acknowledgments}

\appendix

\section{Numerical calculations}
\label{app:num}

Here we describe 
  how we integrated 
  the one-dimensional Langevin equation~(\ref{Langevin.X}),
  and also how we evaluated the statistical quantities,
  including $\Av{R^2}$, $\Dc$, and $S$,  
  from the numerical solution. 
As the system contains $N$ particles 
  in a periodic box of the size $L$,
  the mean density is given by $\rho_0 = N/L$.

The potential $V$ in Eq.~(\ref{Langevin.X})
  was specified as  
\begin{equation}  
  V(r) =
  \begin{cases}
    \Vmax\left(1-\dfrac{|r|}{\sigma}\right)^2 & (|r| \le\sigma) \\
    0                                         & (|r|  > \sigma)
  \end{cases}  
  \label{V=}
\end{equation}  
  with $\Vmax \gg \kT$. 
In the present numerical calculations 
  we adopted the value $\Vmax = 50\,\kT$, 
  which is high enough 
  to forbid the overtaking of the particles completely.

The random forcing is the zero-mean Gaussian noise
  whose variance is given as 
\begin{equation}
  \left\langle{f_i(t) f_j(t')}\right\rangle
   = 2D \delta_{ij} \delta(t-t')  
   \label{f1}.
\end{equation}
Computationally, 
  the delta function in Eq.~(\ref{f1}) 
  was discretized with the time interval $\dtF$,
  and as the values of $(f_0,f_1,\ldots,f_{N-1})$ 
  for each time interval, 
  $N$ independent Gaussian random numbers 
  with the variance $2D/\dtF$
  were generated 
  with the Mersenne twister and the Box--Muller transform.

With $V$ and $f_i$ given as above, 
  in nondimensionalization of the governing equation
  using $\sigma$, $\sigma^2/D$, and $m$
  as the units of length, time, and mass,
  there appears a nondimensional parameter 
  specifiable as the ratio of $\tauB = m/\mu$
  to the time unit $\sigma^2/D$, 
  which we chose as $1:1$.
Then the time integration of Eq.~(\ref{Langevin.X})
  was performed with a Verlet-like scheme.
The time step $\Delta{t}$
  was taken equal to a hundredth of the time unit $\sigma^2/D$,
  and the renewal interval of the random forcing, $\dtF$, 
  was chosen to be the same as the time step:
\begin{equation}
  \Delta{t} = \dtF = 10^{-2}\times\frac{\sigma^2}{D}
  \label{dt=}.
\end{equation}
We also tested 
  some different choices of $\Delta{t}$ and $\dtF$,
  such as 
\[
  \frac{(\Delta{t},\dtF)}{\sigma^2/D}
  = \left({10^{-3}},{10^{-3}}\right)
  \quad
  \text{or} 
  \quad
  \left({10^{-3}},{10^{-2}}\right).  
\]
To bring the system into equilibrium, 
  we started each calculation at $t=-\Tw$, 
  introducing a sufficiently long waiting time $\Tw$
  (typically $\Tw = 10^4 \sigma^2/D$),
  and waited till $t=0$.
Then, from the simulation data 
  recorded for 
  $0 \le t \le \tmax$, 
  we calculated a desired statistical quantity 
  as the average for $n$ samples 
  extracted from the data by a time shift.
For example, 
  $\Av{R^2}$ is calculated as 
\begin{equation}
  \Av{R(t')^2} 
  = {\frac{1}{nN}}
  \sum_{l=0}^{n-1}\sum_{i=0}^{N-1} 
  \left[ X_i(t_l + t') - X_i(t_l) \right]^2
  \label{MSD.av},
\end{equation}
  where $t_l$ denotes 
  the starting time of the $l$-th sample.
Care must be taken 
  so that the maximal value of $t'$ 
  in Eq.~(\ref{MSD.av}),
  which equals the span of each sample, 
  should not exceed the waiting time $\Tw$;
  a result for a longer span 
  will expose insufficiency of equilibration.
Typically 
  we chose $\tmax = 5\Tw$, $n = 100$,
  and $t_l = l(\tmax-\Tw)/n$,
  allowing the samples to overlap.

The computation of $\chiR(\tilde{d},t)$,   
  shown in Fig.~\ref{Fig:RR-vs-init},
  was performed with discretization of Eq.~(\ref{RR.i1})
  in which the delta function was approximated 
  by a statistical bin $0.5\,\sigma$ in width.
After recording $X_i(t_l)$ for every particle, 
  we classified every pair $(i,j)$
  into a statistical bin 
  according to the ``initial'' distance 
  $X_j(t_l) - X_i(t_l)$,
  so that the $\kappa$-th bin contains the pairs 
  for which $| X_j(t_l) - X_i(t_l) - \tilde{d}_\kappa |$
  is smaller than the half width of the bin.
Then we calculated 
  $\chiR(\tilde{d}_\kappa,t)$
  as the average of ${R_i(t)}{R_j(t)}$ 
  for the $\kappa$-th bin,
  where $R_i(t) = X_i(t_l+t) - X_i(t_l)$.
If, instead, 
  the sum of the absolute values of the data in each bin 
  were calculated,
  this would be analogous to the quantity  
  studied by Donati \textit{et al.} \cite{Donati.PRL82}.

The collective diffusion coefficient $\Dc$ 
  is determined from the temporal decay 
  of the dynamical structure factor
  $F(q,t)$ \cite{Lutz.PRL93}.
We computed $F(q,t)$ 
  for $0 < q \ll \rho_0$ 
  and made a linear fit 
  for $\log{F(q,t)} = \log{S(q)} -{\Dc}q^2 t$
  to obtain the values of $S(q)$ and $\Dc$
  as the fitting parameters. 
After taking an average for several small values of $q$,
  the results are summarized 
  in Table~\ref{Tab:Dc}.
These values are used 
  in evaluation of the theoretical predictions,
  such as Eq.~(\ref{R1*R2}),
  and also for rescaling of the horizontal axis 
  in Fig.~\ref{Fig:MSD.NL}.

\begin{table}
  \caption{\label{Tab:Dc}%
    Numerical values of $S(0)$ and $\Dc/D$
    computed for three different values of the density.
  }
  \begin{tabular}{lccc}\hline
    $\rho_0$  &
    $(1/4) \sigma^{-1}$ &
    $(1/8) \sigma^{-1}$ &
    $(1/16)\sigma^{-1}$ \\\hline
    $1-2\rho_0\sigma$  & $0.500$  & $0.750$  & $0.875$ \\
    $S(0)$             & $0.624$  & $0.787$  & $0.888$ \\
    $\Dc/D$            & $1.59$   & $1.27$   & $1.12$  \\ 
    $S(0)\sqrt{\Dc/D}$ & $0.79$   & $0.89$   & $0.94$  \\
    \hline
  \end{tabular} 
\end{table}


\section{Direct-interaction approximation}
\label{app:DIA}

The main idea of DIA~\cite{Kraichnan.JFM5,Goto.PhysicaD117} 
  for evaluation of the triple correlation 
\[
  \Av{\check\psi(-p,t)\check\psi(-q,t)\check\psi(-k,0)}
  = \Av{\check\psi(p,t)\check\psi(q,t)\check\psi(k,0)}^*
  \relax,
\]
  with the asterisk denoting the complex conjugate,
  is to utilize the property 
  of $\Vertex = (\Vertex_\alpha^{\beta\gamma})$
  that, out of its $N^3$ components, 
  almost all are zero 
  because the condition $\alpha+\beta+\gamma = 0$ 
  is not satisfied.
The nonzero components of $\Vertex$
  constitute what we call triad interactions:
  rewriting Eq.~(\ref{*3+}) as 
\begin{equation}
  (\dt + \mu_\alpha)\check\psi(\alpha,t) 
  =
  \smash{\sum_{\beta,\gamma}} \vphantom{\sum}
  {\Vertex_\alpha^{\beta\gamma}} 
  \check\psi^*(\beta,t) \check\psi^*(\gamma,t)
  + \rho_0 \check{f}_{\text{L}}(\alpha,t)
  \label{*3++}\tag{$\ref{*3+}'$}
\end{equation}
  makes it clearer 
  that $\Vertex_\alpha^{\beta\gamma}$ 
  engages in connecting the ``triad'' that consists 
  of $\alpha$, $\beta$, and $\gamma$.
If we visualize each triad interaction
  as a triangle on a graph,
  the property of $\Vertex$ is such 
  that no triangle shares its side with other triangles.
Therefore,
  if we ``switch off'' a single triad, 
  say, $\{p,q,k\}$ (with which we mean 
  $\Vertex_k^{pq}$, $\Vertex_k^{qp}$,
  $\Vertex_p^{qk}$, $\Vertex_p^{kq}$,
  $\Vertex_q^{kp}$, and $\Vertex_q^{pk}$;
  note the symmetry in regard to the interchange 
  of the superscripts,
  $\Vertex_k^{pq} = \Vertex_k^{qp}$ etc.),
  the \emph{direct interactions} between 
  the three modes $p,q,k$ are lost.

To concretize this idea, 
  let us suppose that an artificial forcing term
\[
  I_\alpha
  = 
  -\theta(t-t_0) \times
  \begin{cases}
    2\Vertex_k^{pq} \check\psi^*(p,t)\check\psi^*(q,t)& (\alpha=k)\\
    2\Vertex_p^{qk} \check\psi^*(q,t)\check\psi^*(k,t)& (\alpha=p)\\
    2\Vertex_q^{kp} \check\psi^*(k,t)\check\psi^*(p,t)& (\alpha=q)\\
    0 & (\text{otherwise}),
  \end{cases}
\]
  designed to cancel a single triad $\{p,q,k\}$,
  is applied to the system given by Eq.~(\ref{*3++}).
We denote the solution to this artificial system 
  with $\check\psi_0 = \check\psi_0(\{p,q,k\};\alpha,t)$.
One of the two main assumptions of DIA 
  is that the three selected modes, 
  namely $\check\psi_0(p,t)$, $\check\psi_0(q,t)$,
  and $\check\psi_0(k,t)$ in this case,
  become uncorrelated,
  since the forcing $I$ cancels the direct interactions.
On the other hand, 
  $I$ is regarded as a small perturbation,
  because it cancels only a single triad interaction
  and there remain still a large number
  of triads connecting, say, $p$ and $q$ indirectly.
Therefore 
  the difference 
  $\check\psi_1 = \check\psi - \check\psi_0$
  is assumed to be small,
  which is the second main assumption of DIA.
  
Due to these assumptions, 
  the triple correlation is expanded as 
\begin{align}
  \Av{\check\psi(p,t)\check\psi(q,t)\check\psi(k,0)} 
  &= 
  \Av{\check\psi_0(p,t)\check\psi_0(q,t)\check\psi_0(k,0)}  \notag\\&\quad{}
  + \Av{\check\psi_1(p,t)\check\psi_0(q,t)\check\psi_0(k,0)}\notag\\&\quad{}
  + \Av{\check\psi_0(p,t)\check\psi_1(q,t)\check\psi_0(k,0)}\notag\\&\quad{}
  + \Av{\check\psi_0(p,t)\check\psi_0(q,t)\check\psi_1(k,0)}\notag\\&\quad{}
  + O(\check\psi_1^2) 
  \label{DIA.XXX.0+1},
\end{align}
  and the zero-th term, 
  $\Av{\check\psi_0(p,t)\check\psi_0(q,t)\check\psi_0(k,0)}$,
  vanishes.
Since $I$ is a small perturbation
  and $\psi_1$ is a response to it,
  formally $\psi_1$ can be expressed 
  in terms of the propagator $G$ as
\begin{equation}
  \check\psi_1(\alpha,t) 
  = -\int_{t_0}^t \D{t'}
  \sum_{\alpha'} {G(\alpha,t;\,\alpha',t')} I_{\alpha'}(t')
  \label{DIA.psi1=}
\end{equation}
  for $t > t_0$.
We substitute Eq.~(\ref{DIA.psi1=}) 
  into each term on the right-hand side of Eq.~(\ref{DIA.XXX.0+1})
  to find, to our surprise,
  that the result is \emph{naturally factorized}
  due to the assumption of DIA 
  that $\check\psi_0(p,t)$, $\check\psi_0(q,t)$,
  and $\check\psi_0(k,t)$ are uncorrelated.

By applying the decomposition 
  $\check\psi = \check\psi_0 + \check\psi_1$ 
  to each triple correlation term 
  in the equation for $\dt\C$,
  we are led to Eq.~(\ref{DIA.C}).
Similarly, 
  the equation for $\dt\bar{G}$ 
  contains $\Av{{\check\psi_0}G}$,
  which is evaluated 
  with the aid of the DIA decomposition of $G$,
  resulting in Eq.~(\ref{DIA.G}).  
For more details, 
  see Refs.~\cite{Goto.PhysicaD117,Matsumoto.PTPS195}.


\section{Three-dimensional calculation of $\chiS$}
\label{app:3D}

Here we outline 
  how to calculate $\chiS$
  by phenomenologically extending the present theory 
  to the three-dimensional cases.
The displacement correlation tensor $\ChiR$
  is related to the distribution function 
  $P(\di; \mb{R}_i,\mb{R}_j)$ 
  by
\begin{equation}
  \ChiR(\di)
  = \iint
  {\mb{R}_i}\otimes{\mb{R}_j}
  P(\di; \mb{R}_i,\mb{R}_j)\D^3\mb{R}_i\D^3\mb{R}_j
  \notag;
\end{equation}
in what follows, 
  the $\di$-dependence of $P$ 
  is taken for granted 
  and therefore omitted.
The relation can be inverted 
  if the functional form of $P$ is known.
In particular, 
  if a multivariate Gaussian distribution 
  (which we denote with $P_0$) is assumed,
  $P = P_0$ can be factorized as 
\begin{equation}
 P_0(\mb{R}_i,\mb{R}_j) 
 = 
 {P_\para}(\mb{R}_i^\para, \mb{R}_j^\para) 
 {P_\perp}(\mb{R}_i^\perp, \mb{R}_j^\perp)
 \label{Gauss3D}
\end{equation}
  by splitting $\mb{R}$
  into the longitudinal and transverse components
  as $\mb{R} = {\mb{R}^\para} + {\mb{R}^\perp}$
  (so that ${\mb{R}^\para}\para\di$ and ${\mb{R}^\perp}\perp\di$).
Then we introduce $X^0$ 
  such that 
\[
  X_\para(\diScalar=0)
  = X_\perp(\diScalar=0)
  = X^0
\]
  and write the two factors 
  explicitly as 
\begin{align}
 P_\para
 &= \frac{1}{2\pi\sqrt{\Delta_\para}} 
 \exp\left[ 
 -\frac{{X^0}({R_i^\para}^2 + {R_j^\para}^2) 
 - 2 \Xl(\diScalar) {{R_i^\para}{R_j^\para}}}
 {2\Delta_\para} 
 \right]
 \notag, 
 \\
 P_\perp
 &= \frac{1}{(2\pi)^2{\Delta_\perp}} \times \notag\\&\qquad
 \exp\left[ 
 -\frac{{X^0}({\mb{R}_i^\perp}^2 + {\mb{R}_j^\perp}^2) 
 - 2 \Xtr(\diScalar) 
 {{\mb{R}_i^\perp}\cdot{\mb{R}_j^\perp}}}
 {2\Delta_\perp} 
 \right]
 \notag, 
\end{align}
  where 
  $\Delta_\para
  =
  \left({X^0}\right)^2 
  - {X_\para}(\diScalar)^2
  $ etc.
  
Then the calculation of $\chiS$ 
  will be carried out in a way analogous to the 1D cases.
Subsequently,
  assuming that two correlation lengths,
  $\ld_\para = \ld_\para(t)$ and 
  $\ld_\perp = \ld_\perp(t)$,
  can be introduced 
  so that 
\begin{align*}
 &\Xl  \sim X^0(t)\,\Phi_\para(\diScalar/\ld_\para(t)),\\
 &\Xtr \sim X^0(t)\,\Phi_\perp(\diScalar/\ld_\perp(t)),
\end{align*}
  we estimate the number of the particles
  contributing to the sum 
  as $N_{\text{coll}} \sim \rho_0 \ld_\para \ld_\perp^2 $,
  which leads to 
\begin{equation}
  \chiS 
  \sim \frac{1}{\kT}\times 
  \frac{\ld_\para \ld_\perp^2}%
  {\left( 1 + \dfrac{2X^0}{a^2} \right)^3}
  \label{chiS.Gauss3D}
\end{equation}
  as a three-dimensional counterpart of Eq.~(\ref{chi4s.1D.estimation}).

Of course, Eq.~(\ref{chiS.Gauss3D}) needs to be modified 
  by taking $\alpha$ relaxation into account.
We introduce $\alpha = \alpha(t)$ 
  denoting the relative number of the particles 
  that have hopped by the time $t$,
  and assume the distribution function 
  in the form 
\begin{multline}
 P(\mb{R}_i,\mb{R}_j)  \\ {} 
 = 
 \begin{cases}
   (1-\alpha)^2 P_0 + 2 \alpha(1-\alpha) P_1 + \alpha^2 P_2
   & (i \ne j) \\
   (1-\alpha) P_0 + \alpha P_2
   & (i = j) \relax,
 \end{cases}
 \\ \quad
 \label{P0+P1+P2}
\end{multline}
  where $P_0$, governing the pairs of caged particles,
  is given by Eq.~(\ref{Gauss3D}).
Note that, by integrating $P(\mb{R}_i,\mb{R}_j)$
  in regard to the second argument $\mb{R}_j$,
  Eq.~(\ref{P0+P1+P2}) reduces to the van Hove function 
  in the form 
\[
  P(\mb{R}) 
  = (1-\alpha)P_{\text{cage}}(\mb{R}) 
  + \alpha P_{\text{hop}}(\mb{R})
  \relax.
\]
This implies 
  that $\alpha = \alpha(t)$ can be determined, in principle,
  as a fitting parameter for the van Hove function. 
If we assume, for simplicity, 
  that the correlation of displacements is totally lost 
  after the hopping,
  we have 
\begin{align*}
  P_1(\mb{R}_i,\mb{R}_j) 
  &= 
  \frac{P_{\text{hop}}(\mb{R}_i) P_{\text{cage}}(\mb{R}_j)
  + P_{\text{cage}}(\mb{R}_i) P_{\text{hop}}(\mb{R}_j)
  }{2}
  \relax,
  \\
  P_2(\mb{R}_i,\mb{R}_j) 
  &=   
  P_{\text{hop}}(\mb{R}_i) P_{\text{hop}}(\mb{R}_j)
  \notag.
\end{align*}
Using the distribution function in Eq.~(\ref{P0+P1+P2})
  supplemented with the above expressions,
  we evaluate $\chiS$ and obtain Eq.~(\ref{chiS.cage+hop}).


\providecommand{\newblock}{\relax}
\bibliography{diffusion,sgm,turbulence,statmech,Book,misc,%
              note1308,ref1212,ref1305}

\begin{thebibliography}{104}
\expandafter\ifx\csname natexlab\endcsname\relax\def\natexlab#1{#1}\fi
\expandafter\ifx\csname bibnamefont\endcsname\relax
  \def\bibnamefont#1{#1}\fi
\expandafter\ifx\csname bibfnamefont\endcsname\relax
  \def\bibfnamefont#1{#1}\fi
\expandafter\ifx\csname citenamefont\endcsname\relax
  \def\citenamefont#1{#1}\fi
\expandafter\ifx\csname url\endcsname\relax
  \def\url#1{\texttt{#1}}\fi
\expandafter\ifx\csname urlprefix\endcsname\relax\def\urlprefix{URL }\fi
\providecommand{\bibinfo}[2]{#2}
\providecommand{\eprint}[2][]{\url{#2}}

\bibitem[{\citenamefont{Squires and Quake}(2005)}]{Squires.RMP77}
\bibinfo{author}{\bibfnamefont{T.~M.} \bibnamefont{Squires}} \bibnamefont{and}
  \bibinfo{author}{\bibfnamefont{S.~R.} \bibnamefont{Quake}},
  \bibinfo{journal}{Rev. Mod. Phys.} \textbf{\bibinfo{volume}{77}},
  \bibinfo{pages}{977} (\bibinfo{year}{2005}).

\bibitem[{\citenamefont{Hale and Mitchell}(2001)}]{Hale.NanoL1}
\bibinfo{author}{\bibfnamefont{M.~S.} \bibnamefont{Hale}} \bibnamefont{and}
  \bibinfo{author}{\bibfnamefont{J.~G.} \bibnamefont{Mitchell}},
  \bibinfo{journal}{Nano Letters} \textbf{\bibinfo{volume}{1}},
  \bibinfo{pages}{617} (\bibinfo{year}{2001}).

\bibitem[{\citenamefont{Kusumi et~al.}(1993)\citenamefont{Kusumi, Sako, and
  Yamamoto}}]{Kusumi.BPJ65}
\bibinfo{author}{\bibfnamefont{A.}~\bibnamefont{Kusumi}},
  \bibinfo{author}{\bibfnamefont{Y.}~\bibnamefont{Sako}}, \bibnamefont{and}
  \bibinfo{author}{\bibfnamefont{M.}~\bibnamefont{Yamamoto}},
  \bibinfo{journal}{Biophysical Journal} \textbf{\bibinfo{volume}{65}}
  (\bibinfo{year}{1993}).

\bibitem[{\citenamefont{Alberts et~al.}(2007)\citenamefont{Alberts, Johnson,
  Lewis, Raff, Roberts, and Walter}}]{Cell5.Book2007}
\bibinfo{author}{\bibfnamefont{B.}~\bibnamefont{Alberts}},
  \bibinfo{author}{\bibfnamefont{A.}~\bibnamefont{Johnson}},
  \bibinfo{author}{\bibfnamefont{J.}~\bibnamefont{Lewis}},
  \bibinfo{author}{\bibfnamefont{M.}~\bibnamefont{Raff}},
  \bibinfo{author}{\bibfnamefont{K.}~\bibnamefont{Roberts}}, \bibnamefont{and}
  \bibinfo{author}{\bibfnamefont{P.}~\bibnamefont{Walter}},
  \emph{\bibinfo{title}{Molecular Biology of the Cell}}
  (\bibinfo{publisher}{Garland Science}, \bibinfo{address}{New York},
  \bibinfo{year}{2007}), \bibinfo{edition}{5th} ed., ISBN
  \bibinfo{isbn}{9780815341055}.

\bibitem[{\citenamefont{Sekimoto}(2010)}]{Sekimoto.Book2010}
\bibinfo{author}{\bibfnamefont{K.}~\bibnamefont{Sekimoto}},
  \emph{\bibinfo{title}{Stochastic energetics}}
  (\bibinfo{publisher}{Springer-Verlag}, \bibinfo{year}{2010}).

\bibitem[{\citenamefont{H{\"a}nggi and Marchesoni}(2009)}]{Hanggi.RMP81}
\bibinfo{author}{\bibfnamefont{P.}~\bibnamefont{H{\"a}nggi}} \bibnamefont{and}
  \bibinfo{author}{\bibfnamefont{F.}~\bibnamefont{Marchesoni}},
  \bibinfo{journal}{Rev. Mod. Phys.} \textbf{\bibinfo{volume}{81}},
  \bibinfo{pages}{387} (\bibinfo{year}{2009}).

\bibitem[{\citenamefont{Burada et~al.}(2009)\citenamefont{Burada, H{\"a}nggi,
  Marchesoni, Schmid, and Talkner}}]{Burada.CPhC10}
\bibinfo{author}{\bibfnamefont{P.~S.} \bibnamefont{Burada}},
  \bibinfo{author}{\bibfnamefont{P.}~\bibnamefont{H{\"a}nggi}},
  \bibinfo{author}{\bibfnamefont{F.}~\bibnamefont{Marchesoni}},
  \bibinfo{author}{\bibfnamefont{G.}~\bibnamefont{Schmid}}, \bibnamefont{and}
  \bibinfo{author}{\bibfnamefont{P.}~\bibnamefont{Talkner}},
  \bibinfo{journal}{ChemPhysChem} \textbf{\bibinfo{volume}{10}},
  \bibinfo{pages}{45} (\bibinfo{year}{2009}).

\bibitem[{\citenamefont{Zwanzig}(1992)}]{Zwanzig.JPhysChem96}
\bibinfo{author}{\bibfnamefont{R.}~\bibnamefont{Zwanzig}}, \bibinfo{journal}{J.
  Phys. Chem.} \textbf{\bibinfo{volume}{96}}, \bibinfo{pages}{3926}
  (\bibinfo{year}{1992}).

\bibitem[{\citenamefont{Reguera et~al.}(2006)\citenamefont{Reguera, Schmid,
  Burada, Rub{\'i}, Reimann, and H{\"a}nggi}}]{Reguera.PRL96}
\bibinfo{author}{\bibfnamefont{D.}~\bibnamefont{Reguera}},
  \bibinfo{author}{\bibfnamefont{G.}~\bibnamefont{Schmid}},
  \bibinfo{author}{\bibfnamefont{P.~S.} \bibnamefont{Burada}},
  \bibinfo{author}{\bibfnamefont{J.~M.} \bibnamefont{Rub{\'i}}},
  \bibinfo{author}{\bibfnamefont{P.}~\bibnamefont{Reimann}}, \bibnamefont{and}
  \bibinfo{author}{\bibfnamefont{P.}~\bibnamefont{H{\"a}nggi}},
  \bibinfo{journal}{Phys. Rev. Lett.} \textbf{\bibinfo{volume}{96}},
  \bibinfo{pages}{130603} (\bibinfo{year}{2006}).

\bibitem[{\citenamefont{Reguera et~al.}(2012)\citenamefont{Reguera, Luque,
  Burada, Schmid, Rub{\'i}, and H{\"a}nggi}}]{Reguera.PRL108}
\bibinfo{author}{\bibfnamefont{D.}~\bibnamefont{Reguera}},
  \bibinfo{author}{\bibfnamefont{A.}~\bibnamefont{Luque}},
  \bibinfo{author}{\bibfnamefont{P.~S.} \bibnamefont{Burada}},
  \bibinfo{author}{\bibfnamefont{G.}~\bibnamefont{Schmid}},
  \bibinfo{author}{\bibfnamefont{J.~M.} \bibnamefont{Rub{\'i}}},
  \bibnamefont{and}
  \bibinfo{author}{\bibfnamefont{P.}~\bibnamefont{H{\"a}nggi}},
  \bibinfo{journal}{Phys. Rev. Lett.} \textbf{\bibinfo{volume}{108}},
  \bibinfo{pages}{020604} (\bibinfo{year}{2012}).

\bibitem[{\citenamefont{Berthier and Biroli}(2011)}]{Berthier.RMP83}
\bibinfo{author}{\bibfnamefont{L.}~\bibnamefont{Berthier}} \bibnamefont{and}
  \bibinfo{author}{\bibfnamefont{G.}~\bibnamefont{Biroli}},
  \bibinfo{journal}{Rev. Mod. Phys.} \textbf{\bibinfo{volume}{83}},
  \bibinfo{pages}{587} (\bibinfo{year}{2011}).

\bibitem[{\citenamefont{Liu and Nagel}(2001)}]{Liu.Book2001}
\bibinfo{author}{\bibfnamefont{A.~J.} \bibnamefont{Liu}} \bibnamefont{and}
  \bibinfo{author}{\bibfnamefont{S.~R.} \bibnamefont{Nagel}},
  \emph{\bibinfo{title}{Jamming and Rheology: Constrained Dynamics on
  Microscopic and Macroscopic Scales}} (\bibinfo{publisher}{Taylor \& Francis},
  \bibinfo{address}{London}, \bibinfo{year}{2001}).

\bibitem[{\citenamefont{G{\"o}tze and Sj{\"o}gren}(1992)}]{Gotze.RPP55}
\bibinfo{author}{\bibfnamefont{W.}~\bibnamefont{G{\"o}tze}} \bibnamefont{and}
  \bibinfo{author}{\bibfnamefont{L.}~\bibnamefont{Sj{\"o}gren}},
  \bibinfo{journal}{Rep. Prog. Phys.} \textbf{\bibinfo{volume}{55}},
  \bibinfo{pages}{241} (\bibinfo{year}{1992}).

\bibitem[{\citenamefont{G{\"o}tze}(2009)}]{Goetze.Book2009}
\bibinfo{author}{\bibfnamefont{W.}~\bibnamefont{G{\"o}tze}},
  \emph{\bibinfo{title}{Complex Dynamics of Glass-Forming Liquids: A
  mode-coupling theory}} (\bibinfo{publisher}{Oxford University Press},
  \bibinfo{address}{New York}, \bibinfo{year}{2009}), ISBN
  \bibinfo{isbn}{978-0-19-923534-6}.

\bibitem[{\citenamefont{Reichman and Charbonneau}(2005)}]{Reichman.JStat2005}
\bibinfo{author}{\bibfnamefont{D.~R.} \bibnamefont{Reichman}} \bibnamefont{and}
  \bibinfo{author}{\bibfnamefont{P.}~\bibnamefont{Charbonneau}},
  \bibinfo{journal}{J. Stat. Mech.} p. \bibinfo{pages}{P05013}
  (\bibinfo{year}{2005}).

\bibitem[{\citenamefont{Yamamoto and Onuki}(1998)}]{Yamamoto.PRE58}
\bibinfo{author}{\bibfnamefont{R.}~\bibnamefont{Yamamoto}} \bibnamefont{and}
  \bibinfo{author}{\bibfnamefont{A.}~\bibnamefont{Onuki}},
  \bibinfo{journal}{Phys. Rev. E} \textbf{\bibinfo{volume}{58}},
  \bibinfo{pages}{3515} (\bibinfo{year}{1998}).

\bibitem[{\citenamefont{Berthier et~al.}(2011)\citenamefont{Berthier, Biroli,
  Bouchaud, Cipelletti, and van Saarloos}}]{Berthier.Book2011}
\bibinfo{editor}{\bibfnamefont{L.}~\bibnamefont{Berthier}},
  \bibinfo{editor}{\bibfnamefont{G.}~\bibnamefont{Biroli}},
  \bibinfo{editor}{\bibfnamefont{J.-P.} \bibnamefont{Bouchaud}},
  \bibinfo{editor}{\bibfnamefont{L.}~\bibnamefont{Cipelletti}},
  \bibnamefont{and} \bibinfo{editor}{\bibfnamefont{W.}~\bibnamefont{van
  Saarloos}}, eds., \emph{\bibinfo{title}{Dynamical Heterogeneities in Glasses,
  Colloids, and Granular Media}} (\bibinfo{publisher}{Oxford University Press},
  \bibinfo{address}{Oxford}, \bibinfo{year}{2011}).

\bibitem[{\citenamefont{Toninelli et~al.}(2005)\citenamefont{Toninelli, Wyart,
  Berthier, Biroli, and Bouchaud}}]{Toninelli.PRE71}
\bibinfo{author}{\bibfnamefont{C.}~\bibnamefont{Toninelli}},
  \bibinfo{author}{\bibfnamefont{M.}~\bibnamefont{Wyart}},
  \bibinfo{author}{\bibfnamefont{L.}~\bibnamefont{Berthier}},
  \bibinfo{author}{\bibfnamefont{G.}~\bibnamefont{Biroli}}, \bibnamefont{and}
  \bibinfo{author}{\bibfnamefont{J.-P.} \bibnamefont{Bouchaud}},
  \bibinfo{journal}{Phys. Rev. E} \textbf{\bibinfo{volume}{71}},
  \bibinfo{pages}{041505} (\bibinfo{year}{2005}).

\bibitem[{\citenamefont{Biroli et~al.}(2006)\citenamefont{Biroli, Bouchaud,
  Miyazaki, and Reichman}}]{Biroli.PRL97}
\bibinfo{author}{\bibfnamefont{G.}~\bibnamefont{Biroli}},
  \bibinfo{author}{\bibfnamefont{J.-P.} \bibnamefont{Bouchaud}},
  \bibinfo{author}{\bibfnamefont{K.}~\bibnamefont{Miyazaki}}, \bibnamefont{and}
  \bibinfo{author}{\bibfnamefont{D.~R.} \bibnamefont{Reichman}},
  \bibinfo{journal}{Phys. Rev. Lett.} \textbf{\bibinfo{volume}{97}},
  \bibinfo{pages}{195701} (\bibinfo{year}{2006}).

\bibitem[{\citenamefont{Harris}(1965)}]{Harris.JAP2}
\bibinfo{author}{\bibfnamefont{T.~E.} \bibnamefont{Harris}},
  \bibinfo{journal}{J. Appl. Probab.} \textbf{\bibinfo{volume}{2}},
  \bibinfo{pages}{323} (\bibinfo{year}{1965}).

\bibitem[{\citenamefont{Jepsen}(1965)}]{Jepsen.JMP6}
\bibinfo{author}{\bibfnamefont{D.~W.} \bibnamefont{Jepsen}},
  \bibinfo{journal}{J. Math. Phys.} \textbf{\bibinfo{volume}{6}},
  \bibinfo{pages}{405} (\bibinfo{year}{1965}).

\bibitem[{\citenamefont{Fedders}(1978)}]{Fedders.PRB17}
\bibinfo{author}{\bibfnamefont{P.~A.} \bibnamefont{Fedders}},
  \bibinfo{journal}{Phys. Rev. B} \textbf{\bibinfo{volume}{17}},
  \bibinfo{pages}{40} (\bibinfo{year}{1978}).

\bibitem[{\citenamefont{Alexander and Pincus}(1978)}]{Alexander.PRB18}
\bibinfo{author}{\bibfnamefont{S.}~\bibnamefont{Alexander}} \bibnamefont{and}
  \bibinfo{author}{\bibfnamefont{P.}~\bibnamefont{Pincus}},
  \bibinfo{journal}{Phys. Rev. B} \textbf{\bibinfo{volume}{18}},
  \bibinfo{pages}{2011} (\bibinfo{year}{1978}).

\bibitem[{\citenamefont{van Beijeren et~al.}(1983)\citenamefont{van Beijeren,
  Kehr, and Kutner}}]{van-Beijeren.PRB28}
\bibinfo{author}{\bibfnamefont{H.}~\bibnamefont{van Beijeren}},
  \bibinfo{author}{\bibfnamefont{K.~W.} \bibnamefont{Kehr}}, \bibnamefont{and}
  \bibinfo{author}{\bibfnamefont{R.}~\bibnamefont{Kutner}},
  \bibinfo{journal}{Phys. Rev. B} \textbf{\bibinfo{volume}{28}},
  \bibinfo{pages}{5711} (\bibinfo{year}{1983}).

\bibitem[{\citenamefont{Hahn and K{\"a}rger}(1995)}]{Hahn.JPhA28}
\bibinfo{author}{\bibfnamefont{K.}~\bibnamefont{Hahn}} \bibnamefont{and}
  \bibinfo{author}{\bibfnamefont{J.}~\bibnamefont{K{\"a}rger}},
  \bibinfo{journal}{J. Phys. A: Math. Gen.} \textbf{\bibinfo{volume}{28}},
  \bibinfo{pages}{3061} (\bibinfo{year}{1995}).

\bibitem[{\citenamefont{Kollmann}(2003)}]{Kollmann.PRL90}
\bibinfo{author}{\bibfnamefont{M.}~\bibnamefont{Kollmann}},
  \bibinfo{journal}{Phys. Rev. Lett.} \textbf{\bibinfo{volume}{90}},
  \bibinfo{pages}{180602} (\bibinfo{year}{2003}).

\bibitem[{\citenamefont{Taloni and Marchesoni}(2006)}]{Taloni.PRL96}
\bibinfo{author}{\bibfnamefont{A.}~\bibnamefont{Taloni}} \bibnamefont{and}
  \bibinfo{author}{\bibfnamefont{F.}~\bibnamefont{Marchesoni}},
  \bibinfo{journal}{Phys. Rev. Lett.} \textbf{\bibinfo{volume}{96}},
  \bibinfo{pages}{020601} (\bibinfo{year}{2006}).

\bibitem[{\citenamefont{Barkai and Silbey}(2009)}]{Barkai.PRL102}
\bibinfo{author}{\bibfnamefont{E.}~\bibnamefont{Barkai}} \bibnamefont{and}
  \bibinfo{author}{\bibfnamefont{R.}~\bibnamefont{Silbey}},
  \bibinfo{journal}{Phys. Rev. Lett.} \textbf{\bibinfo{volume}{102}},
  \bibinfo{pages}{050602} (\bibinfo{year}{2009}).

\bibitem[{\citenamefont{Barkai and Silbey}(2010)}]{Barkai.PRE81}
\bibinfo{author}{\bibfnamefont{E.}~\bibnamefont{Barkai}} \bibnamefont{and}
  \bibinfo{author}{\bibfnamefont{R.}~\bibnamefont{Silbey}},
  \bibinfo{journal}{Phys. Rev. E} \textbf{\bibinfo{volume}{81}},
  \bibinfo{pages}{041129} (\bibinfo{year}{2010}).

\bibitem[{\citenamefont{{{Ooshida}~Takeshi}
  et~al.}(2011)\citenamefont{{{Ooshida}~Takeshi}, Goto, Matsumoto, Nakahara,
  and Otsuki}}]{Ooshida.JPSJ80}
\bibinfo{author}{\bibnamefont{{{Ooshida}~Takeshi}}},
  \bibinfo{author}{\bibfnamefont{S.}~\bibnamefont{Goto}},
  \bibinfo{author}{\bibfnamefont{T.}~\bibnamefont{Matsumoto}},
  \bibinfo{author}{\bibfnamefont{A.}~\bibnamefont{Nakahara}}, \bibnamefont{and}
  \bibinfo{author}{\bibfnamefont{M.}~\bibnamefont{Otsuki}},
  \bibinfo{journal}{J. Phys. Soc. Japan} \textbf{\bibinfo{volume}{80}},
  \bibinfo{pages}{074007} (\bibinfo{year}{2011}).

\bibitem[{\citenamefont{Delfau et~al.}(2012)\citenamefont{Delfau, Coste, and
  {Saint Jean}}}]{Delfau.PRE85}
\bibinfo{author}{\bibfnamefont{J.-B.} \bibnamefont{Delfau}},
  \bibinfo{author}{\bibfnamefont{C.}~\bibnamefont{Coste}}, \bibnamefont{and}
  \bibinfo{author}{\bibfnamefont{M.}~\bibnamefont{{Saint Jean}}},
  \bibinfo{journal}{Phys. Rev. E} \textbf{\bibinfo{volume}{85}},
  \bibinfo{pages}{061111} (\bibinfo{year}{2012}).

\bibitem[{\citenamefont{Rallison}(1988)}]{Rallison.JFM186}
\bibinfo{author}{\bibfnamefont{J.~M.} \bibnamefont{Rallison}},
  \bibinfo{journal}{J. Fluid Mech.} \textbf{\bibinfo{volume}{186}},
  \bibinfo{pages}{471} (\bibinfo{year}{1988}).

\bibitem[{\citenamefont{Lef{\`e}vre et~al.}(2005)\citenamefont{Lef{\`e}vre,
  Berthier, and Stinchcombe}}]{Lefevre.PRE72}
\bibinfo{author}{\bibfnamefont{A.}~\bibnamefont{Lef{\`e}vre}},
  \bibinfo{author}{\bibfnamefont{L.}~\bibnamefont{Berthier}}, \bibnamefont{and}
  \bibinfo{author}{\bibfnamefont{R.}~\bibnamefont{Stinchcombe}},
  \bibinfo{journal}{Phys. Rev. E} \textbf{\bibinfo{volume}{72}},
  \bibinfo{pages}{010301(R)} (\bibinfo{year}{2005}).

\bibitem[{\citenamefont{Abel et~al.}(2009)\citenamefont{Abel, Tse, and
  Andersen}}]{Abel.PNAS106}
\bibinfo{author}{\bibfnamefont{S.~M.} \bibnamefont{Abel}},
  \bibinfo{author}{\bibfnamefont{Y.-L.~S.} \bibnamefont{Tse}},
  \bibnamefont{and} \bibinfo{author}{\bibfnamefont{H.~C.}
  \bibnamefont{Andersen}}, \bibinfo{journal}{Proc. Natl. Acad. Sci. USA}
  \textbf{\bibinfo{volume}{106}}, \bibinfo{pages}{15142}
  (\bibinfo{year}{2009}).

\bibitem[{\citenamefont{Lutz et~al.}(2004)\citenamefont{Lutz, Kollmann, and
  Bechinger}}]{Lutz.PRL93}
\bibinfo{author}{\bibfnamefont{C.}~\bibnamefont{Lutz}},
  \bibinfo{author}{\bibfnamefont{M.}~\bibnamefont{Kollmann}}, \bibnamefont{and}
  \bibinfo{author}{\bibfnamefont{C.}~\bibnamefont{Bechinger}},
  \bibinfo{journal}{Phys. Rev. Lett.} \textbf{\bibinfo{volume}{93}},
  \bibinfo{pages}{026001} (\bibinfo{year}{2004}).

\bibitem[{\citenamefont{Lizana et~al.}(2010)\citenamefont{Lizana,
  Ambj{\"o}rnsson, Taloni, Barkai, and Lomholt}}]{Lizana.PRE81}
\bibinfo{author}{\bibfnamefont{L.}~\bibnamefont{Lizana}},
  \bibinfo{author}{\bibfnamefont{T.}~\bibnamefont{Ambj{\"o}rnsson}},
  \bibinfo{author}{\bibfnamefont{A.}~\bibnamefont{Taloni}},
  \bibinfo{author}{\bibfnamefont{E.}~\bibnamefont{Barkai}}, \bibnamefont{and}
  \bibinfo{author}{\bibfnamefont{M.~A.} \bibnamefont{Lomholt}},
  \bibinfo{journal}{Phys. Rev. E} \textbf{\bibinfo{volume}{81}},
  \bibinfo{pages}{051118} (\bibinfo{year}{2010}).

\bibitem[{\citenamefont{Miyazaki and Yethiraj}(2002)}]{Miyazaki.JCP117}
\bibinfo{author}{\bibfnamefont{K.}~\bibnamefont{Miyazaki}} \bibnamefont{and}
  \bibinfo{author}{\bibfnamefont{A.}~\bibnamefont{Yethiraj}},
  \bibinfo{journal}{J. Chem. Phys.} \textbf{\bibinfo{volume}{117}},
  \bibinfo{pages}{10448} (\bibinfo{year}{2002}).

\bibitem[{\citenamefont{Miyazaki}(2007)}]{Miyazaki.Bussei88}
\bibinfo{author}{\bibfnamefont{K.}~\bibnamefont{Miyazaki}},
  \bibinfo{journal}{Bussei Kenky\^u} \textbf{\bibinfo{volume}{88}},
  \bibinfo{pages}{621} (\bibinfo{year}{2007}), \bibinfo{note}{[in Japanese]}.

\bibitem[{\citenamefont{Miyazaki}(2008)}]{Miyazaki.JPS08s}
\bibinfo{author}{\bibfnamefont{K.}~\bibnamefont{Miyazaki}}, in
  \emph{\bibinfo{booktitle}{Meeting Absrtacts of the Physical Society of Japan
  (2008 Annual Meeting) Part 2}} (\bibinfo{year}{2008}), p.
  \bibinfo{pages}{393}, \bibinfo{note}{26pWE-11 [in Japanese]}.

\bibitem[{\citenamefont{Landau and Lifshitz}(1987)}]{Landau.fluid}
\bibinfo{author}{\bibfnamefont{L.~D.} \bibnamefont{Landau}} \bibnamefont{and}
  \bibinfo{author}{\bibfnamefont{E.~M.} \bibnamefont{Lifshitz}},
  \emph{\bibinfo{title}{Fluid Mechanics}}, vol.~\bibinfo{volume}{6} of
  \emph{\bibinfo{series}{Theoretical Physics}}
  (\bibinfo{publisher}{Butterworth-Heinemann}, \bibinfo{address}{Oxford},
  \bibinfo{year}{1987}).

\bibitem[{\citenamefont{Bennett}(2006)}]{Bennett.Book2006}
\bibinfo{author}{\bibfnamefont{A.}~\bibnamefont{Bennett}},
  \emph{\bibinfo{title}{Lagrangian fluid dynamics}}
  (\bibinfo{publisher}{Cambridge University Press},
  \bibinfo{address}{Cambridge}, \bibinfo{year}{2006}), ISBN
  \bibinfo{isbn}{0-521-85310-9}.

\bibitem[{\citenamefont{Muranaka and Hiwatari}(1995)}]{Muranaka.PRE51}
\bibinfo{author}{\bibfnamefont{T.}~\bibnamefont{Muranaka}} \bibnamefont{and}
  \bibinfo{author}{\bibfnamefont{Y.}~\bibnamefont{Hiwatari}},
  \bibinfo{journal}{Phys. Rev. E} \textbf{\bibinfo{volume}{51}},
  \bibinfo{pages}{R2735} (\bibinfo{year}{1995}).

\bibitem[{\citenamefont{Hiwatari and Muranaka}(1998)}]{Hiwatari.JNCS235}
\bibinfo{author}{\bibfnamefont{Y.}~\bibnamefont{Hiwatari}} \bibnamefont{and}
  \bibinfo{author}{\bibfnamefont{T.}~\bibnamefont{Muranaka}},
  \bibinfo{journal}{J. Non-Cryst. Solids} \textbf{\bibinfo{volume}{235-237}},
  \bibinfo{pages}{19} (\bibinfo{year}{1998}).

\bibitem[{\citenamefont{Donati et~al.}(1999)\citenamefont{Donati, Glotzer, and
  Poole}}]{Donati.PRL82}
\bibinfo{author}{\bibfnamefont{C.}~\bibnamefont{Donati}},
  \bibinfo{author}{\bibfnamefont{S.~C.} \bibnamefont{Glotzer}},
  \bibnamefont{and} \bibinfo{author}{\bibfnamefont{P.~H.} \bibnamefont{Poole}},
  \bibinfo{journal}{Phys. Rev. Lett.} \textbf{\bibinfo{volume}{82}},
  \bibinfo{pages}{5064} (\bibinfo{year}{1999}).

\bibitem[{\citenamefont{Doliwa and Heuer}(2000)}]{Doliwa.PRE61}
\bibinfo{author}{\bibfnamefont{B.}~\bibnamefont{Doliwa}} \bibnamefont{and}
  \bibinfo{author}{\bibfnamefont{A.}~\bibnamefont{Heuer}},
  \bibinfo{journal}{Phys. Rev. E} \textbf{\bibinfo{volume}{61}},
  \bibinfo{pages}{6898} (\bibinfo{year}{2000}).

\bibitem[{\citenamefont{Taloni et~al.}(2010)\citenamefont{Taloni, Chechkin, and
  Klafter}}]{Taloni.PRE82}
\bibinfo{author}{\bibfnamefont{A.}~\bibnamefont{Taloni}},
  \bibinfo{author}{\bibfnamefont{A.}~\bibnamefont{Chechkin}}, \bibnamefont{and}
  \bibinfo{author}{\bibfnamefont{J.}~\bibnamefont{Klafter}},
  \bibinfo{journal}{Phys. Rev. E} \textbf{\bibinfo{volume}{82}},
  \bibinfo{pages}{061104} (\bibinfo{year}{2010}).

\bibitem[{\citenamefont{Taloni et~al.}(2012)\citenamefont{Taloni, Chechkin, and
  Klafter}}]{Taloni.EPL97}
\bibinfo{author}{\bibfnamefont{A.}~\bibnamefont{Taloni}},
  \bibinfo{author}{\bibfnamefont{A.}~\bibnamefont{Chechkin}}, \bibnamefont{and}
  \bibinfo{author}{\bibfnamefont{J.}~\bibnamefont{Klafter}},
  \bibinfo{journal}{Europhys. Lett.} \textbf{\bibinfo{volume}{97}},
  \bibinfo{pages}{30001} (\bibinfo{year}{2012}).

\bibitem[{\citenamefont{Dean}(1996)}]{Dean.JPhAMG29}
\bibinfo{author}{\bibfnamefont{D.~S.} \bibnamefont{Dean}}, \bibinfo{journal}{J.
  Phys. A: Math. Gen.} \textbf{\bibinfo{volume}{29}}, \bibinfo{pages}{L613}
  (\bibinfo{year}{1996}).

\bibitem[{\citenamefont{Miyazaki and Reichman}(2005)}]{Miyazaki.JPA38}
\bibinfo{author}{\bibfnamefont{K.}~\bibnamefont{Miyazaki}} \bibnamefont{and}
  \bibinfo{author}{\bibfnamefont{D.~R.} \bibnamefont{Reichman}},
  \bibinfo{journal}{J. Phys. A: Math. Gen.} \textbf{\bibinfo{volume}{38}},
  \bibinfo{pages}{L343} (\bibinfo{year}{2005}).

\bibitem[{\citenamefont{Andreanov et~al.}(2006)\citenamefont{Andreanov, Biroli,
  and Lef{\`e}vre}}]{Andreanov.JStat2006}
\bibinfo{author}{\bibfnamefont{A.}~\bibnamefont{Andreanov}},
  \bibinfo{author}{\bibfnamefont{G.}~\bibnamefont{Biroli}}, \bibnamefont{and}
  \bibinfo{author}{\bibfnamefont{A.}~\bibnamefont{Lef{\`e}vre}},
  \bibinfo{journal}{J. Stat. Mech.} p. \bibinfo{pages}{P07008}
  (\bibinfo{year}{2006}).

\bibitem[{\citenamefont{Kim and Kawasaki}(2007)}]{Kim.JPhA40}
\bibinfo{author}{\bibfnamefont{B.}~\bibnamefont{Kim}} \bibnamefont{and}
  \bibinfo{author}{\bibfnamefont{K.}~\bibnamefont{Kawasaki}},
  \bibinfo{journal}{J. Phys. A: Math. Theor.} \textbf{\bibinfo{volume}{40}},
  \bibinfo{pages}{F33} (\bibinfo{year}{2007}).

\bibitem[{\citenamefont{Kim and Kawasaki}(2008)}]{Kim.JStat2008}
\bibinfo{author}{\bibfnamefont{B.}~\bibnamefont{Kim}} \bibnamefont{and}
  \bibinfo{author}{\bibfnamefont{K.}~\bibnamefont{Kawasaki}},
  \bibinfo{journal}{J. Stat. Mech.} p. \bibinfo{pages}{P02004}
  (\bibinfo{year}{2008}).

\bibitem[{\citenamefont{van Kampen}(2007)}]{van-Kampen.Book2007}
\bibinfo{author}{\bibfnamefont{N.~G.} \bibnamefont{van Kampen}},
  \emph{\bibinfo{title}{Stochastic processes in physics and chemistry}}
  (\bibinfo{publisher}{Elsevier}, \bibinfo{year}{2007}), \bibinfo{edition}{3rd}
  ed.

\bibitem[{\citenamefont{Taloni and Lomholt}(2008)}]{Taloni.PRE78}
\bibinfo{author}{\bibfnamefont{A.}~\bibnamefont{Taloni}} \bibnamefont{and}
  \bibinfo{author}{\bibfnamefont{M.~A.} \bibnamefont{Lomholt}},
  \bibinfo{journal}{Phys. Rev. E} \textbf{\bibinfo{volume}{78}},
  \bibinfo{pages}{051116} (\bibinfo{year}{2008}).

\bibitem[{not({\natexlab{a}})}]{note.fRho}
\bibinfo{note}{When comparing Eq.~(\ref{f2}) with Eq.~(6) in
  Ref.~\cite{Taloni.PRE78}, care should be taken of the difference in the
  definition of the random force: $\dx$ is included in our definition of
  $\fRho$, which corresponds to the divergence of the noise term in
  Ref.~\cite{Taloni.PRE78}.}

\bibitem[{\citenamefont{Kawasaki}(1994)}]{Kawasaki.PhysicaA208}
\bibinfo{author}{\bibfnamefont{K.}~\bibnamefont{Kawasaki}},
  \bibinfo{journal}{Physica A} \textbf{\bibinfo{volume}{208}},
  \bibinfo{pages}{36} (\bibinfo{year}{1994}).

\bibitem[{\citenamefont{Kawasaki}(1998)}]{Kawasaki.JStatPhys93}
\bibinfo{author}{\bibfnamefont{K.}~\bibnamefont{Kawasaki}},
  \bibinfo{journal}{Journal of Statistical Physics}
  \textbf{\bibinfo{volume}{93}}, \bibinfo{pages}{527} (\bibinfo{year}{1998}).

\bibitem[{\citenamefont{Krakoviack}(2009)}]{Krakoviack.PRE79}
\bibinfo{author}{\bibfnamefont{V.}~\bibnamefont{Krakoviack}},
  \bibinfo{journal}{Phys. Rev. E} \textbf{\bibinfo{volume}{79}},
  \bibinfo{pages}{061501} (\bibinfo{year}{2009}).

\bibitem[{\citenamefont{Schnyder et~al.}(2011)\citenamefont{Schnyder,
  H{\"o}fling, Franosch, and Voigtmann}}]{Schnyder.JPhysCM23}
\bibinfo{author}{\bibfnamefont{S.~K.} \bibnamefont{Schnyder}},
  \bibinfo{author}{\bibfnamefont{F.}~\bibnamefont{H{\"o}fling}},
  \bibinfo{author}{\bibfnamefont{T.}~\bibnamefont{Franosch}}, \bibnamefont{and}
  \bibinfo{author}{\bibfnamefont{T.}~\bibnamefont{Voigtmann}},
  \bibinfo{journal}{Journal of Physics: Consensed Matters}
  \textbf{\bibinfo{volume}{23}}, \bibinfo{pages}{234121}
  (\bibinfo{year}{2011}).

\bibitem[{\citenamefont{Kraichnan}(1965)}]{Kraichnan.PhF8}
\bibinfo{author}{\bibfnamefont{R.~H.} \bibnamefont{Kraichnan}},
  \bibinfo{journal}{Physics of Fluids} \textbf{\bibinfo{volume}{8}},
  \bibinfo{pages}{575} (\bibinfo{year}{1965}).

\bibitem[{\citenamefont{Kaneda}(1981)}]{Kaneda.JFM107}
\bibinfo{author}{\bibfnamefont{Y.}~\bibnamefont{Kaneda}}, \bibinfo{journal}{J.
  Fluid Mech.} \textbf{\bibinfo{volume}{107}}, \bibinfo{pages}{131}
  (\bibinfo{year}{1981}).

\bibitem[{\citenamefont{Kida and Goto}(1997)}]{Kida.JFM345}
\bibinfo{author}{\bibfnamefont{S.}~\bibnamefont{Kida}} \bibnamefont{and}
  \bibinfo{author}{\bibfnamefont{S.}~\bibnamefont{Goto}}, \bibinfo{journal}{J.
  Fluid Mech.} \textbf{\bibinfo{volume}{345}}, \bibinfo{pages}{307}
  (\bibinfo{year}{1997}).

\bibitem[{\citenamefont{Spivak}(1965)}]{Spivak.Book1965}
\bibinfo{author}{\bibfnamefont{M.}~\bibnamefont{Spivak}},
  \emph{\bibinfo{title}{Calculus on Manifolds: A Modern Approach to Classical
  Theorems of Advanced Calculus}} (\bibinfo{publisher}{W.A. Benjamin Inc.},
  \bibinfo{address}{New York}, \bibinfo{year}{1965}), ISBN
  \bibinfo{isbn}{0-8053-9021-9}.

\bibitem[{not({\natexlab{b}})}]{note.triad}
\bibinfo{note}{It is sometimes more convenient to regard the triad summation,
  such as the one in Eq.~(\ref{*3+}), simply as $\sum_p\sum_q$ with
  $\Vertex_k^{pq}$ given by Eq.~(\ref{vertex}) if the triad condition is
  satisfied and otherwise $\Vertex_k^{pq} = 0$.}

\bibitem[{not({\natexlab{c}})}]{note.f4}
\bibinfo{note}{Terms containing $\delta_{k,k'\pm{N}}, \delta_{k,k'\pm{2N}},
  \ldots$ are omitted from Eq.~(\ref{f4}), as they are not important for the
  behavior of the long-wave modes.}

\bibitem[{\citenamefont{Delfau et~al.}(2011)\citenamefont{Delfau, Coste, and
  {Saint Jean}}}]{Delfau.PRE84}
\bibinfo{author}{\bibfnamefont{J.-B.} \bibnamefont{Delfau}},
  \bibinfo{author}{\bibfnamefont{C.}~\bibnamefont{Coste}}, \bibnamefont{and}
  \bibinfo{author}{\bibfnamefont{M.}~\bibnamefont{{Saint Jean}}},
  \bibinfo{journal}{Phys. Rev. E} \textbf{\bibinfo{volume}{84}},
  \bibinfo{pages}{011101} (\bibinfo{year}{2011}).

\bibitem[{\citenamefont{Edwards and Wilkinson}(1982)}]{Edwards.PRSLA381}
\bibinfo{author}{\bibfnamefont{S.~F.} \bibnamefont{Edwards}} \bibnamefont{and}
  \bibinfo{author}{\bibfnamefont{D.~R.} \bibnamefont{Wilkinson}},
  \bibinfo{journal}{Proc. R. Soc. London, Ser. A}
  \textbf{\bibinfo{volume}{381}}, \bibinfo{pages}{17} (\bibinfo{year}{1982}).

\bibitem[{\citenamefont{Doi and Edwards}(1986)}]{Doi.Book1986}
\bibinfo{author}{\bibfnamefont{M.}~\bibnamefont{Doi}} \bibnamefont{and}
  \bibinfo{author}{\bibfnamefont{S.~F.} \bibnamefont{Edwards}},
  \emph{\bibinfo{title}{The Theory of Polymer Dynamics}}
  (\bibinfo{publisher}{Oxford}, \bibinfo{year}{1986}).

\bibitem[{not({\natexlab{d}})}]{note.factor-C}
\bibinfo{note}{The factor $L^{-2}$ is included in the definition of $\C$ so
  that $\C$ has the dimension of the square of the density. It is a matter of
  choice whether to include it or not.}

\bibitem[{\citenamefont{Honda}(1997)}]{Honda.PRE55}
\bibinfo{author}{\bibfnamefont{K.}~\bibnamefont{Honda}},
  \bibinfo{journal}{Phys. Rev. E} \textbf{\bibinfo{volume}{55}},
  \bibinfo{pages}{R1235} (\bibinfo{year}{1997}).

\bibitem[{\citenamefont{Ikeda et~al.}(2013)\citenamefont{Ikeda, Berthier, and
  Biroli}}]{Ikeda.JCP138}
\bibinfo{author}{\bibfnamefont{A.}~\bibnamefont{Ikeda}},
  \bibinfo{author}{\bibfnamefont{L.}~\bibnamefont{Berthier}}, \bibnamefont{and}
  \bibinfo{author}{\bibfnamefont{G.}~\bibnamefont{Biroli}},
  \bibinfo{journal}{J. Chem. Phys.} \textbf{\bibinfo{volume}{138}},
  \bibinfo{pages}{12A507} (\bibinfo{year}{2013}).

\bibitem[{\citenamefont{Krug}(1997)}]{Krug.AdvPhys46}
\bibinfo{author}{\bibfnamefont{J.}~\bibnamefont{Krug}}, \bibinfo{journal}{Adv.
  in Phys.} \textbf{\bibinfo{volume}{46}}, \bibinfo{pages}{139}
  (\bibinfo{year}{1997}).

\bibitem[{\citenamefont{Das and Mazenko}(1986)}]{Das.PRA34}
\bibinfo{author}{\bibfnamefont{S.~P.} \bibnamefont{Das}} \bibnamefont{and}
  \bibinfo{author}{\bibfnamefont{G.~F.} \bibnamefont{Mazenko}},
  \bibinfo{journal}{Phys. Rev. A} \textbf{\bibinfo{volume}{34}}
  (\bibinfo{year}{1986}).

\bibitem[{\citenamefont{Schmitz et~al.}(1993)\citenamefont{Schmitz, Dufty, and
  De}}]{Schmitz.PRL71}
\bibinfo{author}{\bibfnamefont{R.}~\bibnamefont{Schmitz}},
  \bibinfo{author}{\bibfnamefont{J.~W.} \bibnamefont{Dufty}}, \bibnamefont{and}
  \bibinfo{author}{\bibfnamefont{P.}~\bibnamefont{De}}, \bibinfo{journal}{Phys.
  Rev. Lett.} \textbf{\bibinfo{volume}{71}}, \bibinfo{pages}{2066}
  (\bibinfo{year}{1993}).

\bibitem[{\citenamefont{Martin et~al.}(1973)\citenamefont{Martin, Siggia, and
  Rose}}]{Martin.PRA8}
\bibinfo{author}{\bibfnamefont{P.~C.} \bibnamefont{Martin}},
  \bibinfo{author}{\bibfnamefont{E.~D.} \bibnamefont{Siggia}},
  \bibnamefont{and} \bibinfo{author}{\bibfnamefont{H.~A.} \bibnamefont{Rose}},
  \bibinfo{journal}{Phys. Rev. A} \textbf{\bibinfo{volume}{8}},
  \bibinfo{pages}{423} (\bibinfo{year}{1973}).

\bibitem[{\citenamefont{Nishino and Hayakawa}(2008)}]{Nishino.PRE78}
\bibinfo{author}{\bibfnamefont{T.~H.} \bibnamefont{Nishino}} \bibnamefont{and}
  \bibinfo{author}{\bibfnamefont{H.}~\bibnamefont{Hayakawa}},
  \bibinfo{journal}{Phys. Rev. E} \textbf{\bibinfo{volume}{78}},
  \bibinfo{pages}{061502} (\bibinfo{year}{2008}).

\bibitem[{\citenamefont{Kraichnan}(1959)}]{Kraichnan.JFM5}
\bibinfo{author}{\bibfnamefont{R.~H.} \bibnamefont{Kraichnan}},
  \bibinfo{journal}{J. Fluid Mech.} \textbf{\bibinfo{volume}{5}},
  \bibinfo{pages}{497} (\bibinfo{year}{1959}).

\bibitem[{\citenamefont{Goto and Kida}(1998)}]{Goto.PhysicaD117}
\bibinfo{author}{\bibfnamefont{S.}~\bibnamefont{Goto}} \bibnamefont{and}
  \bibinfo{author}{\bibfnamefont{S.}~\bibnamefont{Kida}},
  \bibinfo{journal}{Physica D} \textbf{\bibinfo{volume}{117}},
  \bibinfo{pages}{191} (\bibinfo{year}{1998}).

\bibitem[{\citenamefont{{{Ooshida}~Takeshi}
  et~al.}(2012{\natexlab{a}})\citenamefont{{{Ooshida}~Takeshi}, Goto,
  Matsumoto, Nakahara, and Otsuki}}]{Matsumoto.PTPS195}
\bibinfo{author}{\bibnamefont{{{Ooshida}~Takeshi}}},
  \bibinfo{author}{\bibfnamefont{S.}~\bibnamefont{Goto}},
  \bibinfo{author}{\bibfnamefont{T.}~\bibnamefont{Matsumoto}},
  \bibinfo{author}{\bibfnamefont{A.}~\bibnamefont{Nakahara}}, \bibnamefont{and}
  \bibinfo{author}{\bibfnamefont{M.}~\bibnamefont{Otsuki}},
  \bibinfo{journal}{Progress of Theoretical Physics Supplement}
  \textbf{\bibinfo{volume}{195}}, \bibinfo{pages}{157}
  (\bibinfo{year}{2012}{\natexlab{a}}).

\bibitem[{\citenamefont{Marconi et~al.}(2008)\citenamefont{Marconi, Puglisi,
  Rondoni, and Vulpiani}}]{Marconi.FR461}
\bibinfo{author}{\bibfnamefont{U.~M.~B.} \bibnamefont{Marconi}},
  \bibinfo{author}{\bibfnamefont{A.}~\bibnamefont{Puglisi}},
  \bibinfo{author}{\bibfnamefont{L.}~\bibnamefont{Rondoni}}, \bibnamefont{and}
  \bibinfo{author}{\bibfnamefont{A.}~\bibnamefont{Vulpiani}},
  \bibinfo{journal}{Physics Reports} \textbf{\bibinfo{volume}{461}},
  \bibinfo{pages}{111} (\bibinfo{year}{2008}).

\bibitem[{\citenamefont{Risken}(1996)}]{Risken.Book1996}
\bibinfo{author}{\bibfnamefont{H.}~\bibnamefont{Risken}},
  \emph{\bibinfo{title}{The {Fokker-Planck} equation: methods of solution and
  applications}} (\bibinfo{publisher}{Springer}, \bibinfo{year}{1996}),
  \bibinfo{edition}{2nd} ed.

\bibitem[{not({\natexlab{e}})}]{note.self}
\bibinfo{note}{In the case of Navier--Stokes turbulence, the self-interaction
  is assured to vanish due to the incompressibility condition.}

\bibitem[{\citenamefont{Krakoviack}(2007)}]{Krakoviack.PRE75}
\bibinfo{author}{\bibfnamefont{V.}~\bibnamefont{Krakoviack}},
  \bibinfo{journal}{Phys. Rev. E} \textbf{\bibinfo{volume}{75}},
  \bibinfo{pages}{031503} (\bibinfo{year}{2007}).

\bibitem[{\citenamefont{Percus}(1974)}]{Percus.PRA9}
\bibinfo{author}{\bibfnamefont{J.~K.} \bibnamefont{Percus}},
  \bibinfo{journal}{Phys. Rev. A} \textbf{\bibinfo{volume}{9}},
  \bibinfo{pages}{557} (\bibinfo{year}{1974}).

\bibitem[{\citenamefont{Glotzer et~al.}(2000)\citenamefont{Glotzer, Novikov,
  and Schr{\o}der}}]{Glotzer.JCP112}
\bibinfo{author}{\bibfnamefont{S.~C.} \bibnamefont{Glotzer}},
  \bibinfo{author}{\bibfnamefont{V.~N.} \bibnamefont{Novikov}},
  \bibnamefont{and} \bibinfo{author}{\bibfnamefont{T.~B.}
  \bibnamefont{Schr{\o}der}}, \bibinfo{journal}{J. Chem. Phys.}
  \textbf{\bibinfo{volume}{112}}, \bibinfo{pages}{509} (\bibinfo{year}{2000}).

\bibitem[{\citenamefont{{La\v{c}evi\'c}
  et~al.}(2003)\citenamefont{{La\v{c}evi\'c}, Starr, Schr{\o}der, and
  Glotzer}}]{Lacevic.JCP119}
\bibinfo{author}{\bibfnamefont{N.}~\bibnamefont{{La\v{c}evi\'c}}},
  \bibinfo{author}{\bibfnamefont{F.~W.} \bibnamefont{Starr}},
  \bibinfo{author}{\bibfnamefont{T.~B.} \bibnamefont{Schr{\o}der}},
  \bibnamefont{and} \bibinfo{author}{\bibfnamefont{S.~C.}
  \bibnamefont{Glotzer}}, \bibinfo{journal}{J. Chem. Phys.}
  \textbf{\bibinfo{volume}{119}}, \bibinfo{pages}{7372} (\bibinfo{year}{2003}).

\bibitem[{\citenamefont{Berthier}(2004)}]{Berthier.PRE69}
\bibinfo{author}{\bibfnamefont{L.}~\bibnamefont{Berthier}},
  \bibinfo{journal}{Phys. Rev. E} \textbf{\bibinfo{volume}{69}},
  \bibinfo{pages}{020201(R)} (\bibinfo{year}{2004}).

\bibitem[{\citenamefont{Dasgupta et~al.}(1991)\citenamefont{Dasgupta, Indrani,
  Ramaswamy, and Phani}}]{Dasgupta.EPL15}
\bibinfo{author}{\bibfnamefont{C.}~\bibnamefont{Dasgupta}},
  \bibinfo{author}{\bibfnamefont{A.~V.} \bibnamefont{Indrani}},
  \bibinfo{author}{\bibfnamefont{S.}~\bibnamefont{Ramaswamy}},
  \bibnamefont{and} \bibinfo{author}{\bibfnamefont{M.~K.} \bibnamefont{Phani}},
  \bibinfo{journal}{Europhys. Lett.} \textbf{\bibinfo{volume}{15}},
  \bibinfo{pages}{307} (\bibinfo{year}{1991}).

\bibitem[{\citenamefont{Dauchot et~al.}(2005)\citenamefont{Dauchot, Marty, and
  Biroli}}]{Dauchot.PRL95}
\bibinfo{author}{\bibfnamefont{O.}~\bibnamefont{Dauchot}},
  \bibinfo{author}{\bibfnamefont{G.}~\bibnamefont{Marty}}, \bibnamefont{and}
  \bibinfo{author}{\bibfnamefont{G.}~\bibnamefont{Biroli}},
  \bibinfo{journal}{Phys. Rev. Lett.} \textbf{\bibinfo{volume}{95}},
  \bibinfo{pages}{265701} (\bibinfo{year}{2005}).

\bibitem[{\citenamefont{Lechenault et~al.}(2008)\citenamefont{Lechenault,
  Dauchot, Biroli, and Bouchaud}}]{Lechenault.EPL83a}
\bibinfo{author}{\bibfnamefont{F.}~\bibnamefont{Lechenault}},
  \bibinfo{author}{\bibfnamefont{O.}~\bibnamefont{Dauchot}},
  \bibinfo{author}{\bibfnamefont{G.}~\bibnamefont{Biroli}}, \bibnamefont{and}
  \bibinfo{author}{\bibfnamefont{J.~P.} \bibnamefont{Bouchaud}},
  \bibinfo{journal}{Europhys. Lett.} \textbf{\bibinfo{volume}{83}},
  \bibinfo{pages}{46002} (\bibinfo{year}{2008}).

\bibitem[{\citenamefont{Shiba et~al.}()\citenamefont{Shiba, Kawasaki, and
  Onuki}}]{Shiba.arXiv1205}
\bibinfo{author}{\bibfnamefont{H.}~\bibnamefont{Shiba}},
  \bibinfo{author}{\bibfnamefont{T.}~\bibnamefont{Kawasaki}}, \bibnamefont{and}
  \bibinfo{author}{\bibfnamefont{A.}~\bibnamefont{Onuki}},
  \bibinfo{note}{{arXiv:1205.6090}}.

\bibitem[{not({\natexlab{f}})}]{note.chiS}
\bibinfo{note}{Though it may be more appropriate to denote the $\Qs$-based
  $\chi_4$ with $\chi_4^{\Qs}$ following the notation of Dauchot \textit{et
  al.} \cite{Dauchot.PRL95}, here we have chosen to write $\chiS$. This is
  actually a somewhat arbitrary choice, which we made simply because it is more
  friendly to the eyes.}

\bibitem[{\citenamefont{Levitt}(1973)}]{Levitt.PRA8}
\bibinfo{author}{\bibfnamefont{D.~G.} \bibnamefont{Levitt}},
  \bibinfo{journal}{Phys. Rev. A} \textbf{\bibinfo{volume}{8}},
  \bibinfo{pages}{3050} (\bibinfo{year}{1973}).

\bibitem[{\citenamefont{K{\"a}rger}(1992)}]{Kaerger.PRA45}
\bibinfo{author}{\bibfnamefont{J.}~\bibnamefont{K{\"a}rger}},
  \bibinfo{journal}{Phys. Rev. A} \textbf{\bibinfo{volume}{45}},
  \bibinfo{pages}{4173} (\bibinfo{year}{1992}).

\bibitem[{\citenamefont{Glarum}(1960)}]{Glarum.JCP33}
\bibinfo{author}{\bibfnamefont{S.~H.} \bibnamefont{Glarum}},
  \bibinfo{journal}{J. Chem. Phys.} \textbf{\bibinfo{volume}{33}},
  \bibinfo{pages}{639} (\bibinfo{year}{1960}).

\bibitem[{\citenamefont{Sellitto and Arenzon}(2000)}]{Sellitto.PRE62}
\bibinfo{author}{\bibfnamefont{M.}~\bibnamefont{Sellitto}} \bibnamefont{and}
  \bibinfo{author}{\bibfnamefont{J.~J.} \bibnamefont{Arenzon}},
  \bibinfo{journal}{Phys. Rev. E} \textbf{\bibinfo{volume}{62}},
  \bibinfo{pages}{7793} (\bibinfo{year}{2000}).

\bibitem[{\citenamefont{Mori}(1965)}]{Mori.PTP33}
\bibinfo{author}{\bibfnamefont{H.}~\bibnamefont{Mori}}, \bibinfo{journal}{Prog.
  Theor. Phys.} \textbf{\bibinfo{volume}{33}}, \bibinfo{pages}{423}
  (\bibinfo{year}{1965}).

\bibitem[{\citenamefont{Kawasaki and Onuki}()}]{Kawasaki.arXiv1210}
\bibinfo{author}{\bibfnamefont{T.}~\bibnamefont{Kawasaki}} \bibnamefont{and}
  \bibinfo{author}{\bibfnamefont{A.}~\bibnamefont{Onuki}},
  \bibinfo{note}{{arXiv:1210.0369v2}}.

\bibitem[{\citenamefont{Chertkov et~al.}(1999)\citenamefont{Chertkov, Pumir,
  and Shraiman}}]{Chertkov.PhF11}
\bibinfo{author}{\bibfnamefont{M.}~\bibnamefont{Chertkov}},
  \bibinfo{author}{\bibfnamefont{A.}~\bibnamefont{Pumir}}, \bibnamefont{and}
  \bibinfo{author}{\bibfnamefont{B.~I.} \bibnamefont{Shraiman}},
  \bibinfo{journal}{Physics of Fluids} \textbf{\bibinfo{volume}{11}},
  \bibinfo{pages}{2394} (\bibinfo{year}{1999}).

\bibitem[{\citenamefont{{{Ooshida}~Takeshi}
  et~al.}(2012{\natexlab{b}})\citenamefont{{{Ooshida}~Takeshi}, Otsuki, Goto,
  Nakahara, and Matsumoto}}]{Ooshida.Nagare31}
\bibinfo{author}{\bibnamefont{{{Ooshida}~Takeshi}}},
  \bibinfo{author}{\bibfnamefont{M.}~\bibnamefont{Otsuki}},
  \bibinfo{author}{\bibfnamefont{S.}~\bibnamefont{Goto}},
  \bibinfo{author}{\bibfnamefont{A.}~\bibnamefont{Nakahara}}, \bibnamefont{and}
  \bibinfo{author}{\bibfnamefont{T.}~\bibnamefont{Matsumoto}},
  \bibinfo{journal}{Nagare} \textbf{\bibinfo{volume}{31}}
  (\bibinfo{year}{2012}{\natexlab{b}}), \bibinfo{note}{[in Japanese]}.

\bibitem[{\citenamefont{Matsukawa and Fukuyama}(1994)}]{Matsukawa.PRB49}
\bibinfo{author}{\bibfnamefont{H.}~\bibnamefont{Matsukawa}} \bibnamefont{and}
  \bibinfo{author}{\bibfnamefont{H.}~\bibnamefont{Fukuyama}},
  \bibinfo{journal}{Phys. Rev. B} \textbf{\bibinfo{volume}{49}},
  \bibinfo{pages}{17286} (\bibinfo{year}{1994}).

\bibitem[{\citenamefont{Yoshino
  et~al.}(2010{\natexlab{a}})\citenamefont{Yoshino, Nogawa, and
  Kim}}]{Yoshino.PTPS184}
\bibinfo{author}{\bibfnamefont{H.}~\bibnamefont{Yoshino}},
  \bibinfo{author}{\bibfnamefont{T.}~\bibnamefont{Nogawa}}, \bibnamefont{and}
  \bibinfo{author}{\bibfnamefont{B.}~\bibnamefont{Kim}},
  \bibinfo{journal}{Progress of Theoretical Physics Supplement}
  \textbf{\bibinfo{volume}{184}}, \bibinfo{pages}{153}
  (\bibinfo{year}{2010}{\natexlab{a}}).

\bibitem[{\citenamefont{Mondal and Sengupta}(2012)}]{Mondal.PRE85}
\bibinfo{author}{\bibfnamefont{C.}~\bibnamefont{Mondal}} \bibnamefont{and}
  \bibinfo{author}{\bibfnamefont{S.}~\bibnamefont{Sengupta}},
  \bibinfo{journal}{Phys. Rev. E} \textbf{\bibinfo{volume}{85}},
  \bibinfo{pages}{020402(R)} (\bibinfo{year}{2012}).

\bibitem[{\citenamefont{Yoshino
  et~al.}(2010{\natexlab{b}})\citenamefont{Yoshino, Nogawa, and
  Kim}}]{Yoshino.PRL105}
\bibinfo{author}{\bibfnamefont{H.}~\bibnamefont{Yoshino}},
  \bibinfo{author}{\bibfnamefont{T.}~\bibnamefont{Nogawa}}, \bibnamefont{and}
  \bibinfo{author}{\bibfnamefont{B.}~\bibnamefont{Kim}},
  \bibinfo{journal}{Phys. Rev. Lett.} \textbf{\bibinfo{volume}{105}},
  \bibinfo{pages}{257004} (\bibinfo{year}{2010}{\natexlab{b}}).

\end{thebibliography}


\end{document}